\documentclass[seceqn]{elsart}
\usepackage{epic}
\usepackage{amsmath,amssymb}
\usepackage[all]{xy}

\def\l{\left}			
\def\r{\right}			
\def\lang{\left\langle}		
\def\rang{\right\rangle}	
\def\lPB{\left\{}		
\def\rPB{\right\}}		
\def\lpb{\left[\,}		
\def\rpb{\,\right]}		
\def\com{\,{\mathchar"213B}\,}	
\def\dcom{{\mathchar"213B}\,}	
\def\imi{{\rm i}}		
\def\ldef{\mathrel{\raisebox{.069ex}{:}\!\!=}}
\def\rdef{\mathrel{=\!\!\raisebox{.069ex}{:}}}
\def\pd{\partial}		
\def\fd{\delta}			
\def\grad{\nabla}		
\def\lapl{\grad^2}		
\def\invlapl{{\grad^{-2}}}	
\def\heavyside{\Theta}		
\def\matzero{\makebox[1.3em]{0}}
\def\matone{\makebox[1.3em]{1}}

\def\LieA{\mathfrak{g}}		
\def\LieAx{\mathfrak{h}}	
\def\LieAxb{\mathfrak{a}}	
\def\SOthree{SO(3)}		
\def\SOtwoone{SO(2,1)}		
\def\sothree{so(3)}		
\def\vs{V}			
\def\reals{{\Rset}}		
\def\field{K}			
\def\fv{\xi}			
\def\fdomain{\Omega}		
\def\magfcont{\Psi}		
\def\vort{\omega}		
\def\magf{\psi}			
\def\elecp{\phi}		
\def\ecurrent{J}		
\def\streamf{\phi}		
\def\pvel{v}			
\def\pres{p}			
\def\crmhdbeta{{\beta_\mathrm{i}}}
\def\W{W}			
\def\Wt{{\smash{\mbox{$\widetilde\W$}}\!\mskip.8\thinmuskip}}
\def\Wb{{\smash{\mbox{$\overline\W$}}\!\mskip.8\thinmuskip}}
\def\Wn{g}			
\def\Wni{g}			
\def\Wcob{V}			
\def\coW{A}			
\def\Proj{P}			
\def\ww{w}			
\def\Nilb{N}			
\def\M{M}			
\def\Cas{C}			
\def\Casi{\mathcal{C}}		
\def\CasiSD{\mathcal{C}_{\rm sd}}
\def\mm{m}			
\def\zerorone{\theta}		
\def\ev{\Lambda}		
\def\evone{\zerorone}		
\def\xv{{\bf x}}		
\def\ad{{\rm ad}}		
\def\range{{\rm range\, }}	
\def\pmz{\lambda}		
\def\cycperm{\mathrm{cyc.\ perm.}}
\def\ag{\mathcal{D}}		
\def\agc{D}			
\def\ae{\mathcal{E}}		
\def\af{f}			
\def\afi{f}			
\def\afii{g}			
\def\afiii{h}			
\def\afiv{k}			
\def\afsd{f}			

\global\let\ifmypprint\iffalse
\def\mypprint{\global\let\ifmypprint\iftrue}

\newcommand{\comment}[1]{}
\newcommand{\qcomment}[1]{\ifmypprint{({\it *** #1})\marginpar{???}}\else{}\fi}
\newcommand{\inthesis}[1]{}

\newcommand{\eqlabel}[1]{\label{eq:#1}}
\newcommand{\seclabel}[1]{\label{sec:#1}}
\newcommand{\apxlabel}[1]{\label{apx:#1}}

\newcommand{\tablabel}[1]{\label{table:#1}}
\newcommand{\caselabel}[1]{\label{case:#1}}

\renewcommand{\eqref}[1]{(\ref{eq:#1})}
\newcommand{\apxref}[1]{Appendix~\ref{apx:#1}}
\newcommand{\secref}[1]{Section~\ref{sec:#1}}
\newcommand{\secreftwo}[2]{Sections~\ref{sec:#1} and~\ref{sec:#2}}

\newcommand{\tabref}[1]{Table~\ref{table:#1}}
\newcommand{\caseref}[1]{Case~\ref{case:#1}}

\begin{document}

\bibliographystyle{pfa}

\ifmypprint
{\small
\begin{verbatim}
$Id: liepoisson.tex,v 2.1 1998/11/13 08:29:59 jeanluc Exp $
\end{verbatim}
}
\fi


\begin{frontmatter}

\title{Classification and Casimir Invariants of \\ Lie--Poisson Brackets}

\author{Jean-Luc Thiffeault\thanksref{jltemail}}
\author{ P. J. Morrison\thanksref{pjmemail}}

\address{Institute for Fusion Studies and Department of Physics, \\ University
of Texas at Austin, Austin, Texas, 78712-1060}

\thanks[jltemail]{Electronic mail: jeanluc@physics.utexas.edu} 
\thanks[pjmemail]{Electronic mail: morrison@hagar.ph.utexas.edu}

\begin{abstract}
%
%
%
%
%
%
%
%

\ifmypprint
{\small
\begin{verbatim}
$Id: abstract.tex,v 2.1 1998/11/13 08:29:27 jeanluc Exp $
\end{verbatim}
}
\fi

We classify Lie--Poisson brackets that are formed from Lie algebra extensions.
The problem is relevant because many physical systems owe their Hamiltonian
structure to such brackets.  A classification involves reducing all brackets
to a set of normal forms, and is achieved partially through the use of Lie
algebra cohomology.  For extensions of order less than five, the number of
normal forms is small and they involve no free parameters.  We derive a
general method of finding Casimir invariants of Lie--Poisson bracket
extensions.  The Casimir invariants of all low-order brackets are explicitly
computed.  We treat in detail a four field model of compressible reduced
magnetohydrodynamics.

\end{abstract}
\end{frontmatter}

%
%
%
%
%
%
%
%
%
%
%
%
%
%

\section{Introduction}
\seclabel{intro}

\ifmypprint
{\small
\begin{verbatim}
$Id: sec-introduction.tex,v 2.1 1998/11/13 08:37:45 jeanluc Exp $
\end{verbatim}
}
\fi

This paper deals with the classification of Lie--Poisson brackets obtained
from extensions of Lie algebras. A large class of finite- and
infinite-dimensional dynamical equations admit a Hamiltonian formulation using
noncanonical brackets of the Lie--Poisson type.  Finite-dimensional examples
include the Euler equations for the rigid body~\cite{Arnold} and the moment
reduction of the Kida vortex~\cite{Meacham1997} \inthesis{Also, and a
low-order model of atmospheric dynamics cite{Bokhove1996}.}, while
infinite-dimensional examples include the Vlasov
equation~\cite{Morrison1980b,Marsden1982b} and the Euler equation for the
ideal
fluid~\cite{Morrison1980a,Kuznetsov1980,Morrison1982,Olver1982,Marsden1983}.
Lie--Poisson brackets naturally define a Poisson structure (i.e., a symplectic
structure) on the dual of a Lie algebra.  For the rigid body, the Lie algebra
is the one associated with the rotation group,~$\SOthree$, while for the Kida
vortex moment reduction the underlying group is~$\SOtwoone$.  For the
two-dimensional ideal fluid, the relevant Lie algebra corresponds to the group
of volume-preserving diffeomorphisms on the fluid domain.

We will classify low-order bracket extensions and find their Casimir
invariants. An extension is simply a new Lie bracket, derived from a base
algebra (for example, $\SOthree$), and defined on~$n$-tuples of that algebra.
We are ruling out extensions where the brackets that appear are not of the
same form as that of the base algebra.  We are thus omitting some
brackets~\cite{Morrison1980a,Nore1997}, but the brackets we are considering
are amenable to a general classification.

The method of extension yields interesting and physically relevant algebras.
Using this method we can describe finite-dimensional systems of several
variables and infinite-dimensional systems of several fields.  For
finite-dimensional systems an example is the two vector model of the heavy
top~\cite{Holmes1983}.  For infinite-dimensional systems there are models with
two~\cite{Morrison1984,Benjamin1984,McLachlan1997}, three
\cite{Morrison1984,Hazeltine1985,Kuvshinov1994}, and four~\cite{Hazeltine1987}
fields. \inthesis{Cite gyro-MHD Morrison1984b paper in thesis} Knowing the
bracket allows one to find the Casimir invariants of the
system~\cite{Trofimov,Kuroda1991,Hernandez1998}.  These are quantities which
commute with every functional on the Poisson manifold, and thus are conserved
by the dynamics for any Hamiltonian.  They are useful for analyzing the
constraints in the system~\cite{Thiffeault1998} and for establishing stability
criteria~\cite{Hazeltine1984,Holm1985,Morrison1986,Morrison1987,Morrison1998}.

The outline of this paper is as follows. In \secref{LiePoisson}, we review the
general theory behind Lie--Poisson brackets. We give examples of physical
systems of Lie--Poisson type, both finite and infinite-dimensional. We
introduce the concept of Lie algebra extensions and derive some of their basic
properties. \secref{cohoext} is devoted to the more abstract treatment of
extensions through the theory of Lie algebra
cohomology~\cite{Chevalley1948,Knapp,Azcarraga}. We define some terminology
and special extensions such as the semidirect sum and the Leibniz
extension. In \secref{classext}, we use the cohomology techniques to treat the
specific type of extension with which we are concerned, brackets
over~$n$-tuples.  We give an explicit classification of low-order
extensions. By classifying we mean reducing---through coordinate changes---all
possible brackets to independent normal forms. We find that the normal forms
are relatively few and involve no free parameters---at least for low-order
extensions.  In \secref{casinv}, we turn to the problem of finding the Casimir
invariants of the brackets, those functionals that commute with every other
functional in the algebra. We derive some general techniques for doing so that
apply to extensions of any order. Some explicit examples are derived,
including the Casimir invariants of a particular model of magnetohydrodynamics
(MHD). These are also given a physical interpretation. A formula for the
invariants of Leibniz extenions of any order is also derived. Then in
\secref{caslowdim} we use the classification of \secref{classext} to derive
the Casimir invariants for low-order extensions. Finally in
\secref{conclusion} we offer some concluding remarks and discuss future
directions.

%
%
%
%
%
%
%
%
%
%
%
%
%
%
%
%

\section{Lie--Poisson Brackets}
\seclabel{LiePoisson}

\ifmypprint
{\small
\begin{verbatim}
$Id: sec-lpbracket.tex,v 2.1 1998/11/13 08:30:06 jeanluc Exp $
\end{verbatim}
}
\fi

Lie--Poisson brackets define a natural Poisson structure on duals of Lie
algebras. Physically, they often arise in the \emph{reduction} of a system.
For our purposes, a reduction is a mapping of the dynamical variables of a
system to a smaller set of variables, such that the transformed Hamiltonian
and bracket depend only on the smaller set of variables. (For a more detailed
mathematical treatment, see for
example~%
\cite{Marsden1974,AbrahamMarsden,Marsden1982,Guillemin,MarsdenRatiu}\@.)
\inthesis{For a brief overview mention cite{Audin}'s book.} The
simplest example of a reduction is the case in which a cyclic variable is
eliminated, but more generally a reduction exists as a consequence of an
underlying symmetry of the system. For instance, the Lie--Poisson bracket for
the rigid body is obtained from a reduction of the canonical Euler angle
description using the rotational symmetry of the system~\cite{Holmes1983}.
The Euler equation for the two-dimensional ideal fluid is obtained from a
reduction of the Lagrangian description of the fluid, which has a relabeling
symmetry~\cite{Morrison1998,Newcomb1967,Bretherton1970,Padhye1996a}.

Here we shall take a more abstract viewpoint: we do not assume that the
Lie--Poisson bracket is obtained from a reduction, though it is always
possible to do so by the method of Clebsch variables~\cite{Morrison1998}.
Rather we proceed directly from a given Lie algebra to build a Lie--Poisson
bracket. The choice of algebra can be guided by the symmetries of the system.
After deriving the basic theory behind Lie--Poisson brackets in
\secref{lpbasic}, we will show some explicit examples in \secref{lpexample}.
We then describe general Lie algebra extensions in \secref{theproblem}.

\subsection{Lie--Poisson Brackets on Duals of Lie Algebras}
\seclabel{lpbasic}

We begin by taking the Lie algebra~$\LieA$ associated with some Lie group.
The Lie group might be chosen to reflect the symmetries of a physical
system. There will be a Lie bracket~\hbox{$\lpb \com \rpb : \LieA
\times \LieA \rightarrow \LieA$} associated with~$\LieA$. Consider the
dual~$\LieA^*$ of~$\LieA$ with respect to the
pairing\index{pairing}~\hbox{$\lang\ \com\
\rang: \LieA^*
\times \LieA \rightarrow \reals$}. Then for real-valued functionals~$F$
and~$G$, that is,~\hbox{$F,G:\LieA^*\rightarrow\reals$}, and~\hbox{$\fv \in
\LieA^*$}, we can define
\begin{equation}
	{\lPB F\com G \rPB}_\pm(\fv) = \pm\lang\fv\com {\lpb \frac{\fd
		F}{\fd\fv}\com\frac{\fd G}{\fd\fv} \rpb}\rang .  \eqlabel{LPB}
\end{equation}
The sign choice comes from whether we are considering right invariant ($+$) or
left invariant ($-$) functions on the cotangent bundle of the Lie
group~\cite{MarsdenRatiu,Marsden1983}, but for our purposes we simply choose
the sign as needed. The functional derivative~$\fd F/\fd\fv$ is defined by
\begin{equation}
	\fd F[\,\fv;\fd\fv\,] \ldef 
	{\l.\frac{d}{d\epsilon}F[\fv + \epsilon\,
		\fd\fv]\r|}_{\epsilon=0}
	\rdef \lang\fd\fv\com\frac{\fd F}{\fd \fv}\rang .
	\eqlabel{funcder}
\end{equation}
We shall refer to the bracket~$\lpb\com\rpb$ as the inner bracket and to the
bracket~$\lPB\com\rPB$ as the Lie--Poisson bracket. The dual~$\LieA^*$
together with the Lie--Poisson bracket is a Poisson manifold; that is, the
bracket~$\lPB\com\rPB$ is a Lie algebra structure on real-valued functionals
that is a derivation in each of its arguments. For finite-dimensional groups,
Eq.~\eqref{LPB} was first written down by Lie~\cite{Lie} and was rediscovered
by Berezin~\cite{Berezin1967}; it is also closely related to work of
Arnold~\cite{Arnold1966a}, Kirillov~\cite{Kirillov1962},
Kostant~\cite{Kostant1966}, and Souriau~\cite{Souriau}.

The bracket in~$\LieA$ is the same as the adjoint action of~$\LieA$ on itself:
\hbox{$\lpb\alpha\com\beta\rpb = \ad_\alpha\,\beta$}, where~$\alpha$, $\beta
\in \LieA$. From this we define the coadjoint action~$\ad_\alpha^\dagger$
of~$\LieA$ on~$\LieA^*$ by
\comment{Following the convention of Arnold, p.321}
\begin{equation}
	\lang \ad_\alpha^\dagger\,\fv\com \beta\,\rang \ldef
		\lang \,\fv \com\,\ad_\alpha\, \beta\,\rang ,
	\eqlabel{coadj}
\end{equation}
where~$\fv \in \LieA^*$. We also define the coadjoint
bracket~\hbox{$\lpb\com\rpb^{\dag}:\LieA
\times \LieA^* \rightarrow \LieA^*$} 
by \hbox{$\lpb\alpha\com\fv\,\rpb^{\dag} \ldef \ad_\alpha^\dagger\,\fv$},
so that
\begin{equation}
	\lang \lpb\alpha\com\fv\rpb^\dagger\com \beta\,\rang \ldef
		\lang \,\fv \com\,\lpb\alpha \com \beta\rpb\,\rang ;
	\eqlabel{cobracket}
\end{equation}
the bracket~${\lpb\com\rpb}^\dagger$ satisfies the identity
\[
	\lang \lpb\alpha\com\fv\rpb^\dagger\com \beta\,\rang =
	-\lang \lpb\beta\com\fv\rpb^\dagger\com \alpha\,\rang.
\]

Since the inner bracket is Lie, it satisfies the Jacobi identity, and
consequently the form given by~\eqref{LPB} for the Lie--Poisson bracket will
automatically satisfy the Jacobi identity~\cite[p.~614]{AMR}.
\inthesis{Prove this in thesis.}

Given a Hamiltonian~\hbox{$H:\LieA^*\rightarrow\reals$}, the equation of
motion for~$\fv \in \LieA^*$ is
\begin{eqnarray}
	\dot\fv &=& \lPB\fv\com H\rPB 
		= \pm\lang\fv\com\lpb{\Delta}\com
			\frac{\fd H}{\fd \fv}\rpb\rang
		\nonumber\\
	&=& \mp\lang\lpb \frac{\fd H}{\fd \fv}\com\,\fv\rpb^\dagger
		\com\,\Delta\rang
	= \mp\lpb \frac{\fd H}{\fd \fv}\com\,\fv\rpb^\dagger,
	\eqlabel{motion}
\end{eqnarray}
where~$\Delta$ is a Kronecker or Dirac delta, or a combination of both for an
infinite-dimensional system of several fields.

\subsection{Examples of Lie--Poisson Systems}
\seclabel{lpexample}

We will say that a physical systems can be described by a given Lie--Poisson
bracket and Hamiltonian if its equations of motion can be written as
\eqref{motion};  the system is then said to be Hamiltonian of the Lie--Poisson
type. We give four examples: the first is finite-dimensional (the free rigid
body, \secref{rigidbody}) and the second infinite-dimensional (Euler's
equation for the ideal fluid, \secref{twodfluid}). The third and fourth
examples are also infinite-dimensional and serve to introduce the concept of
extension. They are low--beta reduced magnetohydrodynamics (MHD) in
\secref{lowbetaRMHD} and compressible reduced MHD in \secref{CRMHD}. These
last two examples are meant to illustrate the physical relevance of Lie
algebra extensions.

\subsubsection{The Free Rigid Body}
\seclabel{rigidbody}

The classic example of a Lie--Poisson bracket is obtained by taking
for~$\LieA$ the Lie algebra of the rotation group~$\SOthree$. If the~$\hat
{\bf e}_{(i)}$ denote a basis of~$\LieA = \sothree$, the Lie bracket is given
by
\[
	\lpb \hat {\bf e}_{(i)}\com \hat {\bf e}_{(j)}\rpb
		= c_{ij}^k\, \hat {\bf e}_{(k)}\,,
\]
where the~$c_{ij}^k = \varepsilon_{ijk}$ are the structure constants of the
algebra, in this case the totally antisymmetric symbol.  Using as a pairing
the usual contraction between upper and lower indices, with~\eqref{LPB} we are
led to the Lie--Poisson bracket
\[
	\lPB f\com g\rPB = -c_{ij}^k\, \ell_k\,\frac{\pd f}{\pd \ell_i}
		\,\frac{\pd g}{\pd \ell_j}\,,
\]
where the three-vector~$\ell$ is in~$\LieA^*$, and we have chosen the minus
sign in \eqref{LPB}. The coadjoint bracket is obtained using~\eqref{coadj},
\[
	{\lpb \beta\com \ell\rpb}_i^\dagger = -c_{ij}^k\,\beta^j\,\ell_k.
\]
If we use this coadjoint bracket and insert the Hamiltonian
\[
	H = \half {(I^{-1})}^{ij}\,\ell_i\,\ell_j
\]
in~\eqref{motion} we obtain
\[
	\dot \ell_m = \lPB \ell_m\com H\rPB 
		= c_{mj}^k\, {(I^{-1})}^{jp}\,\ell_k\,\ell_p\,.
\]
Notice how the moment of inertia tensor~$I$ plays the role of a metric---it
allows us to build a quadratic form (the Hamiltonian) from two elements
of~$\LieA^*$.  If we take~$I = {\rm diag}(I_1,I_2,I_3)$, we recover Euler's
equations for the motion of the free rigid body
\[
	\dot \ell_1 = \l(\frac{1}{I_2} - \frac{1}{I_3}\r)\,\ell_2\,\ell_3,
\]
and cyclic permutations of 1,2,3. The~$\ell_i$ are the angular momenta about
the axes and the~$I_i$ are the principal moments of inertia. This result is
naturally appealing because we expect the rigid body equations to be invariant
under the rotation group, hence the choice of~$\SOthree$ for~$G$.

\subsubsection{The Two-dimensional Ideal Fluid}
\seclabel{twodfluid}

Consider now an ideal fluid with the flow taking place over a two-dimensional
domain~$\fdomain$. Let~$\LieA$ be the infinite-dimensional Lie algebra
associated with the Lie group of volume-preserving diffeomorphisms
of~$\fdomain$. In two spatial dimensions this is the same as the group of
canonical transformations on~$\fdomain$. The bracket in~$\LieA$ is the
canonical bracket
\begin{equation}
	\lpb a \com b \rpb = \frac{\pd a}{\pd x}\,\frac{\pd b}{\pd y}
		- \frac{\pd b}{\pd x}\,\frac{\pd a}{\pd y}.
	\eqlabel{canibrak}
\end{equation}
We formally identify~$\LieA$ and~$\LieA^*$ and use as the
pairing~$\lang\com\rang$ the usual integral over the fluid domain,
\[
	\lang F\com G \rang = \int_\fdomain F(\xv)\,G(\xv)\d^2x,
\]
where~$\xv \ldef (x,y)$.  For infinite-dimensional spaces, there are
functional analytic issues about whether we can make this identification, and
take~$\LieA^{**}=\LieA$. We will assume here that these relationships hold
formally. See Marsden and Weinstein~\cite{Marsden1982} for references on this
subject and Audin~\cite{Audin} for a treatment of the identification
of~$\LieA$ and~$\LieA^*$.

Assuming appropriate boundary conditions for simplicity, we
get~\hbox{${\lpb\com\rpb}^\dagger = -\lpb\com\rpb$} from
\eqref{cobracket}. (Otherwise the coadjoint bracket would involve extra
boundary terms.) Take the vorticity~$\vort$ as the field variable~$\fv$ and
write for the Hamiltonian
\[
	H[\vort] = -\half\lang \vort\com \invlapl\,\vort\rang ,
\]
where
\[
	(\invlapl\,\vort)(\xv) 
		\ldef \int_\fdomain K(\xv|\xv')\,\vort(\xv')\d^2x',
\]
and~$K$ is Green's function for the Laplacian. The Green's function plays the
role of a metric since it maps an element of~$\LieA^*$ (the vorticity~$\vort$)
into an element of~$\LieA$ to be used in the right slot of the pairing. This
relationship is only weak: the mapping~$K$ is not surjective, and thus the
metric cannot formally inverted (it is called
\emph{weakly nondegenerate}). When we have identified~$\LieA$ and~$\LieA^*$ we
shall often drop the comma in the pairing and write
\[
	H[\vort] = -\half\lang \vort\,\streamf\rang =
		\half\lang|\grad\streamf|^2\rang,
\]
where~$\vort=\lapl\streamf$ defines the streamfunction~$\streamf$. We work
out the evolution equation for~$\vort$ explicitly:
\begin{eqnarray*}
	\dot\vort(\xv) &=& \lPB\vort\com H\rPB 
		= \int_\fdomain \vort(\xv')
			\lpb\frac{\fd\vort(\xv)}{\fd\vort(\xv')}\com
			\frac{\fd H}{\fd\vort(\xv')}\rpb\d^2x'\nonumber\\
	&=& \int_\fdomain\vort(\xv')
			\lpb\delta(\xv-\xv')\com {-\streamf(\xv')}\rpb\d^2x'
			\nonumber\\
	&=& \int_\fdomain\delta(\xv-\xv')
			\lpb\vort(\xv')\com\streamf(\xv')\rpb\d^2x'
			\nonumber\\
	&=& \lpb\vort(\xv)\com\streamf(\xv)\rpb\, .
\end{eqnarray*}
This is Euler's equation for a two-dimensional ideal fluid. We could also
have written this result down directly from \eqref{motion}
using~\hbox{${\lpb\com\rpb}^\dagger = -{\lpb\com\rpb}$}.

\subsubsection{Low--beta Reduced MHD}
\seclabel{lowbetaRMHD}

This example will illustrate the concept of a Lie algebra extension, the main
topic of this paper. Essentially, the idea is to use an algebra of~$n$-tuples,
which we call an extension, to describe a physical system with more than one
dynamical variable. As in \secref{twodfluid} we consider a flow taking place
over a two-dimensional domain~$\fdomain$. The Lie algebra~$\LieA$ is again
taken to be that of volume preserving diffeomorphisms on~$\fdomain$, but now
we consider also the vector space~$\vs$ of real-valued functions on~$\fdomain$
(an Abelian Lie algebra under addition). The \emph{semidirect sum} of~$\LieA$
and~$\vs$ is a new Lie algebra whose elements are two-tuples~$(\alpha,v)$ with
a bracket defined by
\begin{equation}
	\lpb(\alpha,v)\com(\beta,w)\rpb \ldef
	\l(\lpb\alpha\com\beta\rpb\com
	\lpb\alpha\com w\rpb - \lpb\beta\com v\rpb
	\r),
	\eqlabel{RMHDbrak}
\end{equation}
where~ $\alpha$ and~$\beta\in\LieA$, $v$ and~$w\in\vs$. This is a Lie
algebra, so we can use the prescription of \secref{lpbasic} to
build a Lie--Poisson bracket,
\[
	{\lPB F\com G \rPB} = \int_\fdomain
		\l\lgroup
		\vort{\lpb \frac{\fd F}{\fd\vort}\com\frac{\fd G}{\fd\vort}
		\rpb}
	+ \magf \l({\lpb \frac{\fd F}{\fd\vort}\com\frac{\fd G}{\fd \magf}
		\rpb}
	- {\lpb \frac{\fd G}{\fd\vort}\com\frac{\fd F}{\fd \magf}
		\rpb}\r) \r\rgroup\d^2x.
\]
Let $\vort=\lapl\elecp$, where $\elecp$ is the electric potential, $\magf$
is the magnetic flux, and $\ecurrent=\lapl\magf$ is the current. (We use
the same symbol for the electric field as for the streamfunction in
\secref{twodfluid} since they play a similar role.) The pairing used is a dot
product of the vectors followed by an integral over the fluid domain (again
identifying~$\LieA$ and~$\LieA^*$ as in \secref{twodfluid}). The Hamiltonian
\[
	H[\vort;\psi] = \frac{1}{2}\int_\fdomain\,\l\lgroup
		|\grad\elecp|^2+|\grad\magf|^2\r\rgroup\d^2x
\]
with the above bracket leads to the equations of motion
\begin{eqnarray*}
	\dot\vort &=& \l[\vort,\elecp\r] + \l[\magf,\ecurrent\r] \ ,\nonumber\\
	\dot\magf &=& \l[\magf,\elecp\r] \, .
\end{eqnarray*}
This is a model for low-beta reduced
MHD~\cite{Morrison1984,Strauss1977,Zeitlin1992}.  It is obtained by an
expansion in the inverse aspect ratio~$\epsilon$ of a tokamak, with~$\epsilon$
small.  This is called low beta since the plasma beta (the ratio of plasma
pressure to magnetic pressure) is of order~$\epsilon^2$.  With a strong
toroidal magnetic field, the dynamics are then approximately two-dimensional.

Benjamin~\cite{Benjamin1984} used a system with a similar Lie--Poisson
structure, but for waves in a density-stratified fluid. Semidirect sum
structures are ubiquitous in advective systems: one variable (in this
example,~$\elecp$) ``drags'' the others along~\cite{Thiffeault1998}.

\subsubsection{Compressible Reduced MHD}
\seclabel{CRMHD}

In general there are other, more general ways to extend Lie algebras besides
the semidirect sum. The model derived by Hazeltine \emph{et
al.}~\cite{Hazeltine1987,Hazeltine1985b} for two-dimensional compressible
reduced MHD (CRMHD) is an example. This model has four fields, and as for the
system in \secref{lowbetaRMHD} it is also obtained from an expansion in the
inverse aspect ratio of a tokamak. It includes compressibility and finite ion
Larmor radius effects. The Hamiltonian is
\begin{equation}
	H[\vort,\pvel,\pres,\magf] = \frac{1}{2}\int_\fdomain\l\lgroup
		|\grad\elecp|^2 + \pvel^2
		+ \frac{(\pres-2\crmhdbeta\,x)^2}{\crmhdbeta}
		+ |\grad\magf|^2\r\rgroup \d^2x,
	\eqlabel{CRMHDHam}
\end{equation}
where $\pvel$ is the parallel ion velocity, $\pres$ is the pressure, and
$\crmhdbeta$ is a parameter that measures compressibility. The other variables
are as in \secref{lowbetaRMHD}. The coordinate~$x$ points outward from the
center of the tokamak in the horizontal plane and~$y$ is the vertical
coordinate.  The motion is made two-dimensional by the strong toroidal
magnetic field. The bracket we will use is
\begin{eqnarray}
	\lPB F\com G\rPB &=&
		\int_\fdomain\l\lgroup\vort\lpb\frac{\fd F}{\fd \vort}
			\com\frac{\fd G}{\fd \vort}\rpb\r.
			+ \pvel\l(\lpb\frac{\fd F}{\fd \vort}
			\com\frac{\fd G}{\fd \pvel}\rpb
			+ \lpb\frac{\fd F}{\fd \pvel}
			\com\frac{\fd G}{\fd \vort}\rpb\r)\nonumber\\
	&&\mbox{}\!\!\!\!\!\!\!\!\!\! + \pres\l(\lpb\frac{\fd F}{\fd \vort}
			\com\frac{\fd G}{\fd \pres}\rpb
		+ \lpb\frac{\fd F}{\fd \pres}
			\com\frac{\fd G}{\fd \vort}\rpb\r)
		+ \magf\l(\lpb\frac{\fd F}{\fd \vort}
			\com\frac{\fd G}{\fd \magf}\rpb
		+ \lpb\frac{\fd F}{\fd \magf}
			\com\frac{\fd G}{\fd \vort}\rpb\r)\nonumber\\
	&&\mbox{}\!\!\!\!\!\!\!\!\!\! - \l.\crmhdbeta\,\magf\l(
		\lpb\frac{\fd F}{\fd \pres}
		\com\frac{\fd G}{\fd \pvel}\rpb
		+ \lpb\frac{\fd F}{\fd \pvel}
		\com\frac{\fd G}{\fd \pres}\rpb\r)\r\rgroup\d^2x.
	\eqlabel{CRMHDbracket}
\end{eqnarray}
Together this bracket and the Hamiltonian \eqref{CRMHDHam} lead to the
equations
\begin{eqnarray*}
	\dot\vort &=& \lpb \vort\com\elecp\rpb + \lpb\magf\com \ecurrent\rpb 
		+ 2\lpb \pres\com x\rpb\\
	\dot\pvel &=& \lpb\pvel\com\elecp\rpb + \lpb\magf\com\pres \rpb
		+ 2\crmhdbeta\lpb x \com \magf \rpb\\
	\dot\pres &=& \lpb\pres\com\elecp\rpb
		+ \crmhdbeta\lpb\magf\com\pvel\rpb\\
	\dot\magf &=& \lpb \magf\com\elecp\rpb ,
\end{eqnarray*}
which reduce to the example of \secref{lowbetaRMHD} in the
limit~\hbox{$\pvel=\pres=\crmhdbeta=0$} (when compressibility effects are
unimportant).

It is far from clear that the Jacobi identity for~\eqref{CRMHDbracket} is
satisfied. A direct verification is straightforward (if tedious), but we shall
see in \secref{theproblem} that there is an easier way.

\subsection{General Algebra Extensions}
\seclabel{theproblem}

We wish to generalize the types of bracket used in
\secreftwo{lowbetaRMHD}{CRMHD}. We build an algebra extension by forming an~$n$-tuple of elements of a single Lie algebra~$\LieA$,
\begin{equation}
	\alpha \ldef \l(\alpha_1,\dots,\alpha_n\r), \eqlabel{algtuple}
\end{equation}
where~\hbox{$\alpha_i \in \LieA$}. The most general bracket on this~$n$-tuple
space obtained from a linear combination of the one in~$\LieA$ has components
\begin{equation}
	{\lpb\alpha\com\beta\rpb}_\lambda = \sum_{\mu,\nu=1}^n
		{\W_\lambda}^{\mu\nu}\,
		\lpb\alpha_\mu\com\beta_\nu\rpb\,,
		\ \ \ \lambda=1,\dots,n,
	\eqlabel{extbrack}
\end{equation}
where the~${\W_\lambda}^{\mu\nu}$ are constants.  (From now on we will assume
that repeated indices are summed unless otherwise noted.) Since the bracket
in~$\LieA$ is antisymmetric the~$\W$'s must be symmetric in their upper
indices,
\begin{equation}
	{\W_\lambda}^{\mu\nu} = {\W_\lambda}^{\nu\mu}\,.
	\eqlabel{upsym}
\end{equation}
This bracket must also satisfy the Jacobi identity
\[
	{\lpb\alpha\com\lpb\beta\com\gamma\rpb\rpb}_\lambda +
	{\lpb\beta\com\lpb\gamma\com\alpha\rpb\rpb}_\lambda +
	{\lpb\gamma\com\lpb\alpha\com\beta\rpb\rpb}_\lambda = 0, \ \
	\lambda=1,\dots,n.
\]
The first term can be written
\[
	{\lpb\alpha\com\lpb\beta\com\gamma\rpb\rpb}_\lambda = 
	{\W_\lambda}^{\sigma\tau}\,{\W_\sigma}^{\mu\nu}\,
	{\lpb\alpha_\tau\com\lpb\beta_\mu\com\gamma_\nu\rpb\rpb},
\]
which when added to the other two gives
\[
	{\W_\lambda}^{\sigma\tau}\,{\W_\sigma}^{\mu\nu}\,\l(
	{\lpb\alpha_\tau\com\lpb\beta_\mu\com\gamma_\nu\rpb\rpb}
	+ {\lpb\beta_\tau\com\lpb\gamma_\mu\com\alpha_\nu\rpb\rpb}
	+ {\lpb\gamma_\tau\com\lpb\alpha_\mu\com\beta_\nu\rpb\rpb}
	\r) = 0.
\]
We cannot yet make use of the Jacobi identity in~$\LieA$: the subscripts
of~$\alpha$, $\beta$, and~$\gamma$ are different in each term so they
represent different elements of~$\LieA$.  We first relabel the sums and then
make use of the Jacobi identity in~$\LieA$ to obtain
\begin{eqnarray*}
&&\l({\W_\lambda}^{\sigma\tau}\,{\W_\sigma}^{\mu\nu}
	- {\W_\lambda}^{\sigma\nu}\,{\W_\sigma}^{\tau\mu}\r)\,
	{\lpb\alpha_\tau\com\lpb\beta_\mu\com\gamma_\nu\rpb\rpb}\nonumber\\
&&\mbox{} + \l({\W_\lambda}^{\sigma\mu}\,{\W_\sigma}^{\nu\tau}
	- {\W_\lambda}^{\sigma\nu}\,{\W_\sigma}^{\tau\mu}\r)\,
	{\lpb\beta_\mu\com\lpb\gamma_\nu\com\alpha_\tau\rpb\rpb} = 0\,.
\end{eqnarray*}
This identity is satisfied if and only if
\begin{equation}
	{\W_\lambda}^{\sigma\tau}\,{\W_\sigma}^{\mu\nu}
	= {\W_\lambda}^{\sigma\nu}\,{\W_\sigma}^{\tau\mu}\,,
	\eqlabel{Wjacob}
\end{equation}
which together with~\eqref{upsym} implies that the
quantity~${\W_\lambda}^{\sigma\tau}\,{\W_\sigma}^{\mu\nu}$ is symmetric in all
three free upper indices. If we write the~$\W$'s as~$n$
matrices~${\W}^{(\nu)}$ with rows labeled by~$\lambda$ and columns by~$\mu$,
\begin{equation}
	{{\l[{\W}^{(\nu)}\r]}_\lambda}^\mu \ldef {\W_\lambda}^{\mu\nu},
	\eqlabel{Wupdef}
\end{equation}
then~\eqref{Wjacob} says that those matrices pairwise commute:
\begin{equation}
	\W^{(\nu)}\,\W^{(\sigma)} = \W^{(\sigma)}\,\W^{(\nu)}.
	\eqlabel{Wcommute}
\end{equation}
Equations~\eqref{upsym} and~\eqref{Wcommute} form a necessary and sufficient
condition: a set of~$n$ commuting matrices of size~$n\times n$ satisfying the
symmetry given by~\eqref{upsym} can be used to make a good Lie algebra
bracket. From this Lie bracket we can build a Lie--Poisson bracket using the
prescription of~\eqref{LPB} to obtain
\[
	{\lPB F\com G \rPB}_\pm(\fv) = \pm\sum_{\lambda,\mu,\nu=1}^n
		{\W_\lambda}^{\mu\nu}\lang\fv^\lambda\com
		{\lpb \frac{\fd F}{\fd\fv^\mu}\com\frac{\fd G}{\fd\fv^\nu}
		\rpb}\rang .
\]

We now return to the two extension examples of \secreftwo{lowbetaRMHD}{CRMHD}
and examine them in light of the general extension concept introduced here.

\subsubsection{Low-beta Reduced MHD}
\seclabel{matlowbetaRMHD}

For this example we have~$(\fv^0,\fv^1)=(\vort,\magf)$, with
\[
	\W^{(0)} = \begin{pmatrix} \matone & \matzero \\
			\matzero & \matone
		\end{pmatrix},\ \ \ \
	\W^{(1)} = \begin{pmatrix} \matzero & \matzero \\
			\matone & \matzero
		\end{pmatrix}.
\]
The reason why we start labeling at~$0$ will become clearer in
\secref{semisimple}. The two~$\W^{(\mu)}$ must commute since~$\W^{(0)}=I$, the
identity. The tensor~$\W$ also satisfies the symmetry
property~\eqref{upsym}. Hence, the bracket is a good Lie algebra bracket.

\subsubsection{Compressible Reduced MHD}
\seclabel{matCRMHD}

We have~$n=4$ and take~$(\fv^0,\fv^1,\fv^2,\fv^3)=(\vort,\pvel,\pres,\magf)$,
so the tensor~$\W$ is given by
\begin{equation}
\begin{array}{rclrcl}
	\W^{(0)} &=& \l(\begin{array}{cccc}
				\matone & \matzero & \matzero & \matzero\\
				\matzero & \matone & \matzero & \matzero\\
				\matzero & \matzero & \matone & \matzero\\
				\matzero & \matzero & \matzero & \matone\\
		\end{array}\r), & \ \
	\W^{(1)} &=& \l(\begin{array}{cccc}
				\matzero & \matzero & \matzero & \matzero\\
				\matone & \matzero & \matzero & \matzero\\
				\matzero & \matzero & \matzero & \matzero\\
				\matzero & \matzero & -\crmhdbeta & \matzero\\
		\end{array}\r),\\
	\\
	\W^{(2)} &=& \l(\begin{array}{cccc}
				\matzero & \matzero & \matzero & \matzero\\
				\matzero & \matzero & \matzero & \matzero\\
				\matone & \matzero & \matzero & \matzero\\
				\matzero & -\crmhdbeta & \matzero & \matzero\\
		\end{array}\r), & \ \
	\W^{(3)} &=& \l(\begin{array}{cccc}
				\matzero & \matzero & \matzero & \matzero\\
				\matzero & \matzero & \matzero & \matzero\\
				\matzero & \matzero & \matzero & \matzero\\
				\matone & \matzero & \matzero & \matzero\\
		\end{array}\r).
\end{array}
	\eqlabel{matCRMHD}
\end{equation}
It is easy to verify that these matrices commute and that the tensor~$\W$
satisfies the symmetry property, so that the Lie--Poisson bracket given
by~\eqref{CRMHDbracket} satisfies the Jacobi identity. (See
\secref{semisimple} for an explanation of why the labeling is chosen to begin
at zero.)


%
%
%
%
%
%
%
%
%
%
%
%
%
%
%
%

\section{Extension of a Lie Algebra}
\seclabel{cohoext}

\ifmypprint
{\small
\begin{verbatim}
$Id: sec-lieaextension.tex,v 2.1 1998/11/13 08:36:57 jeanluc Exp $
\end{verbatim}
}
\fi

In this section we review the theory of Lie algebra cohomology and its
application to extensions. This is useful for shedding light on the methods
used in \secref{classext} for classifying extensions. However, the
mathematical details presented in this section can be skipped without
seriously compromising the flavor of the classification scheme of
\secref{classext}.

%
%
%
%
%
%
%
%
%
%
%
%

\subsection{Cohomology of Lie Algebras}
\seclabel{cohoalgebra}

\ifmypprint
{\small
\begin{verbatim}
$Id: sec-cohoalgebra.tex,v 2.1 1998/11/13 08:37:04 jeanluc Exp $
\end{verbatim}
}
\fi

We now introduce the abstract formalism of Lie algebra cohomology.
Historically there were two different reasons for the development of this
theory. One, known as the Chevalley--Eilenberg
formulation~\cite{Chevalley1948}, was developed from de Rham cohomology\@. de
Rham cohomology concerns the relationship between exact and closed
differential forms, which is determined by the global properties (topology) of
a differentiable manifold. A Lie group is a differentiable manifold and so has
an associated de Rham cohomology. If invariant differential forms are used in
the computation, one is led to the cohomology of Lie algebras presented in
this section~\cite{Knapp,Azcarraga,CecileII}.  The second motivation is the
one that concerns us: we will show in \secref{extension} that the extension
problem---the problem of enumerating extensions of a Lie algebra---can be
related to the cohomology of Lie algebras.

Let~$\LieA$ be a Lie algebra, and let the vector space~$\vs$ over the
field~$\field$ (which we take to be the real numbers later) be a
left~$\LieA$-module,%
\footnote{When~$\vs$ is a right~$\LieA$-module, we have~\hbox{$\rho_{\lpb
\alpha\com\alpha'\rpb} = -\lpb \rho_\alpha \com \rho_{\alpha'} \rpb$}.
The results of this section can be adapted to a right action by changing the
sign every time a commutator appears.}  that is, there is an operator~$\rho:
\LieA \times \vs \rightarrow \vs$ such that
\begin{eqnarray}
	\rho_\alpha\, (v + v') &=& \rho_\alpha\,v + \rho_\alpha\,v',\nonumber\\
	\rho_{\alpha + \alpha'}\, v &=& \rho_\alpha\,v + \rho_{\alpha'}\,v,
		\nonumber\\
	\rho_{\lpb \alpha\com\alpha'\rpb}v &=& 
		\lpb \rho_\alpha \com \rho_{\alpha'} \rpb\,v\,,
	\eqlabel{rhomo}
\end{eqnarray}
for~$\alpha, \alpha' \in \LieA$ and~$v,v' \in \vs$.  The operator~$\rho$ is
known as a left action.  A~$\LieA$-module gives a representation of~$\LieA$
on~$\vs$.

An $n$-dimensional $\vs$-valued cochain~$\omega_n$ for~$\LieA$, or
just~$n$-cochain for short, is a skew-symmetric~$n$-linear mapping
\[
	\omega_n :\ \stackrel{\longleftarrow n \longrightarrow}
	{\LieA \times \LieA \times \dots \times \LieA}\
	\longrightarrow \vs.
\]
Cochains are Lie algebra cohomology analogues of differential forms on a
manifold.  Addition and scalar multiplication of~$n$-cochains are defined in
the obvious manner by
\begin{eqnarray*}
	(\omega_n + \omega_n')(\alpha_1,\dots,\alpha_n)
	&\ldef& \omega_n(\alpha_1,\dots,\alpha_n)
	+ \omega_n'(\alpha_1,\dots,\alpha_n),\nonumber\\
	(a\,\omega_n)(\alpha_1,\dots,\alpha_n)
	&\ldef& a\,\omega_n(\alpha_1,\dots,\alpha_n),
\end{eqnarray*}
where~\hbox{$\alpha_1,\dots,\alpha_n \in \LieA$} and~\hbox{$a \in \field$}.
The set of all~$n$-cochains thus forms a vector space over the field~$\field$
and is denoted by~$C^n(\LieA,\vs)$.  The $0$-cochains are defined to be just
elements of~$\vs$, so that~\hbox{$C^0(\LieA,\vs) = \vs$}.

The coboundary operator is the map between cochains,
\[
	s_n : C^n(\LieA,\vs) \longrightarrow C^{n+1}(\LieA,\vs),
\]
defined by
\begin{eqnarray*}
	(s_n\,\omega_n)(\alpha_1,\dots,\alpha_{n+1}) &\ldef&
	\sum_{i=1}^{n+1}(-)^{i+1}\rho_{\alpha_i}\omega_n(\alpha_1,\dots,
	\hat\alpha_i,\dots,\alpha_{n+1})\nonumber\\ &&\mbox{} 
	\!\!\!\!\!\!\!\!\!\!\!\!\!\!\!\!\!\!\!\!\!\!\!\!\!\!\!\!\!\!\!\!
	+ \sum_{\scriptscriptstyle{j,k=1}\atop\scriptscriptstyle{j<k}}^{n+1}
	(-)^{j+k}\omega_n(\lpb\alpha_j\com\alpha_k\rpb,\alpha_1,\dots,
	\hat\alpha_j,\dots,\hat\alpha_k,\dots,\alpha_{n+1}),
\end{eqnarray*}
where the caret means an argument is omitted.  We shall often drop the~$n$
subscript on~$s_n$, deducing it from the dimension of the cochain on which~$s$
acts.

We shall make use mostly of the first few cases
\begin{eqnarray}
	(s\,\omega_0)(\alpha_1) &=& \rho_{\alpha_1}\,\omega_0,
	\eqlabel{1cobound}\\
	(s\,\omega_1)(\alpha_1,\alpha_2) &=&
		\rho_{\alpha_1}\,\omega_1(\alpha_2)
		- \rho_{\alpha_2}\,\omega_1(\alpha_1)
		- \omega_1(\lpb\alpha_1\com\alpha_2\rpb),
	\eqlabel{2cobound}\\
	(s\,\omega_2)(\alpha_1,\alpha_2,\alpha_3) &=&
		\rho_{\alpha_1}\,\omega_2(\alpha_2,\alpha_3)
		+ \rho_{\alpha_2}\,\omega_2(\alpha_3,\alpha_1)
		+ \rho_{\alpha_3}\,\omega_2(\alpha_1,\alpha_2)\nonumber\\
	&&\mbox{}
		\!\!\!\!\!\!\!\!\!\!\!\!\!\!\!\!\!\!\!\!\!\!\!\!\!\!\! 
		- \omega_2(\lpb\alpha_1\com\alpha_2\rpb,\alpha_3)
		- \omega_2(\lpb\alpha_2\com\alpha_3\rpb,\alpha_1) 
		- \omega_2(\lpb\alpha_3\com\alpha_1\rpb,\alpha_2)\,.
	\eqlabel{3cobound}
\end{eqnarray}
It is easy to verify that~$s\,\omega_n$ defines an $(n+1)$-cochain, and it is
straightforward (if tedious) to show that~\hbox{$s_{n+1}s_n = s^2=0$}.  For
this to be true, The homomorphism property of~$\rho$ is crucial.

An~$n$-cocycle is an element~$\omega_n$ of~$C^n(\LieA,\vs)$ such
that~$s_n\,\omega_n = 0$.  An~$n$-coboundary $\omega_{\rm cob}$ is an element
of~$C^n(\LieA,\vs)$ for which there exists an element~$\omega_{n-1}$
of~$C^{n-1}(\LieA,\vs)$ such that~\hbox{$\omega_{\rm cob} = s\omega_{n-1}$}.
Note that all coboundaries are cocycles, but not vice-versa.

Let
\[
	Z^{n}_\rho(\LieA,\vs) \ldef \ker s_n
\]
be the vector subspace of all~$n$-cocycles,~\hbox{$Z^{n}_\rho(\LieA,\vs)
\subset C^n(\LieA,\vs)$}, and let
\[
	B^{n}_\rho(\LieA,\vs) \ldef \range s_{n-1}
\]
be the vector subspace of all~$n$-coboundaries,~\hbox{$B^{n}_\rho(\LieA,\vs)
\subset C^n(\LieA,\vs)$}.  The~$n$th cohomology group
of~$\LieA$ with coefficients in~$\vs$ is defined to be the quotient vector
space
\begin{equation}
	H^n_\rho(\LieA,\vs) \ldef Z^{n}_\rho(\LieA,\vs)/B^{n}_\rho(\LieA,\vs).
\end{equation}
Note that for~\hbox{$n > \dim \LieA$}, we have~\hbox{$H^n_\rho(\LieA,\vs) =
Z^{n}_\rho(\LieA,\vs) = B^{n}_\rho(\LieA,\vs) = 0$}.

\subsection{Application of Cohomology to Extensions}
\seclabel{extension}

In \secref{theproblem} we gave a definition of extension that is specific to
our problem.  We will now define extensions in a more abstract manner.  We
then show how the cohomology of Lie algebras of \secref{cohoalgebra} is
related to the problem of classifying extensions.  In \secref{classext} we
will return to the more concrete concept of extension, of the form given in
\secref{theproblem}.

Let~$f_i: \LieA_i \rightarrow
\LieA_{i+1}$ be a collection of Lie algebra homomorphisms,
\inthesis{define homomorphism.}
\[
\xymatrix@M=4pt{
	\dots \ar[r] & {\LieA_i} \ar[r]^-{f_i} &
	{\LieA_{i+1}} \ar[r]^-{f_{i+1}} & {\LieA_{i+2}} \ar[r] & \dots\
} .
\]
The sequence~$f_i$ is called an exact sequence of Lie algebra homomorphisms if
\[
	\range f_i = \ker f_{i+1}\,.
\]

Let~$\LieA$, $\LieAx$, and~$\LieAxb$ be Lie algebras. The algebra~$\LieAx$ is
said to be an~\emph{extension} of~$\LieA$ by~$\LieAxb$ if there is a short
exact sequence of Lie algebra homomorphisms
\begin{equation}\xymatrix@M=4pt{
	0 \ar[r] & {\LieAxb} \ar[r]^{i} & {\LieAx} \ar[r]^{\pi}
	& {\LieA} \ar[r] & 0
	}.
	\eqlabel{extdef}
\end{equation}
The homomorphism~$i$ is an insertion (injection), and~$\pi$ is a projection
(surjection). We shall distinguish brackets in the different algebras by
appropriate subscripts.  We also define~\hbox{$\tau:\LieA\rightarrow\LieAx$}
to be a linear mapping such that~\hbox{$\pi\circ\tau = 1_{|\LieA}$} (the
identity mapping in~$\LieA$).  Note that~$\tau$ is not unique, since the
kernel of~$\pi$ is not trivial.  Let~$\beta \in
\LieAx$,~$\eta \in \LieAxb$; then
\[
	\pi{\lpb\beta\com i\,\eta\rpb}_\LieAx 
		= {\lpb\pi\,\beta\com\pi\,i\,\eta\rpb}_\LieA
		= 0,
\]
using the homomorphism property of~$\pi$ and~\hbox{$\pi\circ i=0$}, a
consequence of the exactness of the sequence. Thus~\hbox{${\lpb\beta\com
i\,\eta\rpb}_\LieAx \in
\ker \pi = \range i$}, and~$i\,\LieAxb$ is an ideal in~$\LieAx$
since~\hbox{$\lpb\beta\com i\eta\rpb \in i \LieAxb$}.
Hence, we can form the quotient algebra~$\LieAx/\LieAxb$, with equivalence
classes denoted by~\hbox{$\beta + \LieAxb$}. By exactness~\hbox{$\pi(\beta +
\LieAxb) = \pi\,\beta$}, so~$\LieA$ is isomorphic to~$\LieAx/\LieAxb$ and we
write~\hbox{$\LieA = \LieAx/\LieAxb$}.

Though~$i\,\LieAxb$ is a subalgebra of~$\LieAx$,~$\tau\,\LieA$ is not
necessarily a subalgebra of~$\LieAx$, for in general
\[
	{\lpb\tau\,\alpha\com\tau\,\beta\rpb}_\LieAx
	\ne \tau\,{\lpb\alpha\com\beta\rpb}_\LieA,
\]
for~\hbox{$\alpha,\beta \in \LieA$}; that is,~$\tau$ is not necessarily a
homomorphism. The classification problem essentially resides in the
determination of how much~$\tau$ differs from a homomorphism. The cohomology
machinery of \secref{cohoalgebra} is the key to quantifying this difference,
and we proceed to show this.

To this end, we use the algebra~$\LieAxb$ as the vector space~$\vs$ of
\secref{cohoalgebra}, so that~$\LieAxb$ will be a left $\LieA$-module. We
define the left action as
\begin{equation}
	\rho_\alpha\,\eta \ldef
		i^{-1}{\lpb \tau\,\alpha\com i\,\eta\rpb}_\LieAx
	\eqlabel{rhodef}
\end{equation}
for~\hbox{$\alpha \in \LieA$} and~\hbox{$\eta \in \LieAxb$}. For~$\LieAxb$ to
be a left~$\LieA$-module, we need~$\rho$ to be a homomorphism, i.e.,~$\rho$
must satisfy~\eqref{rhomo}.  Therefore consider
\begin{eqnarray*}
	{\lpb\rho_\alpha\com\rho_\beta\rpb}\,\eta &=&
		(\rho_\alpha\rho_\beta - \rho_\beta\rho_\alpha)\,\eta
		\nonumber\\
	&=& \rho_\alpha\,i^{-1}{\lpb\tau\,\beta\com i\,\eta\rpb}_\LieAx
		- \rho_\beta\,i^{-1}{\lpb\tau\,\alpha\com i\,\eta\rpb}_\LieAx
		\nonumber\\
	&=& i^{-1}{\lpb\tau\,\alpha\com
		{\lpb\tau\,\beta\com i\,\eta\rpb}_\LieAx\rpb}_\LieAx
		- i^{-1}{\lpb\tau\,\beta\com
		{\lpb\tau\,\alpha\com i\,\eta\rpb}_\LieAx\rpb}_\LieAx,
\end{eqnarray*}
which upon using the Jacobi identity in~$\LieAx$ becomes
\begin{eqnarray}
	{\lpb\rho_\alpha\com\rho_\beta\rpb}\,\eta &=& 
		i^{-1}{\lpb{\lpb\tau\,\alpha\com \tau\,\beta\rpb}_\LieAx\com
		i\,\eta\rpb}_\LieAx\nonumber\\
	&=& i^{-1}{\lpb\tau\,{\lpb\alpha\com \beta\rpb}_\LieA\com
		i\,\eta\rpb}_\LieAx
		+ i^{-1}{\lpb\l(
		{\lpb\tau\,\alpha\com \tau\,\beta\rpb}_\LieAx
		- \tau\,{\lpb\alpha\com \beta\rpb}_\LieA\r)\com
		i\,\eta\rpb}_\LieAx\nonumber\\
	&=& \rho_{{\lpb\alpha\com\beta\rpb}_\LieA}\,\eta
		+ i^{-1}{\lpb\l(
		{\lpb\tau\,\alpha\com \tau\,\beta\rpb}_\LieAx
		- \tau\,{\lpb\alpha\com \beta\rpb}_\LieA\r)\com
		i\,\eta\rpb}_\LieAx.
	\eqlabel{mismatch}
\end{eqnarray}
By applying~$\pi$ on the expression in parentheses of the last term
of~\eqref{mismatch}, we see that it vanishes and so is in~$\ker \pi$, and by
exactness it is also in~$i\,\LieAxb$. Thus the $\LieAx$ commutator above
involves two elements of~$i\,\LieAxb$. We define~\hbox{$\omega:
\LieA \times \LieA \rightarrow \LieAxb$} by
\begin{equation}
	\omega(\alpha,\beta) \ldef i^{-1}\l(
	{\lpb\tau\,\alpha\com\tau\,\beta\rpb}_\LieAx
	- \tau\,{\lpb\alpha\com\beta\rpb}_\LieA\r).
	\eqlabel{cocycle}
\end{equation}
The mapping~$i^{-1}$ is well defined on~$i\,\LieAxb$. Equation
\eqref{mismatch} becomes
\begin{equation}
	{\lpb\rho_\alpha\com\rho_\beta\rpb}\,\eta =
		\rho_{{\lpb\alpha\com\beta\rpb}_\LieA}\,\eta + {\lpb
		\omega(\alpha,\beta) \com
		\eta\rpb}_\LieAxb.
	\eqlabel{rhohomo}
\end{equation}
Therefore,~$\rho$ satisfies the homomorphism property if either of the
following is true:
\begin{enumerate}
\item[(i)] $\LieAxb$ is Abelian,
\item[(ii)] $\tau$ is a homomorphism,
\end{enumerate}
Condition~(i) implies~$\lpb\com\rpb_\LieAxb=0$, while condition~(ii)
means
\[
	{\lpb\tau\,\alpha\com \tau\,\beta\rpb}_\LieAx =
		\tau\,{\lpb\alpha\com\beta\rpb}_\LieA,
\]
which implies~$\omega\equiv 0$. If either of these conditions is
satisfied,~$\LieAxb$ with the action~$\rho$ is a left~$\LieA$-module.  We
treat these two cases separately in \secreftwo{abelianext}{semidext},
respectively.

\subsection{Extension by an Abelian Lie Algebra}
\seclabel{abelianext}

In this section we assume that the homomorphism condition~(i) at the end of
\secref{extension} is met.  Therefore~$\LieAxb$ is a left~$\LieA$-module, and
we can define~$\LieAxb$-valued cochains on~$\LieA$.  In particular,~$\omega$
defined by~\eqref{cocycle} is a 2-cochain,~\hbox{$\omega
\in C^2(\LieA,\LieAxb)$}, that measures the ``failure''
of~$\tau$ to be a homomorphism. We now show, moreover, that~$\omega$ is a
2-cocycle,~\hbox{$\omega \in Z^2_\rho(\LieA,\LieAxb)$}.  By
using~\eqref{3cobound},
\begin{eqnarray*}
	\!\!\!\!\!\!\!
	(s\,\omega)(\alpha,\beta,\gamma) &=&
		\rho_{\alpha}\,\omega(\beta,\gamma)
		+ \rho_{\beta}\,\omega(\gamma,\alpha)
		+ \rho_{\gamma}\,\omega(\alpha,\beta)\nonumber\\
	&&\mbox{}- \omega({\lpb\alpha\com\beta\rpb}_\LieA,\gamma)
		- \omega({\lpb\beta\com\gamma\rpb}_\LieA,\alpha) 
		- \omega({\lpb\gamma\com\alpha\rpb}_\LieA,\beta)\,,
		\nonumber\\
	&=& i^{-1}\l({\lpb\tau\,\alpha\com
		{\lpb\tau\,\beta\com\tau\,\gamma\rpb}_\LieAx\rpb}_\LieAx
		+ \cycperm\r)\nonumber\\
	&&\mbox{} + i^{-1}\tau\l({\lpb{\lpb
		\alpha\com\beta\rpb}_\LieA\com\gamma\rpb}_\LieA
		+ \cycperm\r) = 0.
\end{eqnarray*}
\inthesis{Include all commented out steps.}
The first parenthesis vanishes by the Jacobi identity in~$\LieAx$, the second
by the Jacobi identity in~$\LieA$, and the other terms were canceled in
pairs. Hence,~$\omega$ is a 2-cocycle.

Two extensions~$\LieAx$ and~$\LieAx'$ are equivalent if there exists a Lie
algebra isomorphism~$\sigma$ such that the diagram
\begin{equation}
	\xymatrix@M=4pt{
	& & {\LieAx} \ar[dr]^{\pi} \ar[dd]^{\sigma} & & \\
	0 \ar[r] & {\LieAxb} \ar[ur]^i \ar[dr]_{i'} & & {\LieA} \ar[r] & 0 \\
	& & {\LieAx'} \ar[ur]_{\pi'} & &
	}
	\eqlabel{equivext}
\end{equation}
is commutative, that is if~\hbox{$\sigma\circ i=i'$ and~$\pi =
\pi'\circ\sigma$}.

There will be an injection~$\tau$ associated with~$\pi$ and a~$\tau'$
associated with~$\pi'$, such that~\hbox{$\pi\circ\tau = 1_{|\LieA} =
\pi'\circ\tau'$.} The linear map~\hbox{$\nu = \sigma^{-1}\tau' - \tau$} must
be from~$\LieA$ to~$i\,\LieAxb$, so~\hbox{$i^{-1}\nu \in C^1(\LieA,\LieAxb)$}.
Consider~$\rho$ and~$\rho'$ respectively defined using~$\tau,i$ and~$\tau',i'$
by~\eqref{rhodef}. Then
\begin{eqnarray}
	(\rho_\alpha - {\rho'}_\alpha)\,\eta 
		&=& i^{-1}{\lpb\tau\,\alpha\com i\,\eta\rpb}_\LieAx -
		{i'}^{-1}{\lpb\tau'\,\alpha\com i'\,\eta\rpb}_{\LieAx'}\ ,
		\nonumber\\
	&=& i^{-1}{\lpb\tau\,\alpha\com i\,\eta\rpb}_\LieAx -
		i^{-1}{\lpb(\nu + \tau)\,\alpha\com
		i\,\eta\rpb}_\LieAx\ ,\nonumber\\
	&=& -i^{-1}{\lpb\nu\,\alpha\com i\,\eta\rpb}_\LieAx = 0,
	\eqlabel{samerho}
\end{eqnarray}
\inthesis{Include all commented out steps.}
since~$\LieAxb$ is Abelian. Hence $\tau$ and~$\tau'$ define the same~$\rho$.
Now consider the 2-cocycles~$\omega$ and~$\omega'$ defined from~$\tau$
and~$\tau'$ by~\eqref{cocycle}. We have
\begin{eqnarray*}
	\!\!\!\!\!\!\!\!\!
	\omega'(\alpha,\beta) - \omega(\alpha,\beta)
		&=& {i'}^{-1}
		\l({\lpb\tau'\,\alpha\com\tau'\,\beta\rpb}_{\LieAx'}
		- \tau'\,{\lpb\alpha\com\beta\rpb}_\LieA\r)\nonumber\\
	&&\mbox{} - i^{-1}\l({\lpb\tau\,\alpha\com\tau\,\beta\rpb}_\LieAx
		- \tau\,{\lpb\alpha\com\beta\rpb}_\LieA\r),\nonumber\\
	&=& i^{-1}\l({\lpb\tau\,\alpha\com\nu\,\beta\rpb}_\LieAx
		+ {\lpb\nu\,\alpha\com\tau\,\beta\rpb}_\LieAx
		- \nu\,{\lpb\alpha\com\beta\rpb}_\LieA\r),\nonumber\\
	&=& \rho_\alpha\,(i^{-1}\nu\,\beta) - \rho_\beta\,(i^{-1}\nu\,\alpha)
		- i^{-1}\nu\,{\lpb\alpha\com\beta\rpb}_\LieA.
\end{eqnarray*}
\inthesis{Include all commented out steps.}
Comparing this with~\eqref{2cobound}, we see that
\begin{equation}
	\omega' - \omega = s\,(i^{-1}\nu),
	\eqlabel{cobdiff}
\end{equation}
so~$\omega$ and~$\omega'$ differ by a coboundary. Hence they represent the
same element in~$H^2_\rho(\LieA,\LieAxb)$. Equivalent extensions uniquely
define an element of the second cohomology group~$H^2_\rho(\LieA,\LieAxb)$.
Note that this is true in particular for~$\LieAx=\LieAx'$,~$\sigma=1$, so that
the element of~$H^2_\rho(\LieA,\LieAxb)$ is independent of the choice
of~$\tau$.

We are now ready to write down explicitly the bracket in~$\LieAx$. We can
represent an element~\hbox{$\alpha \in \LieAx$} as a two-tuple:~\hbox{$\alpha
= (\alpha_1, \alpha_2)$} where~\hbox{$\alpha_1 \in \LieA$} and~\hbox{$\alpha_2
\in \LieAxb$} (\hbox{$\LieAx = \LieA \oplus \LieAxb$} as a vector space). The
injection~$i$ is then~\hbox{$i\,\alpha_2 = (0,\alpha_2)$}, the
projection~$\pi$ is~\hbox{$\pi\,(\alpha_1,\alpha_2) = \alpha_1$}, and since
the extension is independent of the choice of~$\tau$ we
take~\hbox{$\tau\,\alpha_1 = (\alpha_1,0)$}. By linearity
\begin{eqnarray*}
	{\lpb\alpha,\beta\rpb}_\LieAx &=&
		{\lpb(\alpha_1,0),(\beta_1,0)\rpb}_\LieAx
		+ {\lpb(0,\alpha_2),(0,\beta_2)\rpb}_\LieAx\nonumber\\
	&&\mbox{} + {\lpb(\alpha_1,0),(0,\beta_2)\rpb}_\LieAx
		+ {\lpb(0,\alpha_2),(\beta_1,0)\rpb}_\LieAx.
\end{eqnarray*}
We know that~${\lpb(0,\alpha_2),(0,\beta_2)\rpb}_\LieAx = 0$ since~$\LieAxb$
is Abelian. By definition of the cocycle~$\omega$, Eq.~\eqref{cocycle}, we
have
\begin{eqnarray*}
	{\lpb(\alpha_1,0),(\beta_1,0)\rpb}_\LieAx &=&
		{\lpb\tau\,\alpha_1\com\tau\,\beta_1\rpb}_\LieAx \nonumber\\
	&=& i\,\omega(\alpha_1,\beta_1) +
		\tau\,{\lpb\alpha_1\com\beta_1\rpb}_\LieA\nonumber\\
	&=& ({\lpb\alpha_1\com\beta_1\rpb}_\LieA\,,\,\omega(\alpha_1,\beta_1)).
\end{eqnarray*}
Finally, by the definition of~$\rho$, Eq.~\eqref{rhodef}, 
\[
	{\lpb(\alpha_1,0),(0,\beta_2)\rpb}_\LieAx
		= {\lpb\tau\,\alpha_1,i\,\beta_2\rpb}_\LieAx
		= \rho_{\alpha_1}\,\beta_2,
\]
and similarly for~${\lpb(0,\alpha_2),(\beta_1,0)\rpb}_\LieAx$, with opposite
sign. So the bracket is
\begin{equation}
	{\lpb\alpha,\beta\rpb}_\LieAx =
		\Bigl({\lpb\alpha_1\com\beta_1\rpb}_\LieA\, , \,
		\rho_{\alpha_1}\,\beta_2 - \rho_{\beta_1}\,\alpha_2
		+ \omega(\alpha_1,\beta_1)\Bigr).
	\eqlabel{abelianbracket}
\end{equation}
As a check we work out the Jacobi identity in~$\LieAx$:
\begin{eqnarray*}
	\!\!\!\!\!\!\!\!
	{\lpb\alpha\com{\lpb\beta\com\gamma\rpb}_\LieAx\rpb}_\LieAx &=& 
		\l({\lpb\alpha_1\com{\lpb\beta\com\gamma\rpb}_1\rpb}_\LieA
		\,,\, \rho_{\alpha_1}\,{\lpb\beta\com\gamma\rpb}_2
		- \rho_{{\lpb\beta\com\gamma\rpb}_1}\,\alpha_2
		+ \omega(\alpha_1,{\lpb\beta\com\gamma\rpb}_1)\r)\nonumber\\
	&=& \Bigl({\lpb\alpha_1\com{\lpb\beta_1
		\com\gamma_1\rpb}_\LieA\rpb}_\LieA
		\,,\, \rho_{\alpha_1}(\rho_{\beta_1}\,\gamma_2
		- \rho_{\gamma_1}\,\beta_2 + \omega(\beta_1,\gamma_1))
		\nonumber\\
	&&\mbox{}\ \ \ \ \ \ \ \ \ \ \ \ \ \ \ \ \ \ \ \ \ \ \ \ \
		- \rho_{{\lpb\beta_1\com\gamma_1\rpb}_\LieA}\,\alpha_2
		+ \omega(\alpha_1,{\lpb\beta_1\com\gamma_1\rpb}_\LieA)\Bigr).
\end{eqnarray*}
Upon adding permutations, the first component will vanish by the Jacobi
identity in~$\LieA$. We are left with
\begin{eqnarray*}
	{\lpb\alpha\com{\lpb\beta\com\gamma\rpb}_\LieAx\rpb}_\LieAx 
		+ \cycperm &=& \Bigl(0\,,\,
	\l(\rho_{\alpha_1}\rho_{\beta_1} - \rho_{\beta_1}\rho_{\alpha_1}
		- \rho_{{\lpb\alpha_1\com\beta_1\rpb}_\LieA}\r)
		\gamma_2\nonumber\\
	&&\mbox{} \!\!\!\!\!\!\!\!\!\!\!\!\!\!
		+ \rho_{\alpha_1}\,\omega(\beta_1,\gamma_1)
		- \omega({\lpb\alpha_1\com\beta_1\rpb}_\LieA,\gamma_1)\Bigr)
		+ \cycperm,
\end{eqnarray*}
which vanishes by the the homomorphism property of~$\rho$ and the fact
that~$\omega$ is a 2-cocycle, Eq.~\eqref{3cobound}.

Equation~\eqref{abelianbracket} is the most general form of the Lie bracket
for extension by an Abelian Lie algebra. It turns out that the theory of
extension by a non-Abelian algebra can be reduced to the study of extension by
the center of~$\LieAxb$, which is Abelian~\cite{Azcarraga}. We will not need
this fact here, as the only extensions by non-Abelian algebras we will deal
with are of the simpler type of \secref{semidext}.

We have thus shown that equivalent extensions are enumerated by the second
cohomology group~$H^2_\rho(\LieA,\LieAxb)$.  The coordinate
transformation~$\sigma$ used in~\eqref{equivext} to define equivalence of
extensions preserves the form of~$\LieA$ and~$\LieAxb$ as subsets of~$\LieAx$.
However, we have the freedom to choose coordinate transformations which do
transform these subsets. All we require is that the isomorphism~$\sigma$
between~$\LieAx$ and~$\LieAx'$ be a Lie algebra isomorphism.  We can represent
this by the diagram
\begin{equation}\xymatrix@M=4pt{
	0 \ar[r] & {\LieAxb} \ar[r]^{i} & {\LieAx} \ar[r]^{\pi}
		\ar[d]^\sigma
		& {\LieA} \ar[r] & 0 \\
	0 \ar[r] & {\LieAxb'} \ar[r]^{i} & {\LieAx'} \ar[r]^{\pi}
		& {\LieA'} \ar[r] & 0. \\
	}
	\eqlabel{equivext2}
\end{equation}
The primed and the unprimed extensions are not equivalent, but they are
isomorphic~\cite{Weiss}.  Cohomology for us is not the whole story, since we
are interested in isomorphic extensions, but it will guide our classification
scheme.  We discuss this point further in \secref{furthertrans}.

\subsection{Semidirect and Direct Extensions}
\seclabel{semidext}


Assume now that~$\omega$ defined by~\eqref{cocycle} is a coboundary.
By~\eqref{cobdiff} there exists an equivalent extension with~\hbox{$\omega
\equiv 0$}.  For that equivalent extension,~$\tau$ is a homomorphism and
condition~(ii) at the end of \secref{extension} is satisfied. Thus the
sequence
\[\xymatrix@M=4pt{
	{\LieAx} & {\LieA} \ar[l]_{\tau} & 0 \ar[l]
	}
\]
is an exact sequence of Lie algebra homomorphisms, as well as the sequence
given by~\eqref{extdef}. We then say that the extension is a semidirect
extension (or a semidirect sum of algebras) by analogy with the group
case. More generally, we say that~$\LieAx$ splits if it is isomorphic to a
semidirect sum, which corresponds to~$\omega$ being a coboundary, not
necessarily zero. If~$\LieAxb$ is not Abelian, then~\eqref{samerho} is not
satisfied and two equivalent extensions (or two different choices of~$\tau$)
do not necessarily lead to the same~$\rho$.

Representing elements of~$\LieAx$ as 2-tuples, as in \secref{abelianext}, we
can derive the bracket in~$\LieAx$ for a semidirect sum,
\inthesis{Explicitely show the differences, that is, vanishing of cocycle and
nonabelianiticity.}
\begin{equation}
	{\lpb\alpha,\beta\rpb}_\LieAx =
		\Bigl({\lpb\alpha_1\com\beta_1\rpb}_\LieA\, , \,
		\rho_{\alpha_1}\,\beta_2 - \rho_{\beta_1}\,\alpha_2
		+ {\lpb\alpha_2\com\beta_2\rpb}_\LieAxb\Bigr),
	\eqlabel{sdbracket}
\end{equation}
where we have not assumed~$\LieAxb$ Abelian. Verifying Jacobi
for~\eqref{sdbracket} we find the~$\rho$ must also satisfy
\[
	\rho_{\alpha_1}\,{\lpb\beta_2\com\gamma_2\rpb}_\LieAxb
		= {\lpb\rho_{\alpha_1}\,\beta_2\com\gamma_2\rpb}_\LieAxb
		+ {\lpb\beta_2\com\rho_{\alpha_1}\,\gamma_2\rpb}_\LieAxb\, ,
\]
which is trivially satisfied if~$\LieAxb$ is Abelian, but in general this
condition states that~$\rho_\alpha$ is a derivation on~$\LieAxb$.

\inthesis{Relation to~$H^1_\rho$.}

Now consider the case where~$i^{-1}$ is a homomorphism and~\hbox{$\ker i^{-1}
= \range \tau$}. Then the sequence
\[\xymatrix@M=4pt{
	0 \ar@_{->}@<-.3ex>[r]
	& {\LieAxb} \ar@_{->}@<-.3ex>[l] \ar@_{->}@<-.3ex>[r]_{i}
	& {\LieAx} \ar@_{->}@<-.3ex>[l]_{i^{-1}} \ar@_{->}@<-.3ex>[r]_{\pi}
	& {\LieA} \ar@_{->}@<-.3ex>[l]_{\tau} \ar@_{->}@<-.3ex>[r]
	& 0 \ar@_{->}@<-.3ex>[l]
	}
\]
is exact in both directions and, hence, both~$i$ and~\hbox{$\pi=\tau^{-1}$}
are bijections. The action of~$\LieA$ on~$\LieAxb$ is
\[
	\rho_{\alpha}\,\eta = i^{-1}{\lpb\tau\,\alpha\com i\eta\rpb}_\LieAx
		 = {\lpb i^{-1}\tau\,\alpha\com \eta\rpb}_\LieAxb = 0
\]
since by exactness~\hbox{$i^{-1}\circ\tau = 0$}. This is called a direct
sum. Note that in this case the role of~$\LieA$ and~$\LieAxb$ is
interchangeable and they are both ideals in~$\LieAx$. The bracket in~$\LieAx$
is easily obtained from~\eqref{sdbracket} by letting~\hbox{$\rho=0$},
\begin{equation}
	{\lpb\alpha,\beta\rpb}_\LieAx =
		\Bigl({\lpb\alpha_1\com\beta_1\rpb}_\LieA\com
		{\lpb\alpha_2\com\beta_2\rpb}_\LieAxb\Bigr).
	\eqlabel{dbracket}
\end{equation}
Semidirect and direct extensions play an important role in physics. A simple
example of a semidirect extension structure is when $\LieA$ is the Lie
algebra~$\sothree$ associated with the rotation group $\SOthree$ and $\LieAxb$
is $\reals^3$. Their semidirect sum is the algebra of the six parameter
Euclidean group of rotations and translations. That algebra can be used in a
Lie--Poisson bracket to describe the dynamics of the heavy top (see for
example~\cite{Holmes1983,Marsden1984}). We have already discussed the
semidirect sum in \secref{lowbetaRMHD}. The bracket \eqref{RMHDbrak} is a
semidirect sum, with~$\LieA$ the algebra of the group of volume-preserving
diffeomorphisms and~$\LieAxb$ the Abelian Lie algebra of functions
on~$\reals^2$. The action is just the adjoint action~\hbox{$\rho_\alpha\,v
\ldef \lpb\alpha\com v\rpb$} obtained by identifying~$\LieA$ and~$\LieAxb$.

A Lie--Poisson bracket built from a direct extension is just a sum of the
separate brackets. The interaction between the variables can only come from
the Hamiltonian or from constitutive equations.  For example in the baroclinic
instability model of two superimposed two fluid layers with different
potential vorticities the two layers are coupled through the potential
vorticity relation~\cite{McLachlan1997}.

\section{Classification of Extensions of a Lie Algebra}
\seclabel{classext}

\ifmypprint
{\small
\begin{verbatim}
$Id: sec-classext.tex,v 2.1 1998/11/13 08:37:16 jeanluc Exp $
\end{verbatim}
}
\fi

In this section we return to the main problem introduced in
\secref{theproblem}: the classification of algebra extensions built by
forming~$n$-tuples of elements of a single Lie algebra~$\LieA$.  The elements
of this Lie algebra~$\LieAx$ are written as~\hbox{$\alpha \ldef
\l(\alpha_1,\dots,\alpha_n\r)$},~\hbox{$\alpha_i \in \LieA$}, with a
bracket defined by
\begin{equation}
	{\lpb\alpha\com\beta\rpb}_\lambda = {\W_\lambda}^{\mu\nu}\,
		\lpb\alpha_\mu\com\beta_\nu\rpb,
	\tag{\ref{eq:extbrack}}
\end{equation}
where~${\W_\lambda}^{\mu\nu}$ are constants. We will call~$n$ the \emph{order}
of the extension. Recall (see \secref{theproblem}) the~$\W$'s are symmetric in
their upper indices,
\begin{equation}
	{\W_\lambda}^{\mu\nu} = {\W_\lambda}^{\nu\mu}\,,
	\tag{\ref{eq:Wcommute}}
\end{equation}
and commute,
\begin{equation}
	\W^{(\nu)}\,\W^{(\sigma)} = \W^{(\sigma)}\,\W^{(\nu)},
	\tag{\ref{eq:Wcommute}}
\end{equation}
where the~\hbox{$n\times n$} matrices~$\W^{(\nu)}$ are defined
by~${{[\W^{(\nu)}]}_\lambda}^{\mu} := {\W_\lambda}^{\nu\mu}$.  Since
the~$\W$'s are 3-tensors we can also represent their elements by matrices
obtained by fixing the lower index,
\begin{equation}
	\W_{(\lambda)}\ :\ {\l[\W_{(\lambda)}\r]}^{\mu\nu} :=
	{\W_{\lambda}}^{\mu\nu},
	\eqlabel{lowindex}
\end{equation}
which are symmetric but do not commute.  Either collection of
matrices,~\eqref{Wupdef} or~\eqref{lowindex}, completely describes the Lie
bracket, and which one we use will be understood by whether the parenthesized
index is up or down.

What do we mean by a classification?  A classification is achieved if we
obtain a set of normal forms for the extensions that are independent, that is,
not related by linear transformations.  We use linear transformations because
they preserve the Lie--Poisson structure---they amount to transformations of
the~$\W$ tensor.  We thus begin by assuming the most general~$\W$ possible.

We first show in \secref{directprod} how an extension can be broken down into
a direct sum of degenerate subblocks (degenerate in the sense that the
eigenvalues have multiplicity greater than unity). The classification scheme
is thus reduced to the study of a single degenerate subblock. In
\secref{classcoho} we couch our particular extension problem in terms of the
Lie algebra cohomology language of \secref{extension} and apply the techniques
therein. The limitations of this cohomology approach are investigated in
\secref{furthertrans}, and we look at other coordinate transformations
that do not necessarily preserve the extension structure of the algebra, as
expressed in diagram~\eqref{equivext2}. In \secref{Leibniz} we introduce a
particular type of extension, called the Leibniz extension, which is in a sense
the ``maximal'' extension. Finally, in \secref{lowdimext} we give an explicit
classification of solvable extensions up to order four.

\subsection{Direct Sum Structure}
\seclabel{directprod}

A set of commuting matrices can be put into simultaneous block-diagonal form
by a coordinate transformation, each block corresponding to a degenerate
eigenvalue~\cite{Suprunenko}. Let us denote the change of basis by a
matrix~${\M_{\beta}}^{\bar\alpha}$, with
inverse~${\l(M^{-1}\r)_{\bar\alpha}}^{\beta}$, such that the
matrix~\hbox{${\Wt}^{(\nu)}$}, whose components are given by
\[
	{\Wt_{\bar\beta}}{}^{\bar\alpha\nu} =
		{(M^{-1})_{\bar\beta}}^{\lambda} \,
		{\W_{\lambda}}^{\mu\nu} \, {\M_{\mu}}^{\bar\alpha}\ ,
\]
is in block-diagonal form for all~$\nu$~\cite{Suprunenko}.
However,~${\W_\lambda}^{\mu\nu}$ is a 3-tensor and so the third index is
also subject to the coordinate change:
\[
	{\Wb_{\bar\beta}}^{\bar\alpha\bar\gamma} =
	{\Wt_{\bar\beta}}\,{}^{\bar\alpha\nu}
		{\M_{\nu}}^{\bar\gamma}\, .
\]
This last step adds linear combinations of the~$\Wt^{(\nu)}$'s together, so
the~$\Wt^{(\nu)}$'s and the~$\Wb^{(\bar\gamma)}$'s have the same
block-diagonal structure.  Note that the~${\Wb}$ tensors are still symmetric
in their upper indices, since this property is preserved by a change of basis.
So from now on we just assume that we are working in a basis where
the~$\W^{(\nu)}$'s are block-diagonal and symmetric in their upper indices;
this symmetry means that if we look at a~$\W$ as a cube, then in the
block-diagonal basis it consists of smaller cubes along the main
diagonal. This is the 3-tensor equivalent of a block-diagonal matrix.

Block-diagonalization is the first step in the classification: each block
of~$\W$ is associated with an ideal (hence, a subalgebra) in the
full~$n$-tuple algebra~$\LieA$. \inthesis{Show this.} Hence, by the definition
of \secref{semidext} the algebra~$\LieA$ is a direct sum of the
subalgebras associated with each block. Each of these subalgebras can be
studied independently, so from now on we assume that we have~$n$ commuting
matrices, each with~$n$-fold degenerate eigenvalues. The eigenvalues can,
however, be different for each matrix. \inthesis{Could say something about
reducibility here.}

Such a set of commuting matrices can be put into lower-triangular form by a
coordinate change, and again the transformation of the third index preserves
this structure (though it changes the eigenvalue of each matrix).  The
eigenvalue of each matrix lies on the diagonal; we denote the eigenvalue
of~$\W^{(\mu)}$ by~$\ev^{(\mu)}$. The matrix~${\W_{(1)}}$, which as prescribed
by~\eqref{lowindex} consists of the first row of the lower-triangular
matrices~$\W^{(\mu)}$, is given by
\[
	{\W_{(1)}} = \l(\begin{array}{ccccc}
		\ev^{(1)} & 0 & 0 & \cdots & 0 \\
		\ev^{(2)} & 0 & 0 & \cdots & 0 \\
		\vdots  &   &   &        & \vdots \\
		\ev^{(n)} & 0 & 0 & \cdots & 0
		\end{array}\r).
\]
Evidently, the symmetry of~${\W_{(1)}}$ requires
\[
	\ev^{(\nu)} = \evone\,{\delta_1}^\nu\, ;
\]
that is, all the matrices~$\W^{(\mu)}$ are nilpotent (their eigenvalues
vanish) except for~$\W^{(1)}$ when~\hbox{$\evone\ne 0$}.  If this first
eigenvalue is nonzero then it can be scaled to~$\evone=1$ by the coordinate
transformation~${\M_{\nu}}^{\bar\alpha} =
\evone^{-1}~{\delta_\nu}^{\bar\alpha}$. We will use the
symbol~$\zerorone$ to mean a variable that can take the value 0 or 1.

\subsection{Connection to Cohomology}
\seclabel{classcoho}

We now bring together the abstract notions of \secref{cohoext} with
the~$n$-tuple extensions of \secref{theproblem}.  It is shown in
\secref{prelimsplit} that we need only classify the case
of~\hbox{$\zerorone=0$}.  This case will be seen to correspond to solvable
extensions, which we classify in \secref{solvext}.

\subsubsection{Preliminary Splitting}
\seclabel{prelimsplit}

Assume we are in the basis described at the end of \secref{directprod} and,
for now, suppose~$\evone = 1$.  The set of elements of the form~$\beta =
\l(0,\beta_2,\dots,\beta_n\r)$ is a nilpotent ideal in~$\LieAx$ that we
denote by~$\LieAxb$ ($\LieAxb$ is thus a solvable subalgebra~\cite{Jacobson}).
Hence, we can construct the algebra~$\LieA =
\LieAx/\LieAxb$, so that~$\LieAx$ is an extension of~$\LieA$ by~$\LieAxb$.
If~$\LieA$ is semisimple, then~$\LieAxb$ is the radical of~$\LieAx$ (the
maximal solvable ideal).  It is easy to see that the elements of~$\LieA$ are
of the form~$\alpha = \l(\alpha_1,0,\dots,0\r)$.  We will now see
that~$\LieAx$ splits; that is, there exist coordinates in which~$\LieAx$ is
manifestly the semidirect sum of~$\LieA$ and the (in general non-Abelian)
algebra~$\LieAxb$.

In~\apxref{woneident} we give a lower-triangular coordinate transformation
that makes~$\W^{(1)}=I$, the identity matrix.  Assuming we have effected this
transformation, the mappings~$i$,~$\pi$, and~$\tau$ of \secref{extension} are
given by
\begin{alignat*}{3}
	i & :\LieAxb\ &\longrightarrow \LieAx,\ \ \ 
		&i(\alpha_2,\dots,\alpha_n) = (0,\alpha_2,\dots,\alpha_n)
		\nonumber\\
	\pi & :\LieAx\ &\longrightarrow \LieA,\ \ \  
		&\pi(\alpha_1,\alpha_2,\dots,\alpha_n) 
		= \alpha_1,\\
	\tau & :\LieA\ &\longrightarrow \LieAx,\ \ \  
		&\tau(\alpha_1) 
		= (\alpha_1,0,\dots,0),\nonumber
\end{alignat*}
and the cocycle of Eq.~\eqref{cocycle} is
\begin{eqnarray*}
	i\,\omega(\alpha,\beta) &=& 
	{\lpb\tau\,\alpha\com\tau\,\beta\rpb}_\LieAx
	- \tau\,{\lpb\alpha\com\beta\rpb}_\LieA\nonumber\\
	&=& {\lpb(\alpha_1,0,\dots,0)\com(\beta_1,0,\dots,0)\rpb}_\LieAx
	- (\lpb\alpha_1\com\beta_1\rpb,0,\dots,0)\nonumber\\
	&=& 0.
\end{eqnarray*}
\inthesis{Extra steps?}  Since~\hbox{$\omega\equiv 0$}, the extension is a
semidirect sum (see \secref{semidext}).  The coordinate transformation that
made~$\W^{(1)}=I$ removed a coboundary, making the above cocycle vanish
identically.  For the case where~$\LieA$ is finite-dimensional and semisimple,
we have an explicit demonstration of the Levi decomposition theorem: any
finite-dimensional\footnote{The inner bracket can be infinite dimensional, but
the order of the extension is finite.} Lie algebra~$\LieAx$ (of characteristic
zero) with radical~$\LieAxb$ is the semidirect sum of a semisimple Lie
algebra~$\LieA$ and~$\LieAxb$~\cite{Jacobson}.

\subsubsection{Solvable Extensions}
\seclabel{solvext}

Above we assumed the eigenvalue~$\evone$ of the first matrix was unity;
however, if this eigenvalue vanishes, then we have a solvable algebra
of~$n$-tuples to begin with.  Since~$n$ is arbitrary we can study these two
solvable cases together.

Thus, we now suppose~$\LieAx$ is a solvable Lie algebra of~$n$-tuples (we
reuse the symbols~$\LieAx$,~$\LieA$, and~$\LieAxb$ to parallel the notation of
\secref{cohoalgebra}), where all of the the~$\W^{(\mu)}$'s are
lower-triangular with zeros along the diagonal.  Note
that~$\W^{(n)}=0$, so the set of elements of the form~$\alpha =
(0,\dots,0,\alpha_n)$ forms an Abelian subalgebra of~$\LieAx$.  In fact, this
subalgebra is an ideal.  Now assume~$\LieAx$ contains an Abelian ideal of
order~\hbox{$n-m$} (the order of this ideal is at least~$1$), which we denote
by~$\LieAxb$.  The elements of~$\LieAxb$ can always be cast in the form
\[
	\alpha = (0,\dots,0,\alpha_{m+1},\dots,\alpha_n)
\]
via a coordinate transformation that preserves the lower-triangular,
nilpotent form of the~${\W}^{(\mu)}$ \comment{Proved in Lie--Poisson II ring
notepad}.

We also denote by~$\LieA$ the algebra of~$m$-tuples with the bracket
\[
	{{\lpb(\alpha_1,\dots,\alpha_{m})
	\com(\beta_1,\dots,\beta_{m})\rpb}_\LieA}_\lambda
	= \sum_{\mu,\nu = 1}^{m} {\W_\lambda}^{\mu\nu}\,
	\lpb\alpha_\mu\com\beta_\nu\rpb\,,\ \ \lambda=1,\dots,m.
\]
It is trivial to show that~$\LieA = \LieAx/\LieAxb$, so that~$\LieAx$ is an
extension of~$\LieA$ by~$\LieAxb$.  Since~$\LieAxb$ is Abelian we can use the
formalism of \secref{cohoalgebra} (the other case we used above was
for~$\LieAxb$ non-Abelian but where the extension was semidirect).  The
injection and projection maps are given by
\begin{alignat*}{3}
	i& : \LieAxb\ &\longrightarrow \LieAx,\ \ \ 
		&i(\alpha_{m+1},\dots,\alpha_n)
		= (0,\dots,0,\alpha_{m+1},\dots,\alpha_n),\nonumber\\
	\pi& : \LieAx\ &\longrightarrow \LieA,\ \ \  
		&\pi(\alpha_1,\alpha_2,\dots,\alpha_n) 
		= (\alpha_1,\dots,\alpha_{m}),\\
	\tau & :\LieA\ &\longrightarrow \LieAx,\ \ \  
		&\tau(\alpha_1,\dots,\alpha_{m}) 
		= (\alpha_1,\dots,\alpha_{m},0,\dots,0).\nonumber
\end{alignat*}
From the definition of the action, Eq.~\eqref{rhodef}, we have
for~\hbox{$\alpha \in \LieA$} and~\hbox{$\eta \in \LieAxb$},
\begin{eqnarray}
i\,\rho_\alpha\,\eta &=& {\lpb \tau\,\alpha\com i\,\eta\rpb}_\LieAx\nonumber\\
	&=& {\lpb (\alpha_1,\dots,\alpha_{m},0,\dots,0)
		\com (0,\dots,0,\eta_{m+1},\dots,\eta_n)\rpb}_\LieAx
		\nonumber\\
	&=& \sum_{\mu=1}^{m}\,\sum_{\nu=m+1}^{n-1}(0,\dots,0,
		\W_{m+2}^{\,\,\mu\nu}{\lpb\alpha_\mu\com\eta_\nu\rpb}
		,\dots,
		{\W_n}^{\mu\nu}{\lpb\alpha_\mu\com\eta_\nu\rpb}).
	\eqlabel{solvaction}
\end{eqnarray}
In addition to the action, the solvable extension is also characterized by the
cocycle defined in Eq.~\eqref{cocycle},
\begin{eqnarray}
	\!\!\!\!\!\!\!\!\!
	i\,\omega(\alpha,\beta) &=& 
	{\lpb\tau\,\alpha\com\tau\,\beta\rpb}_\LieAx
	- \tau\,{\lpb\alpha\com\beta\rpb}_\LieA\nonumber\\
	&=& {\lpb(\alpha_1,\dots,\alpha_{m},0,\dots,0)
		\com(\beta_1,\dots,\beta_{m},0,\dots,0)\rpb}_\LieAx
		\nonumber\\
		&&\mbox{} - \tau\,{\lpb(\alpha_1,\dots,\alpha_{m})
		\com(\beta_1,\dots,\beta_{m})\rpb}_\LieA\nonumber\\
	&=& \sum_{\mu,\nu=1}^{m}(0,\dots,0,
		\W_{m+1}^{\!\!\mu\nu}{\lpb\alpha_\mu\com\beta_\nu\rpb}
		,\dots,{\W_n}^{\mu\nu}{\lpb\alpha_\mu\com\beta_\nu\rpb}).
	\eqlabel{solvcocycle}
\end{eqnarray}
We can illustrate which parts of the~$\W$'s contribute to the action and which
to the cocycle by writing
\begin{equation}
	\W_{(\lambda)} = \l(\begin{array}{c|c}
		{\bf \ww}_\lambda\ & \ {\bf r}_\lambda \\ \hline
		{\bf r}_\lambda^T\   & {\bf 0}
	\end{array}\r),\ \ \lambda=m+1,\dots,n,
	\eqlabel{Wform}
\end{equation}
where the~${\bf\ww}_\lambda$'s are~\hbox{$m\times m$} symmetric matrices that
determine the cocycle~$\omega$ and the~${\bf r}_\lambda$'s are~\hbox{$m\times
(n-m)$} matrices that determine the action~$\rho$.  The \hbox{$(n-m)\times(n-m)$}
zero matrix on the bottom right of the~$\W_{(\lambda)}$'s is a consequence
of~$\LieAxb$ being Abelian.

The algebra~$\LieA$ is completely characterized by the~$\W_{(\lambda)}$,
$\lambda = 1,\dots,m$.  Hence we can look for the maximal Abelian ideal
of~$\LieA$ and repeat the procedure we used for the full~$\LieAx$.  It is
straightforward to show that although coordinate transformations of~$\LieA$
might change the cocycle~$\omega$ and the action~$\rho$, they will not alter
the \emph{form} of~\eqref{Wform} \comment{Proved in Lie--Poisson II ring
notepad}.


Recall that in \secref{cohoalgebra} we defined 2-coboundaries as 2-cocycles
obtained from 1-cochains by the coboundary operator,~$s$.  The 2-coboundaries
turned out to be removable obstructions to a semidirect sum structure.  Here
the coboundaries are associated with the parts of the~$\W_{(\lambda)}$ that
can be removed by (a restricted class of) coordinate transformations, as shown
below.

Let us explore the connection between 1-cochains and coboundaries in the
present context.  Since a 1-cochain is just a linear mapping from~$\LieA$
to~$\LieAxb$, for~\hbox{$\alpha = (\alpha_1,\dots,\alpha_{m}) \in \LieA$} we
can write this as
\begin{equation}
	\omega^{(1)}_\mu(\alpha) =
		-\sum_{\lambda=1}^{m}{k_\mu}^\lambda\, \alpha_\lambda\,,
		\ \ \mu=m+1,\dots,n,
	\eqlabel{thecobound}
\end{equation}
where the~${k_\mu}^\lambda$ are arbitrary constants.  To find the form of a
2-coboundary we act on the 1-cochain~\eqref{thecobound} with the coboundary
operator; using~\eqref{2cobound} and~\eqref{solvaction} we obtain
\begin{eqnarray}
	\omega^{\rm cob}_\lambda(\alpha,\beta)
		&=& (s\,\omega^{(1)})(\alpha,\beta),
		\nonumber\\
	&=& \rho_\alpha\omega^{(1)}(\beta) + \rho_\beta\omega^{(1)}(\alpha)
		- \omega^{(1)}({\lpb\alpha\com\beta\rpb}_\LieA),
		\nonumber\\
	&=& \sum_{\mu=1}^{m}\,\sum_{\nu=m+1}^{n}\,{\W_\lambda}^{\mu\nu}
		\lpb\alpha_\mu\com\omega^{(1)}_\nu(\beta)\rpb\eqlabel{cob1}\\
	&&\mbox{} - \sum_{\mu=1}^{m}\,\sum_{\nu=m+1}^{n}\,{\W_\lambda}^{\mu\nu}
		\lpb\beta_\mu\com\omega^{(1)}_\nu(\alpha)\rpb
		+ \sum_{\mu,\nu,\sigma=1}^{m}{k_\lambda}^\sigma\,
		{\W_\sigma}^{\mu\nu}
		\lpb\alpha_\mu\com\beta_\nu\rpb.\nonumber
\end{eqnarray}
After inserting~\eqref{thecobound} into~\eqref{cob1} and relabeling, we obtain
the general form of a 2-coboundary
\[
	\omega^{\rm cob}_\lambda(\alpha,\beta) = \sum_{\mu,\nu=1}^{m}\,
		{\Wcob_\lambda}^{\mu\nu}
		\lpb\alpha_\mu\com\beta_\nu\rpb,
		\ \ \ \lambda=m+1,\dots,n,
\]
where
\begin{equation}
	{\Wcob_\lambda}^{\mu\nu} \ldef
		\sum_{\tau=1}^{m}\,
		{k_\lambda}^\tau\,{\W_\tau}^{\mu\nu}
		- \sum_{\sigma=m+1}^{n}\,
		\l(
			{k_\sigma}^\mu\,{\W_\lambda}^{\nu\sigma}
			+ {k_\sigma}^\nu\,{\W_\lambda}^{\mu\sigma}
		\r).
	\eqlabel{cob}
\end{equation}

To see how coboundaries are removed, consider the lower-triangular coordinate
transformation
\[
	\l[{\M_\sigma}^{\bar \tau}\r] = \l(\begin{array}{c|c}
		{\bf I}\ \  & \,\ {\bf 0} \\ \hline
		{\bf k}\ \ & \ c\,{\bf I}
	\end{array}\r),
\]
where~$\sigma$ labels rows.  This transformation subtracts~$\Wcob_{(\lambda)}$
from~$\W_{(\lambda)}$ for \hbox{$\lambda>m$} and leaves the first~$m$ of the
$\W_{(\lambda)}$'s unchanged\comment{Proved in Lie--Poisson II ring notepad}.
In other words, if~${\Wb}$ is the transformed~$\W$,
\begin{equation}
	{\Wb_{(\lambda)}} = \l\{\begin{array}{ll}
		\W_{(\lambda)}\,,\ \ &\lambda = 1,\dots,m;\\
		\l(\begin{array}{c|c}
			c^{-1}\,({\bf w}_\lambda - {\bf \Wcob}_\lambda)\ 
				& \ {\bf r}_\lambda \\ \hline
			{\bf r}_\lambda^T\   & {\bf 0}
		\end{array}\r),\ \ &\lambda=m+1,\dots,n.
	       \end{array}\r.
	\eqlabel{newW}
\end{equation}
We have also included in this transformation an arbitrary scale factor~$c$.
Since by~\eqref{solvcocycle} the block in the upper-left characterizes the
cocycle, we see that the transformed cocycle is the cocycle characterized
by~${\bf\ww}_\lambda$ minus the coboundary characterized
by~${\bf\Wcob}_\lambda$.

The special case we will encounter most often is when the maximal Abelian
ideal of~$\LieAx$ simply consists of elements of the
form~\hbox{$(0,\dots,0,\alpha_n)$}.  For this case~$m=n-1$, and the action
vanishes since~\hbox{${\W_n}^{\mu n}=0$} (the extension is central).  The
cocycle~$\omega$ is entirely determined by~$\W_{(n)}$.  The form of the
coboundary is reduced to
\begin{equation}
	{\Wcob_n}^{\mu\nu} =
		\sum_{\tau=1}^{n-1}\,
		{k_n}^\tau\,{\W_\tau}^{\mu\nu},
	\eqlabel{cobnoaction}
\end{equation}
that is, a linear combinations of the first~$(n-1)$ matrices. Thus it is easy
to see at a glance which parts of the cocycle characterized $\W_{(n)}$ can be
removed by lower-triangular coordinate transformations.

\subsection{Further Coordinate Transformations}
\seclabel{furthertrans}

In the previous section we restricted ourselves to lower-triangular coordinate
transformations, which in general preserve the lower-triangular structure of
the~$\W^{(\mu)}$.  But when the matrices are relatively sparse, there exist
non-lower-triangular coordinate transformations that nonetheless preserve the
lower-triangular structure.  As alluded to in \secref{abelianext}, these
transformations are outside the scope of cohomology theory, which is
restricted to transformations that preserve the exact form of the action and
the algebras~$\LieA$ and~$\LieAxb$, as shown by~\eqref{newW}.  In other words,
cohomology theory classifies extensions \emph{given} $\LieA$, $\LieAxb$,
and~$\rho$.  We need not obey this restriction.  We can allow
non-lower-triangular coordinate transformations as long as they preserve the
lower-triangular structure of the~$\W^{(\mu)}$'s.

We now discuss a particular class of such transformations that will be useful
in \secref{lowdimext}.  Consider the case where both the algebra of
$(n-1)$-tuples~$\LieA$ and that of $1$-tuples~$\LieAxb$ are Abelian.  Then the
possible (solvable) extensions, in lower triangular form, are characterized
by~$\W_{(\lambda)}=0$, $\lambda=1,\dots,n-1$, with $\W_{(n)}$ arbitrary
(except for~${\W_n}^{\mu n}=0$).  Let us apply a coordinate change of the form
\[
	\M = \l(\begin{array}{c|c} {\bf \mm} \ & {\bf 0} \\ \hline
		{\bf 0} & \ c\,
	\end{array}\r),
\]
where~${\bf \mm}$ is an~$(n-1)\times (n-1)$ nonsingular matrix and~$c$ is
again a nonzero scale factor.  Denoting by~$\Wb$ the transformed~$\W$,
we have
\begin{equation}
	{\Wb_{(\lambda)}} = \l\{\begin{array}{ll}
		0\,,\ \ &\lambda = 1,\dots,n-1;\\
		\l(\begin{array}{c|c}
			c^{-1}\,{\bf m}^T\,{\bf \ww}_\lambda\,{\bf m}\ 
				& \ {\bf 0} \\ \hline
			{\bf 0}\   & \ \ 0\
		\end{array}\r),\ \ &\lambda=n.
	       \end{array}\r.
	\eqlabel{newW2}
\end{equation}

This transformation does not change the lower-triangular form of the
extension, even if~${\bf\mm}$ is not lower-triangular.  The manner in
which~${\bf\ww}_{n}$ is transformed by~$\M$ is very similar to that of a
(possibly singular) metric tensor: it can be diagonalized and rescaled such
that all its eigenvalues are~$0$ or~$\pm 1$.  We can also change the overall
sign of the eigenvalues using~$c$ (something that cannot be done for a metric
tensor).  Hence, we shall order the eigenvalues such that the~$+1$'s come
first, followed by the~$-1$'s, and finally by the~$0$'s.  We will show in
\secref{lowdimext} how the negative eigenvalues can be eliminated to
harmonize the notation.

\inthesis{
Another useful transformation simply exchanges the position of two
matrices~$\W_{()}$.
}

\subsection{Appending a Semisimple Part}
\seclabel{semisimple}

In \secref{classcoho} we showed that because of the Levi decomposition theorem
we only needed to classify the solvable part of the extension for a given
degenerate block. Most physical applications have a semisimple part
($\evone=1$); when this is so, we shall label the matrices
by~$\W^{(0)},\W^{(1)},\dots,\W^{(n)}$, where they are now of
size~\hbox{$n+1$} and~$\W^{(0)}$ is the identity.%
\footnote{The term semisimple is not quite precise: if the base algebra is not
semisimple then neither is the extension. However we will use the term to
distinguish the different cases.} Thus the matrices labeled
by~$\W^{(1)},\dots,\W^{(n)}$ will always form a solvable subalgebra. This
explains the labeling in \secreftwo{matlowbetaRMHD}{matCRMHD}.

If the extension has a semisimple part ($\zerorone=1$, or
equivalently~$\W^{(0)}=I$), we shall refer to it as \emph{semidirect}.  This
was the case treated in \secref{prelimsplit}.  If the extension is not
semidirect, then it is solvable (and contains~$n$ matrices instead of~$n+1$).

Given a solvable algebra of~$n$-tuples we can carry out in some sense the
inverse of the Levi decomposition and append a semisimple part to the
extension.  Effectively, this means that the~\hbox{$n\times n$}
matrices~$\W^{(1)},\dots,\W^{(n)}$ are made~\hbox{$n+1 \times n+1$} by adding
a row and column of zeros.  Then we simply append the matrix~$\W^{(0)}=I$ to
the extension.  In this manner we construct a semisimple extension from a
solvable one.  This is useful since we will be classifying solvable
extensions, and afterwards we will want to recover their semidirect
counterpart.

The extension obtained by appending a semisimple part to the completely
Abelian algebra of~$n$-tuples will be called \emph{pure semidirect}.  It is
characterized by~$\W^{(0)}=I$, and~${\W_\lambda}^{\mu\nu}=0$ for~$\mu,\nu>0$.

\subsection{Leibniz Extension}
\seclabel{Leibniz}

A particular extension that we shall consider is called the Leibniz
extension~\cite{Parthasarathy1976}.  For the solvable case this extension has
the form
\begin{equation}
	{\W}^{(1)} \rdef \Nilb = \l(\begin{array}{ccccc}
	\matzero & & & & \\
	\matone & \matzero & & & \\
	  & \matone & \matzero & & \\
	  &   & \cdots & \cdots & \\
	  &   & & \matone & \matzero
	\end{array}\r)
	\eqlabel{Nilb}
\end{equation}
or~${\W_\lambda}^{\mu\,1} = {\delta_{\lambda-1}}^{\mu}$,~$\lambda>1$; i.e. the
first matrix is an~\hbox{$n \times n$} Jordan block.  In this case the other
matrices, in order to commute with~$\W^{(1)}$, must be in striped
lower-triangular form~\cite{Suprunenko}.  After using the symmetry of the
upper indices the matrices can be reduced to
\inthesis{Add a few steps here.}
\begin{equation}
	\W^{(\nu)} = (\Nilb)^\nu,
	\eqlabel{Nilbmu}
\end{equation}
where on the right-hand side the~$\nu$ denotes an exponent, not a superscript.
An equivalent way of characterizing the Leibniz extension is
\begin{equation}
	{\W_\lambda}^{\mu\nu} = {\delta_\lambda}^{\mu+\nu}\,,
	\ \ \ \mu,\nu, \lambda = 1,\dots,n.
	\eqlabel{sLeib}
\end{equation}
The tensor~$\delta$ is an ordinary Kronecker delta.  Note that
neither~\eqref{Nilbmu} nor~\eqref{sLeib} are covariant expressions, reflecting
the coordinate-dependent nature of the Leibniz extension.

The Leibniz extension is in some sense a ``maximal'' extension: it is the only
extension that has~\hbox{$\W_{(\lambda)} \ne 0$} for
\emph{all}~$\lambda=2,\dots,n$ (up to coordinate transformations).  Its
uniqueness will become clear in \secref{lowdimext}, and is discussed in
Thiffeault~\cite{Thiffeault1998diss}.

To construct the semidirect Leibniz extension, we append~$\W^{(0)}=I$, a
square matrix of size~$n+1$, to the solvable Leibniz extension above, as
described in \secref{semisimple}.


\subsection{Low-order Extensions}
\seclabel{lowdimext}

We now classify the algebra extensions of low order. As demonstrated in
\secref{classcoho} we only need to classify solvable algebras, which
means that~$\W^{(n)}=0$ for all cases. We will do the classification up to
order~$n=4$.  For each case we first write down the most general set of
lower-triangular matrices~$\W^{(\nu)}$ (we have already used the fact that a
set of commuting matrices can be lower-triangularized) with the
symmetry~${\W_\lambda}^{\mu\nu}={\W_\lambda}^{\nu\mu}$ built in. Then we look
at what sort of restrictions the commutativity of the matrices places on the
elements. Finally, we eliminate coboundaries for each case by the methods of
\secreftwo{classcoho}{furthertrans}. This requires coordinate
transformations, but we usually will not bother using new symbols and just
assume the transformation was effected.

Note that, due to the lower-triangular structure of the extensions, the
classification found for an~$m$-tuple algebra applies to the first~$m$
elements of an~$n$-tuple algebra,~\hbox{$n>m$}.  Thus,~$\W_{(n)}$ is the
cocycle that contains all of the new information not included in the
previous~\hbox{$m=n-1$} classification.  These comments will become clearer as
we proceed.

We shall call an order~$n$ extension \emph{trivial} if~\hbox{$\W_{(n)} \equiv
0$}, so that the cocycle appended to the order~$n-1$ extension contributes
nothing to the bracket.

\subsubsection{n=1}
\seclabel{neqone}

This case is Abelian, with the only possible element~${W_1}^{11}=0$.

\subsubsection{n=2}
\seclabel{neqtwo}

The most general lower-triangular form for the matrices is
\[
\W^{(1)} = \l(\begin{array}{cc} \matzero & \matzero \\
	{\W_2}^{11} & \matzero \end{array}\r),
\ \ \ \
\W^{(2)} = \l(\begin{array}{cc} \matzero & \matzero \\
	\matzero & \matzero \end{array}\r).
\]
If~${\W_2}^{11} \ne 0$, then we can rescale it to unity.  Hence we
let~${\W_2}^{11} \ldef \zerorone_1$, where~$\zerorone_1 = 0$ or $1$.  The
case~$\zerorone_1 = 0$ is the Abelian case, while for~$\zerorone=1$ we have
the~$n=2$ Leibniz extension (\secref{Leibniz}).  Thus for~$n=2$ there are only
two possible algebras.  The cocycle which we have added at this stage is
characterized by~$\zerorone_1$.

\subsubsection{n=3}
\seclabel{neqthree}

Using the result of \secref{neqtwo}, the most general lower-triangular form is
\[
\W^{(1)} = \l(\begin{array}{ccc}
	\matzero & \matzero & \matzero \\
	\zerorone_1 & \matzero & \matzero \\
	{\W_3}^{11} & {\W_3}^{21} & \matzero \end{array}\r),
\ \ \ \
\W^{(2)} = \l(\begin{array}{ccc}
	\matzero & \matzero & \matzero \\
	\matzero & \matzero & \matzero \\
	{\W_3}^{21} & {\W_3}^{22} & \matzero \end{array}\r),
\]
and~$\W^{(3)} = 0$.  These satisfy the symmetry condition~\eqref{upsym}, and
the requirement that the matrices commute leads to the condition
\[
	\zerorone_1\,{\W_3}^{22} = 0.
\]
The symmetric matrix representing the cocycle is
\begin{equation}
W_{(3)} = \l(\begin{array}{ccc}
	{\W_3}^{11} & {\W_3}^{21} & \matzero \\
	{\W_3}^{21} & {\W_3}^{22} & \matzero \\ 
	\matzero & \matzero & \matzero \end{array}\r).
	\eqlabel{Wiii}
\end{equation}
If~$\zerorone_1 = 1$, then~${\W_3}^{22}$ must vanish.  Then,
by~\eqref{cobnoaction} we can remove from~$\W_{(3)}$ a multiple of~$\W_{(2)}$,
and therefore we may assume~${\W_3}^{11}$ vanishes.  A suitable rescaling
allows us to write~${\W_3}^{21}=\zerorone_2$, where~$\zerorone_2 = 0$ or $1$.
The cocycle for the case~$\zerorone_1=1$ is thus
\[
W_{(3)} = \l(\begin{array}{ccc}
	\matzero & \zerorone_2 & \matzero \\
	\zerorone_2 & \matzero & \matzero \\ 
	\matzero & \matzero & \matzero \end{array}\r).
\]
For~$\zerorone_2=1$ we have the Leibniz extension (\secref{Leibniz}).

If~$\zerorone_1 = 0$, we have the case discussed in \secref{furthertrans}.
For this case we can diagonalize and rescale~$\W_{(3)}$ such that
\[
W_{(3)} = \l(\begin{array}{ccc}
	\pmz_1 & \matzero & \matzero \\
	\matzero & \pmz_2 & \matzero \\ 
	\matzero & \matzero & \matzero \end{array}\r),
\]
where~$(\pmz_1,\pmz_2)$ can be~$(1,1)$, $(1,0)$, $(0,0)$, or $(1,-1)$.  This
last case, as alluded to at the end of \secref{furthertrans}, can be
transformed so that it corresponds to~$\zerorone_1=0$, $\zerorone_2=1$.  The
choice~$(1,0)$ can be transformed to the~$\zerorone_1=1$, $\zerorone_2=0$
case.
\comment{with $M = \{\{1,0,0\},\{0,0,1\},\{0,1,0\}\}$}  Finally
for~$(\pmz_1,\pmz_2)=(1,1)$ we can use the complex transformation
\[
	\fv^1\rightarrow\frac{1}{\sqrt{2}}(\fv^1+\fv^2),\ \ \
	\fv^2\rightarrow-\frac{\imi}{\sqrt{2}}(\fv^1-\fv^2),\ \ \
	\fv^3\rightarrow\fv^3,
\]
to transform to the~$\zerorone_1=0$, $\zerorone_2=1$ case.

We allow complex transformations in our classification because we are chiefly
interested in finding Casimir invariants for Lie--Poisson brackets.  If we
disallowed complex transformations, the final classification would contain a
few more members.  The use of complex transformations will be noted as we
proceed.


There are thus four independent extensions for~$n=3$, corresponding to
\[
	(\zerorone_1\com\zerorone_2)
	\in \l\{(0\com 0)\com(0 \com 1)\com(1 \com 0)\com(1 \com 1)\r\}.
\]
These will be referred to as Cases~$1$--$4$, respectively.
Cases~\ref{case:n4-00} and~\ref{case:n4-10} have~$\zerorone_2=0$, and so are
trivial ($\W_{(3)}=0$). \caseref{n4-01} is the solvable part of the
compressible reduced MHD bracket (\secref{matCRMHD}).  \caseref{n4-11} is the
solvable Leibniz extension.

\subsubsection{n=4}
\seclabel{neqfour}

Proceeding as before and using the result of \secreftwo{neqtwo}{neqthree}, we
now know that we need only write
\begin{equation}
\W_{(4)} = \l(\begin{array}{cccc}
	{\W_4}^{11} &  {\W_4}^{21} & {\W_4}^{31} & \matzero \\
	{\W_4}^{21} &  {\W_4}^{22} & {\W_4}^{32} & \matzero \\
	{\W_4}^{31} &  {\W_4}^{32} & {\W_4}^{33} & \matzero \\
	\matzero & \matzero & \matzero & \matzero
\end{array}\r).
	\eqlabel{Wfour}
\end{equation}
The matrices~$\W_{(1)}$, $\W_{(2)}$, and~$\W_{(3)}$ are given by their~$n=3$
analogues padded with an extra row and column of zeros (owing to the
lower-triangular form of the matrices).  The requirement that the
matrices~$\W^{(1)}\dots\W^{(4)}$ commute leads to the conditions%
\begin{equation}
\begin{split}
	\zerorone_2\,{\W_4}^{33} &= 0, \\
	\zerorone_2\,{\W_4}^{31} &= \zerorone_1\,{\W_4}^{22}, \\
	\zerorone_2\,{\W_4}^{32} &= 0, \\
	\zerorone_1\,{\W_4}^{32} &= 0.
	\eqlabel{commrel4}
\end{split}
\end{equation}
There are four cases to look at, corresponding to the possible values
of~$\zerorone_1$ and~$\zerorone_2$.

\begin{case}
$\zerorone_1=0$, $\zerorone_2=0$.
\caselabel{n4-00}
\end{case}

This is the unconstrained case discussed in \secref{furthertrans}, that is,
all the commutation relations \eqref{commrel4} are automatically satisfied.
We can diagonalize to give
\[
\W_{(4)} = \l(\begin{array}{cccc}
	\pmz_1' &  \matzero  & \matzero & \matzero \\
	\matzero &  \pmz_2' & \matzero & \matzero \\
	\matzero & \matzero & \pmz_3' & \matzero \\
	\matzero & \matzero & \matzero & \matzero
\end{array}\r),
\]
where
\[
(\pmz_1',\pmz_2',\pmz_3')
	\in \l\{(1,1,1),(1,1,0),(1,0,0),(0,0,0),(1,1,-1),(1,-1,0)\r\},
\]
so there are six distinct cases.  The exact form of the transformation is
unimportant, but the~$(1,1,0)$ extension can be mapped to \caseref{n4-01} (the
transformation is complex\comment{First switch matrices 3 and 4, then use the
same complex transformations for $n=3$.}),~$(1,0,0)$ can be mapped to
\caseref{n4-10}a, and~$(1,-1,0)$ can be mapped to
\caseref{n4-01}.  Finally the~$(1,1,1)$ extension can be mapped to
the~$(1,1,-1)$ case by a complex transformation.  After transforming
that $(1,1,-1)$ case, we are left with
\[
\W_{(4)} = \l(\begin{array}{cccc}
	\matzero & \matzero & \matzero & \matzero \\
	\matzero & \matzero & \matzero & \matzero \\
	\matzero & \matzero & \matzero & \matzero \\
	\matzero & \matzero & \matzero & \matzero
\end{array}\r),
\l(\begin{array}{cccc}
	\matzero & \matzero & \matone & \matzero \\
	\matzero & \matone & \matzero & \matzero \\
	1 & \matzero & \matzero & \matzero \\
	\matzero & \matzero & \matzero & \matzero
\end{array}\r).
\]
These will be called Cases~\ref{case:n4-00}a and~\ref{case:n4-00}b.

\begin{case}
$\zerorone_1=0$, $\zerorone_2=1$.
\caselabel{n4-01}
\end{case}

The commutation relations~\eqref{commrel4} reduce to ${\W_{4}}^{31} =
{\W_{4}}^{32} = {\W_{4}}^{33} = 0$, and we have
\[
\W_{(4)} = \l(\begin{array}{cccc}
	{\W_4}^{11} &  {\W_4}^{21}  & \matzero & \matzero \\
	{\W_4}^{21} &  {\W_4}^{22} & \matzero & \matzero \\
	\matzero & \matzero & \matzero & \matzero \\
	\matzero & \matzero & \matzero & \matzero
	\eqlabel{fff}
\end{array}\r).
\]
We can remove~${\W_4}^{21}$ because it is a coboundary (in this case a
multiple of~${\W_{(3)}}$).  We can also rescale appropriately to obtain four
possible extensions:~$\W_{(4)}=0$, and
\[
\W_{(4)} = 
\l(\begin{array}{cccc}
	1 & \matzero & \matzero & \matzero \\
	\matzero & \matzero & \matzero & \matzero \\
	\matzero & \matzero & \matzero & \matzero \\
	\matzero & \matzero & \matzero & \matzero
\end{array}\r),
\l(\begin{array}{cccc}
	1 & \matzero & \matzero & \matzero \\
	\matzero & \matone & \matzero & \matzero \\
	\matzero & \matzero & \matzero & \matzero \\
	\matzero & \matzero & \matzero & \matzero
\end{array}\r),
\l(\begin{array}{cccc}
	1 & \matzero & \matzero & \matzero \\
	\matzero & -1 & \matzero & \matzero \\
	\matzero & \matzero & \matzero & \matzero \\
	\matzero & \matzero & \matzero & \matzero
\end{array}\r).
\]
Again, the form of the transformation is unimportant, but it turns out that
the first of the above extensions can be mapped to \caseref{n4-10}c, and the
second and third to \caseref{n4-10}b.  This last transformation is complex.
Thus there is only one independent possibility, the trivial
extension~$\W_{(4)}=0$.

\begin{case}
$\zerorone_1=1,\zerorone_2=0$.
\caselabel{n4-10}
\end{case}

We can remove~~${\W_4}^{11}$ using a coordinate transformation.  From the
commutation requirement \eqref{commrel4} we obtain~${\W_4}^{22} =
{\W_4}^{32} = 0$.  We are left with~$\W_{(3)}=0$ and
\[
\W_{(4)} = \l(\begin{array}{cccc}
	\matzero    & {\W_4}^{21} & {\W_4}^{31} & \matzero \\
	{\W_4}^{21} & \matzero    & \matzero    & \matzero \\
	{\W_4}^{31} & \matzero    & {\W_4}^{33} & \matzero \\
	\matzero    & \matzero    & \matzero    & \matzero
\end{array}\r).
\]
Using the fact that elements of the form~$(0,\alpha_2,0,\alpha_4)$ are an
Abelian ideal of this bracket, we find
that~${\W_4}^{33}{\W_4}^{31}=0$. \comment{This involves transforming to move
the Abelian ideal to the end with\\ $M =
\{\{1,0,0,0\},\{0,0,1,0\},\{0,1,0,0\},\{0,0,0,1\}\}$,\\ making use of the
cohomology (see NB II p.69), then transforming back.}  Using an
upper-triangular transformation we can also make~${\W_4}^{21}{\W_4}^{31}=0$.
After suitable rescalings we find there are five cases. One of these,
\[
\W_{(4)} = \l(\begin{array}{cccc}
	\matzero & \matone & \matzero & \matzero \\
	1 & \matzero & \matzero & \matzero \\
	\matzero & \matzero & \matzero & \matzero \\
	\matzero & \matzero & \matzero & \matzero
\end{array}\r),
\]
may be mapped to \caseref{n4-11} (below) with~$\zerorone_3=0$. We are thus
left with four cases: the trivial extensions,~$\W_{(4)}=0$, and
\[
\W_{(4)} = \l(\begin{array}{cccc}
	\matzero & \matzero & \matzero & \matzero \\
	\matzero & \matzero & \matzero & \matzero \\
	\matzero & \matzero & \matone & \matzero \\
	\matzero & \matzero & \matzero & \matzero
\end{array}\r),
\l(\begin{array}{cccc}
	\matzero & \matzero & \matone & \matzero \\
	\matzero & \matzero & \matzero & \matzero \\
	1 & \matzero & \matzero & \matzero \\
	\matzero & \matzero & \matzero & \matzero
\end{array}\r),
\l(\begin{array}{cccc}
	\matzero & \matone & \matzero & \matzero \\
	1 & \matzero & \matzero & \matzero \\
	\matzero & \matzero & \matone & \matzero \\
	\matzero & \matzero & \matzero & \matzero
\end{array}\r).
\]
We will refer to these four extensions as Cases~\ref{case:n4-10}a--d,
respectively (Case~\ref{case:n4-10}a is the trivial extension).

\begin{case}
$\zerorone_1=1$, $\zerorone_2=1$.
\caselabel{n4-11}
\end{case}

The elements~${\W_4}^{11}$ and~${\W_4}^{21}$ are coboundaries that can be
removed by a coordinate transformation.  From \eqref{commrel4} we
have~${\W_4}^{33} = {\W_4}^{32} = 0, {\W_4}^{22} = {\W_4}^{31}
\rdef \zerorone_3$, so that
\[
\W_{(4)} = \l(\begin{array}{cccc}
	\matzero & \matzero & \zerorone_3 & \matzero \\
	\matzero & \zerorone_3 & \matzero & \matzero \\
	\zerorone_3 & \matzero & \matzero & \matzero \\
	\matzero & \matzero & \matzero & \matzero
\end{array}\r).
\]
For~$\zerorone_3=1$ we have the Leibniz extension.  The two cases will be
referred to as \caseref{n4-11}a for~$\zerorone_3=0$
and~\ref{case:n4-11}b for~$\zerorone_3=1$.

\tabref{n=4extensions} summarizes the results.  There are are total
of nine independent~$n=4$ extensions, four of which are trivial
($\W_{(4)}=0$). As noted in \secref{Leibniz} only the Leibniz extension,
\caseref{n4-11}b, has nonvanishing~$\W_{(i)}$ for all~\hbox{$1<i\le n$}.

\begin{table}
\caption{Enumeration of the independent extensions up to~$n=4$.  We
have~$\W_{(1)}=0$ for all the cases, and we have left out a row and a column
of zeros at the end of each matrix. We have also omitted cases 1--4a, for
which~$\W_{(4)}=0$.}
\tablabel{n=4extensions}
\vskip 1em

\begin{center}

\begin{tabular}{lccc@{}c@{}c} \hline

Case & $\W_{(2)}$ & $\W_{(3)}$ & \multicolumn{3}{c}{$\W_{(4)}$} \\
 &  &  & b & c & d \\[3pt] \hline

%
%
1 &
$\l(\matzero\r)$ &
$\l({\begin{array}{cc} \matzero & \matzero \\ \matzero & \matzero
	\end{array}}\r)$ &
$\l({\begin{array}{ccc} \matzero & \matzero & \matone \\ \matzero & \matone
	& \matzero \\ \matone & \matzero & \matzero
\end{array}}\r)$& &
\rule[-3em]{0cm}{6.5em}
\\

%
%
2 &
$\l(\matzero\r)$ &
$\l({\begin{array}{cc} \matzero & \matone \\ \matone & \matzero
	\end{array}}\r)$
&&&
\rule[-3em]{0cm}{6.5em}
\\

%
%
3 &
$\l(\matone\r)$ &
$\l({\begin{array}{cc} \matzero & \matzero \\ \matzero & \matzero
	\end{array}}\r)$ &
$\l({\begin{array}{ccc} \matzero & \matzero & \matzero \\
	\matzero & \matzero & \matzero \\
	\matzero & \matzero & \matone \end{array}}\r)$&
$\l({\begin{array}{ccc} \matzero & \matzero & \matone \\
	\matzero & \matzero & \matzero \\
	\matone & \matzero & \matzero \end{array}}\r)$&
$\l({\begin{array}{ccc} \matzero & \matone & \matzero \\
	\matone & \matzero & \matzero \\
	\matzero & \matzero & \matone \end{array}}\r)$
\rule[-3em]{0cm}{6.5em}
\\

%
%
4 &
$\l(\matone\r)$ &
$\l({\begin{array}{cc} \matzero & \matone \\
	\matone & \matzero \end{array}}\r)$ &
$\l({\begin{array}{ccc} \matzero & \matzero & \matone \\
	\matzero & \matone & \matzero \\
	\matone & \matzero & \matzero \end{array}}\r)$&&
\rule[-3em]{0cm}{6.5em}
\\ \hline

\end{tabular}

\end{center}
\end{table}

The surprising fact is that even to order four the normal forms of the
extensions involve no free parameters: all entries in the coefficients of the
bracket are either zero or one. There is no obvious reason this should hold
true if we try to classify extensions of order~$n>4$. It would be interesting
to find out, but the classification scheme used in this paper becomes
prohibitive at such high order. The problem is that some of the
transformations used to relate extensions cannot be systematically derived and
were obtained by educated guessing.



%
%
%
%
%
%
%
%
%
%
%
%
%
%
%
%
%

\section{Casimir Invariants for Extensions}
\seclabel{casinv}

\ifmypprint
{\small
\begin{verbatim}
$Id: sec-casimirext.tex,v 2.1 1998/11/13 08:37:28 jeanluc Exp $
\end{verbatim}
}
\fi

In this section we will use the bracket extensions of
\secref{classext} to make Lie--Poisson brackets, following the
prescription of \secref{LiePoisson}. In \secref{cascond} we write down the
general form of the Casimir condition (the condition under which a functional
is a Casimir invariant) for a general class of inner brackets. Then in
\secref{Casdirprod} we see how the Casimirs separate for a direct sum
of algebras, the case discussed in \secref{directprod}. \secref{localcas}
discusses the particular properties of Casimirs of solvable extensions. In
\secref{cassoln} we give a general solution to the Casimir problem and
introduce the concept of \emph{coextension}. Finally, in \secref{Casex} we
work out the Casimir invariants for some specific examples, including CRMHD
and the Leibniz extension.

\subsection{Casimir Condition}
\seclabel{cascond}

A generalized Casimir invariant (or Casimir for short) is a
function~$\Cas:\LieA^* \rightarrow \reals$ for which
\[
	\lPB F \com\, \Cas \rPB \equiv 0,
\]
for all~$F:\LieA^* \rightarrow \reals$.  Using~\eqref{LPB}
and~\eqref{cobracket}, we can write this as
\[
	\lang \,\fv \com\,\lpb\frac{\delta F}{\delta \fv} \com
		\frac{\delta \Cas}{\delta \fv}\rpb\,\rang
	= -\lang \lpb\frac{\delta \Cas}{\delta \fv}\com\,\fv\rpb^\dagger \com\,
		\frac{\delta F}{\delta \fv}\,\rang.
\]
Since this vanishes for all~$F$ we conclude
\begin{equation}
	\lpb\frac{\delta \Cas}{\delta \fv}\com\,\fv\rpb^\dagger = 0.
	\eqlabel{cascond}
\end{equation}
To figure out the coadjoint bracket corresponding to~\eqref{extbrack}, we
write
\[
	\lang \,\fv \com\,\lpb\alpha \com \beta\rpb\,\rang = 
	\lang \,\fv^\lambda \com\,{\W_\lambda}^{\mu\nu}
		{\lpb\alpha_\mu \com \beta_\nu\rpb}\,\rang,
\]
which after using the coadjoint bracket of~$\LieA$ becomes
\[
	\lang \lpb\beta\com\fv\rpb^\dagger\com \alpha\,\rang =
	\lang {\W_\lambda}^{\mu\nu}\lpb\beta_\nu\com\fv^\lambda\rpb^\dagger
		\com \alpha_\mu\,\rang
\]
so that
\[
	\lpb\beta\com\fv\rpb^{\dagger\,\nu} = {\W_\lambda}^{\mu\nu}
		\lpb\beta_\mu\com\fv^\lambda\rpb^\dagger.
\]
We can now write the Casimir condition~\eqref{cascond} for the bracket by
extension as
\begin{equation}
	{\W_\lambda}^{\mu\nu} \lpb\frac{\delta \Cas}{\delta\fv^\mu}
		\com\fv^\lambda\rpb^\dagger = 0,\ \ \ \ \nu=0,\dots,n.
	\eqlabel{cascond2}
\end{equation}

We now specialize the bracket to the case of most interested to us, where the
inner bracket is of canonical form~\eqref{canibrak}.  As we saw in
\secref{LiePoisson}, this is the bracket for 2-D fluid flows.  The
construction we give here has a finite-dimensional analogue, where one uses
the Cartan--Killing form to map vectors to covectors, but we will not pursue
this here (see Thiffeault~\cite{Thiffeault1998diss}).  Further, we assume that
the form of the Casimir invariants is
\begin{equation}
	\Cas[\fv] = \int_\fdomain \Casi(\fv(\xv))\d^2x,
	\eqlabel{formcas}
\end{equation}
and thus, since~$\Casi$ does not contain derivatives of~$\fv$, functional
derivatives of~$\Cas$ can be written as ordinary partial derivatives
of~$\Casi$. \inthesis{I took a stab at this in RNB III p. 136 but found no
cases that weren't exact derivatives. Also on p. 138 I show that at least for
2--D Euler we cannot have a dependence on the spatial coordinates.} We can
then rewrite~\eqref{cascond2} as
\begin{equation}
	{\W_\lambda}^{\mu\nu}
		\frac{\partial^2\Casi}{\partial\fv^\mu\partial\fv^\sigma}
		\lpb\fv^\sigma\com\fv^\lambda\rpb
	= 0,\ \ \ \ \nu=0,\dots,n.
	\eqlabel{cascond5}
\end{equation}
In the canonical case where the inner bracket is like~\eqref{canibrak}
the~$\lpb\fv^\sigma\com\fv^\lambda\rpb$ are independent and antisymmetric
in~$\lambda$ and~$\sigma$.\inthesis{Does this hold for other brackets, for
instance finite-dim semisimple?} Thus a necessary and sufficient condition for
the Casimir condition to be satisfied is
\begin{equation}
	{\W_\lambda}^{\mu\nu}
		\frac{\partial^2\Casi}{\partial\fv^\mu\partial\fv^\sigma}
	= {\W_\sigma}^{\mu\nu}
		\frac{\partial^2\Casi}{\partial\fv^\mu\partial\fv^\lambda}\ ,
	\eqlabel{cascond3}
\end{equation}
for~$\lambda,\sigma,\nu=0,\dots,n$.  Sometimes we shall abbreviate this as
\begin{equation}
	{\W_\lambda}^{\mu\nu} \Casi_{\dcom\mu\sigma}
	= {\W_\sigma}^{\mu\nu} \Casi_{\dcom\mu\lambda}\ ,
	\eqlabel{cascond4}
\end{equation}
that is, any subscript~$\mu$ on~$\Casi$ following a comma indicates
differentiation with respect to~$\fv^\mu$.  Equation \eqref{cascond4} is
trivially satisfied when~$\Casi$ is a linear function of the~$\fv$'s.  That
solution usually follows from special cases of more general solutions, and we
shall only mention it in \secref{singWn} where it is the only solution.

An important result is immediate from \eqref{cascond4} for a semidirect
extension.  Whenever the extension is semidirect we shall label the
variables~$\fv^0,\fv^1,\dots,\fv^n$, because the subset~$\fv^1,\dots,\fv^n$
then forms a solvable subalgebra (see \secref{semisimple} for terminology).
For a semidirect extension,~$\W^{(0)}$ is the identity matrix, and
thus~\eqref{cascond4} gives
\begin{eqnarray*}
	{\delta_\lambda}^{\mu} \Casi_{\dcom\mu\sigma}
		&=& {\delta_\sigma}^{\mu} \Casi_{\dcom\mu\lambda}\ ,\nonumber\\
	\Casi_{\dcom\lambda\sigma} &=& \Casi_{\dcom\sigma\lambda}\ ,
\end{eqnarray*}
which is satisfied because we can interchange the order of differentiation.
Hence,~$\nu=0$ does not lead to any conditions on the Casimir. However, the
variables~$\mu,\lambda,\sigma$ still take values from~$0$ to~$n$ in
\eqref{cascond4}.

\subsection{Direct Sum}
\seclabel{Casdirprod}

For the direct sum we found in \secref{directprod} that if we look at the
3-tensor~$\W$ as a cube, then it ``blocks out'' into smaller cubes, or
subblocks, along its main diagonal, each subblock representing a subalgebra.
We denote each subblock of~${\W_\lambda}^{\mu\nu}$
by~${{\W_i}_{\lambda}}^{\mu\nu}$,
\hbox{$i=1,\dots,r$}, where~$r$ is the number of subblocks.  We can
rewrite~\eqref{LPB} as
\begin{eqnarray*}
	\lPB A\com B\rPB &=& \sum_{i=1}^r \lang\fv_i^{\lambda}\com
		{{\W_i}_\lambda}^{\mu\nu}\,
		\lpb {\frac{\fd A}{\fd\fv_i^{\mu}}}\com
		{\frac{\fd B}{\fd\fv_i^{\nu}}}\rpb\rang	\nonumber \\
	&\rdef& \sum_{i=1}^r {\lPB A\com B\rPB}_i\, ,
\end{eqnarray*}
where~$i$ labels the different subblocks and the greek indices run over the
size of the~$i$th subblock.  Each of the subbrackets~\hbox{${\lPB \com
\rPB}_i$} depends on different fields.  In particular, if the
functional~$\Cas$ is a Casimir, then, for any functional~$F$
\[
	\lPB F\com \Cas\rPB = \sum_{i=1}^r {\lPB F\com \Cas\rPB}_i = 0 
		\ \ \ \Longrightarrow
	\ \ \ {\lPB F\com \Cas\rPB}_i = 0,\ \ i=1,\dots,r\, .
\]
The solution for this is
\[
	\Cas[\fv] = \Cas_1[\fv_1] + \cdots
		+ \Cas_r[\fv_r]\, ,\ \ \
	{\rm where}\ {\lPB F\com \Cas_i\rPB}_i =0,\ i=1,\dots,r\, ,
\]
that is, the Casimir is just the sum of the Casimir for each subbracket.
Hence, the question of finding the Casimirs can be treated separately for each
component of the direct sum.  We thus assume we are working on a single
degenerate subblock, as we did for the classification in
\secref{classext}, and henceforth we drop the subscript~$i$.

There is a complication when a single (degenerate) subblock has more that one
simultaneous eigenvector.  By this we mean~$k$ vectors~$u^{(a)}$,
$a=1,\dots,k$, such that
\[
	{\W_\lambda}^{\mu(\nu)}\,u^{(a)}_\mu = \ev^{(\nu)}\,\,u^{(a)}_\lambda.
\]
Note that lower-triangular matrices always have at least the simultaneous
eigenvector \hbox{$u_\mu={\delta_\mu}^n$}.  Let~$\eta^{(a)} \ldef
u^{(a)}_\rho\xi^\rho$, and consider a
form~$\Casi(\eta^{(1)},\dots,\eta^{(k)})$ for the Casimir.  Then
\begin{eqnarray*}
	{\W_\lambda}^{\mu(\nu)}
		\frac{\partial^2\Casi}{\partial\fv^\mu\partial\fv^\sigma}
	&=& {\W_\lambda}^{\mu(\nu)} \sum_{a,b=1}^k
		u^{(a)}_\mu u^{(b)}_\sigma
		\frac{\partial^2\Casi}{\partial\eta^{(a)}\partial\eta^{(b)}}\,,
		\nonumber\\
	&=& \ev^{(\nu)} \sum_{a,b=1}^k
		u^{(a)}_\lambda u^{(b)}_\sigma
		\frac{\partial^2\Casi}{\partial\eta^{(a)}\partial\eta^{(b)}}.
\end{eqnarray*}
Because the eigenvalue~$\ev^{(\nu)}$ does not depend on~$a$ (the block was
assumed to have degenerate eigenvalues), the above expression is symmetric
in~$\lambda$ and~$\sigma$.  Hence, the Casimir condition~\eqref{cascond3}
is satisfied.

The reason this is introduced here is that if a degenerate block splits into a
direct sum, then it will have several simultaneous eigenvectors.  The Casimir
invariants~$\Casi^{(a)}(\eta^{(a)})$ and~$\Casi^{(b)}(\eta^{(b)})$
corresponding to each eigenvector, instead of adding
as~$\Casi^{(a)}(\eta^{(a)}) + \Casi^{(b)}(\eta^{(b)})$, will combine into one
function,~$\Casi{(\eta^{(a)},\eta^{(b)})}$, a more general functional
dependence. However, these situations with more than one eigenvector are not
limited to direct sums.  For instance, they occur in semidirect sums.  In
\secref{caslowdim} we will see examples of both cases.

\subsection{Local Casimirs for Solvable Extensions}
\seclabel{localcas}

In the solvable case, when all the~$\W^{(\mu)}$'s are lower-triangular with
vanishing eigenvalues, a special situation occurs.  If we consider the Casimir
condition \eqref{cascond5}, we notice that derivatives with respect to~$\fv^n$
do not occur at all, since~$\W^{(n)}=0$.  Hence the functional
\[
	\Cas[\fv] = \int_\fdomain\,\fv^n(\xv')\,\delta(\xv-\xv')\d^2x'
		= \fv^n(\xv)
\]
is conserved.  The variable~$\fv^n(\xv)$ is \emph{locally} conserved.  It
cannot have any dynamics associated with it.  This holds true for any other
simultaneous null eigenvectors the extension happens to have, but for the
solvable case~$\fv^n$ is always such a vector (provided the matrices have been
put in lower-triangular form, of course).

Hence there are at most~$n-1$ dynamical variables in an order~$n$ solvable
extension.  An interesting special case occurs when the only
nonvanishing~$\W_{(\mu)}$ is for~$\mu=n$.  Then the Lie--Poisson bracket is
\[
	\lPB F\com G\rPB = \sum_{\mu,\nu=1}^{n-1}{\W_n}^{\mu\nu}
		\int_\fdomain\,\fv^n(\xv)
		\,\lpb \frac{\fd F}{\fd \fv^\mu(\xv)}\com
		\frac{\fd G}{\fd \fv^\nu(\xv)}\rpb\d^2x,
\]
where~$\fv^n(\xv)$ is some function of our choosing.  This bracket is not what
we would normally call Lie--Poisson because~$\fv^n(\xv)$ is not dynamical.
It gives equations of motion of the form
\[
	\dot\fv^\nu = {\W_n}^{\nu\mu}\,\lpb \frac{\fd H}{\fd \fv^\mu}
		\com \fv^n \rpb,
\]
which can be used to model, for example, advection of scalars in a specified
flow given by~$\fv^n(\xv)$.  This bracket occurs naturally when a Lie--Poisson
bracket is linearized~\cite{Morrison1998,MarsdenRatiu}.

\subsection{Solution of the Casimir Problem}
\seclabel{cassoln}

We now proceed to find the solution to~\eqref{cascond5}.  We assume that all
the~$\W^{(\mu)}$, $\mu=0,\dots,n$, are in lower-triangular form, and that the
matrix~$\W^{(0)}$ is the identity matrix.  Although this is the semidirect
form of the extension, we will see that we can also recover the Casimir
invariants of the solvable part.  We assume~$\nu>0$ in~\eqref{cascond5},
since~$\nu=0$ does not lead to a condition on the Casimir (\secref{cascond}).
Therefore~${\W_{\lambda}}^{n\nu}=0$.  Thus, we separate the Casimir condition
into a part involving indices ranging from~$0,\dots,n-1$ and a part that
involves only~$n$. The condition
\[
\sum_{\mu,\sigma,\lambda=0}^{n} {\W_\lambda}^{\mu\nu} \Casi_{\dcom\mu\sigma}
	\lpb\fv^\lambda\com\fv^\sigma\rpb = 0, \ \ \ \nu > 0,
\]
becomes
\[
	\sum_{\lambda=0}^{n} \l\lgroup
	\sum_{\mu,\sigma=0}^{n-1}{\W_\lambda}^{\mu\nu} \Casi_{\dcom\mu\sigma}
	\lpb\fv^\lambda\com\fv^\sigma\rpb
	+ \sum_{\mu=0}^{n-1}{\W_\lambda}^{\mu\nu} \Casi_{\dcom\mu n}
	\lpb\fv^\lambda\com\fv^n\rpb
	\r\rgroup = 0,
\]
where we have used~${\W_{\lambda}}^{n\nu}=0$ to limit the sum on~$\mu$.
Separating the sum on~$\lambda$ gives
\begin{eqnarray*}
	\sum_{\lambda=0}^{n-1} \l\lgroup
	\sum_{\mu,\sigma=0}^{n-1}{\W_\lambda}^{\mu\nu} \Casi_{\dcom\mu\sigma}
	\lpb\fv^\lambda\com\fv^\sigma\rpb
	+ \sum_{\mu=0}^{n-1}{\W_\lambda}^{\mu\nu} \Casi_{\dcom\mu n}
	\lpb\fv^\lambda\com\fv^n\rpb
	\r\rgroup\nonumber\\
	+ \sum_{\mu,\sigma=0}^{n-1}{\W_n}^{\mu\nu} \Casi_{\dcom\mu\sigma}
	\lpb\fv^n\com\fv^\sigma\rpb
	+ \sum_{\mu=0}^{n-1}{\W_n}^{\mu\nu} \Casi_{\dcom\mu n}
	\lpb\fv^n\com\fv^n\rpb
	= 0.
\end{eqnarray*}
The last sum vanishes because~$\lpb\fv^n\com\fv^n\rpb=0$. Now we separate the
condition into semisimple and solvable parts,
\begin{eqnarray*}
	\sum_{\mu=1}^{n-1} \Biggl\lgroup
	\sum_{\lambda,\sigma=0}^{n-1}{\W_\lambda}^{\mu\nu}
		\Casi_{\dcom\mu\sigma}
	\lpb\fv^\lambda\com\fv^\sigma\rpb
	- \sum_{\sigma=0}^{n-1}{\W_\sigma}^{\mu\nu} \Casi_{\dcom\mu n}
	\lpb\fv^n\com\fv^\sigma\rpb
	\nonumber\\
	\mbox{} + 
	\sum_{\sigma=0}^{n-1}{\W_n}^{\mu\nu} \Casi_{\dcom\mu\sigma}
	\lpb\fv^n\com\fv^\sigma\rpb
	\Biggr\rgroup
	+ \sum_{\lambda,\sigma=0}^{n-1}{\W_\lambda}^{0\nu}
		\Casi_{\dcom 0\sigma}
	\lpb\fv^\lambda\com\fv^\sigma\rpb\nonumber\\
	- \sum_{\sigma=0}^{n-1}{\W_\sigma}^{0\nu} \Casi_{\dcom 0 n}
	\lpb\fv^n\com\fv^\sigma\rpb
	+ \sum_{\sigma=0}^{n-1}{\W_n}^{0\nu} \Casi_{\dcom 0\sigma}
	\lpb\fv^n\com\fv^\sigma\rpb
	= 0.
\end{eqnarray*}
Using~${\W_\sigma}^{0\nu} = {\delta_\sigma}^\nu$, we can separate the
conditions into a part for~$\nu=n$ and one for~\hbox{$0<\nu<n$}.  For~$\nu=n$,
the only term that survives is the last sum
\[
	\sum_{\sigma=0}^{n-1} \Casi_{\dcom 0\sigma}
	\lpb\fv^n\com\fv^\sigma\rpb = 0.
\]
Since the commutators are independent, we have the conditions,
\begin{equation}
	\Casi_{\dcom 0\sigma} = 0, \ \ \ \sigma=0,\dots,n-1.
	\eqlabel{C0seq0}
\end{equation}
and for~\hbox{$0<\nu<n$},
\begin{eqnarray*}
	\sum_{\mu=1}^{n-1} \Biggl\lgroup
	\sum_{\lambda,\sigma=1}^{n-1}{\W_\lambda}^{\mu\nu}
		\Casi_{\dcom\mu\sigma}
	\lpb\fv^\lambda\com\fv^\sigma\rpb
	- \sum_{\sigma=1}^{n-1}{\W_\sigma}^{\mu\nu} \Casi_{\dcom\mu n}
	\lpb\fv^n\com\fv^\sigma\rpb
	\nonumber\\
	\mbox{} + 
	\sum_{\sigma=1}^{n-1}{\W_n}^{\mu\nu} \Casi_{\dcom\mu\sigma}
	\lpb\fv^n\com\fv^\sigma\rpb
	\Biggr\rgroup
	- \Casi_{\dcom 0 n} \lpb\fv^n\com\fv^\nu\rpb
	= 0,
\end{eqnarray*}
where we have used~\eqref{C0seq0}.  Using independence of the inner
brackets gives
\begin{eqnarray}
	{\Wt_\lambda}^{\mu\nu}
		\Casi_{\dcom\mu\sigma} &=&
	{\Wt_\sigma}^{\mu\nu}
		\Casi_{\dcom\mu\lambda},
	\eqlabel{cascondsubext} \\
	{\Wn}^{\nu\mu} \Casi_{\dcom\mu\sigma} &=&
		{\Wt_\sigma}^{\nu\mu} \Casi_{\dcom\mu n}
		+ {\delta^\nu}_\sigma\, \Casi_{\dcom 0 n},
	\eqlabel{Axeqb}
\end{eqnarray}
for~$0< \sigma,\lambda,\nu,\mu < n$.  From now on in this section repeated
indices are summed, and all greek indices run from~$1$ to~$n-1$ unless
otherwise noted.  We have written a tilde over the~$\W$'s to stress the fact
that the indices run from~$1$ to~$n-1$, so that the~$\Wt$ represent a solvable
order~$(n-1)$ subextension of~$\W$.  This subextension does not
include~$\W_{(n)}$.  We have also made the definition
\begin{equation}
	\Wn^{\mu\nu} \ldef {\W_n}^{\mu\nu}.
	\eqlabel{Wndef}
\end{equation}
Equation~\eqref{cascondsubext} is a Casimir condition: it says that~$\Casi$ is
also a Casimir of~$\Wt$. We now proceed to solve~\eqref{Axeqb} for the case
where~$\Wn$ is nonsingular. In \secref{singWn} we will solve the
singular~$\Wn$ case. We will see that in both cases \eqref{cascondsubext}
follows from \eqref{Axeqb}.

\subsubsection{Nonsingular $\Wn$}
\seclabel{nsingWn}

The simplest case occurs when~$\Wn$ has an inverse, which we will
call~$\Wni_{\mu\nu}$.  Then~\eqref{Axeqb} has the solution
\begin{equation}
	\Casi_{\dcom\tau\sigma} =
		\coW^\mu_{\tau\sigma}\, \Casi_{\dcom\mu n}
		+ \Wni_{\tau\sigma}\, \Casi_{\dcom 0 n}\, ,
	\eqlabel{nonsingsol}
\end{equation}
where
\begin{equation}
	\coW^\mu_{\tau\sigma} \ldef \Wni_{\tau\nu}\,{\Wt_\sigma}^{\nu\mu}.
	\eqlabel{coextdef}
\end{equation}
It is easily verified that~$\coW^\mu_{\tau\sigma} = \coW^\mu_{\sigma\tau}$,
as required by the symmetry of the left-hand side of~\eqref{nonsingsol}.

In~\eqref{nonsingsol}, it is clear that the~$n$th variable is ``special'';
this suggests that we try the following form for the Casimir:
\begin{equation}
	\Casi(\fv^0,\fv^1,\dots,\fv^n) =
	\sum_{i \ge 0}
	\ag^{(i)}(\fv^0,\fv^1,\dots,\fv^{n-1})\,\af_{i}(\fv^n),
	\eqlabel{Casform}
\end{equation}
where~$\af$ is arbitrary and~$\af_i$ is the $i$th derivative of~$\af$ with
respect to its argument.  One immediate advantage of this form is that
\eqref{cascondsubext} follows from \eqref{Axeqb}.  Indeed, taking a derivative
of \eqref{Axeqb} with respect to~$\fv^\lambda$, inserting~\eqref{Casform}, and
equating derivatives of~$\af$ leads to
\[
	{\Wn}^{\nu\mu}\, \ag^{(i)}_{\dcom\mu\sigma\lambda} =
		{\Wt_\sigma}^{\nu\mu}\, \ag^{(i+1)}_{\dcom\mu\lambda},
\]
where we have used \eqref{C0seq0}.  Since the left-hand side is symmetric
in~$\lambda$ and~$\sigma$ then so is the right-hand side, and
\eqref{cascondsubext} is satisfied.

Now, inserting the form of the Casimir~\eqref{Casform} into the
solution~\eqref{nonsingsol}, we can equate derivatives of~$\af$ to obtain,
for~\hbox{$\tau,\sigma=1,\dots,n-1$},
\begin{eqnarray}
	\ag^{(0)}_{\dcom\tau\sigma} &=& 0, \ \ \ \ \ \ \ \
		\tau,\sigma=1,\dots,n-1;\\ \ag^{(i)}_{\dcom\tau\sigma} &=&
		\coW^\mu_{\tau\sigma}\, \ag^{(i-1)}_{\dcom\mu} +
		\Wni_{\tau\sigma}\,\ag^{(i-1)}_{\dcom 0}, \ \ \ i \ge 1.
		\eqlabel{gcondi}
\end{eqnarray}
The first condition, together with~\eqref{C0seq0}, says that~$\ag^{(0)}$
is linear in~$\fv^0,\dots\fv^{n-1}$.  There are no other conditions
on~$\ag^{(0)}$, so we can obtain~$n$ independent solutions by choosing
\begin{equation}
	\ag^{(0)\nu} = \fv^\nu, \ \ \ \nu=0,\dots,n-1.
	\eqlabel{nsingsolzero}
\end{equation}
The equation for~$\ag^{(1)\nu}$ is
\begin{equation}
	\ag^{(1)\nu}_{\dcom\tau\sigma} = \l\{\begin{array}{l}
		\Wni_{\tau\sigma},\ \ \ \nu = 0; \\
		\coW^\nu_{\tau\sigma},\ \ \ \nu = 1,\dots,n-1.
	\eqlabel{nsingsolone}
	\end{array}\r.
\end{equation}
Thus~$\ag^{(1)\nu}$ is a quadratic polynomial (the arbitrary linear part does
not yield an independent Casimir, so we set it to zero).  Note
that~$\ag^{(1)\nu}$ does not depend on~$\fv^0$
since~$\tau,\sigma=1,\dots,n-1$.  Hence, for~$i>1$ we can drop
the~$\ag^{(i-1)}_{\dcom 0}$ term in~\eqref{gcondi}.  Taking derivatives
of~\eqref{gcondi}, we obtain
\begin{equation}
	\ag^{(i)\nu}_{\dcom\tau_1\tau_2\dots\tau_{(i+1)}} =
		\coW^{\mu_1}_{\tau_1\tau_2}\,
		\coW^{\mu_2}_{\mu_1\tau_3}\cdots
		\coW^{\mu_{(i-1)}}_{\mu_{(i-2)}\tau_{i}}
		\,\ag^{(1)\nu}_{\dcom\mu_{(i-1)}\tau_{(i+1)}}.
	\eqlabel{nonsingsolni}
\end{equation}
We know the series will terminate because the~$\Wt^{(\mu)}$, and hence
the~$\coW_{(\mu)}$, are nilpotent.  The solution to \eqref{nonsingsolni} is
\begin{equation}
	\ag^{(i)\nu} = \frac{1}{(i+1)!}\,\,
		\agc^{(i)\nu}_{\tau_1\tau_2\dots\tau_{(i+1)}}\,
		\fv^{\tau_1}\fv^{\tau_2}\cdots\fv^{\tau_{(i+1)}}\,,
		\ \ \ \ i > 1,
	\eqlabel{Cascoeff}
\end{equation}
where the constants~$\agc$ are defined by
\begin{equation}
	\agc^{(i)\nu}_{\tau_1\tau_2\dots\tau_{(i+1)}} \ldef
		\coW^{\mu_1}_{\tau_1\tau_2}\,
		\coW^{\mu_2}_{\mu_1\tau_3}\cdots
		\coW^{\mu_{(i-1)}}_{\mu_{(i-2)}\tau_{i}}
		\,\ag^{(1)\nu}_{\dcom\mu_{(i-1)}\tau_{(i+1)}}.
	\eqlabel{agcdef}
\end{equation}
In summary, the~$\ag^{(i)}$'s of~\eqref{Casform} are
given by~\eqref{nsingsolzero},~\eqref{nsingsolone}, and~\eqref{Cascoeff}.

Because the left-hand side of~\eqref{nonsingsolni} is symmetric in all its
indices, we require
\begin{equation}
	\coW^{\mu}_{\tau\sigma}\,\coW^{\nu}_{\mu\lambda} =
	\coW^{\mu}_{\tau\lambda}\,\coW^{\nu}_{\mu\sigma}, \qquad i>1.
	\eqlabel{coextcond}
\end{equation}
This is automatically satisfied for the nonsingular~$\Wn$
case~\cite{Thiffeault1998diss}.  Comparing this to~\eqref{Wjacob}, we see that
the~$\coW$'s satisfy all the properties of an extension, except with the dual
indices.  Thus we call the~$\coW$'s the \emph{coextension} of~$\Wt$ with
respect to~$\Wn$.  Essentially~$\Wn$ serves the role of a metric that allows
us to raise and lower indices.

For a solvable extension we simply restrict~$\nu > 0$ and the above treatment
still holds.  We conclude that the Casimirs of the solvable part of a
semidirect extension are Casimirs of the full extension.  We have also shown,
for the case of nonsingular~$\Wn$, that the number of independent Casimirs is
equal to the order of the extension.

\subsubsection{Singular $\Wn$}
\seclabel{singWn}

In general,~$\Wn$ is singular and thus has no inverse.  However, it always has
a (symmetric and unique) pseudoinverse~$\Wni_{\mu\nu}$ such that
\begin{eqnarray}
	\Wni_{\mu\sigma}\,\Wn^{\sigma\tau}\,\Wni_{\tau\nu}
		&=& \Wni_{\mu\nu},\eqlabel{pseudoinv1}\\
	\Wn^{\mu\sigma}\,\Wni_{\sigma\tau}\,\Wn^{\tau\nu}
		&=& \Wn^{\mu\nu}.
	\eqlabel{pseudoinv2}
\end{eqnarray}
The pseudoinverse is also known as the strong generalized inverse or the
Moore--Penrose inverse~\cite{Osta}. It follows from~\eqref{pseudoinv1}
and~\eqref{pseudoinv2} that the matrix operator
\[
	{\Proj^\nu}_\tau \ldef \Wn^{\nu\kappa}\,\Wni_{\kappa\tau}
\]
projects onto the range of~$\Wn$.  The system~\eqref{Axeqb} only has a
solution if the following solvability condition is satisfied:
\begin{equation}
	{\Proj^\nu}_\tau\,
		\l({\Wt_\sigma}^{\tau\mu} \Casi_{\dcom\mu n}
			+ {\delta^\tau}_\sigma\, \Casi_{\dcom 0 n}\r)
	= {\Wt_\sigma}^{\nu\mu} \Casi_{\dcom\mu n}
		+ {\delta^\nu}_\sigma\, \Casi_{\dcom 0 n};
	\eqlabel{solvcondz}
\end{equation}
that is, the right-hand side of~\eqref{Axeqb} must live in the range of~$\Wn$.

If~\hbox{$\Casi_{\dcom 0 n}\ne 0$}, the quantity~${\Wt_\sigma}^{\nu\mu}\,
\Casi_{\dcom\mu n} + {\delta^\nu}_\sigma\, \Casi_{\dcom 0 n}$ has rank equal
to~$n$, because the quantity~${\Wt_\sigma}^{\nu\mu}\,\Casi_{\dcom\mu n}$ is
lower-triangular (it is a linear combination of lower-triangular matrices).
Hence the projection operator must also have rank~$n$.  But then this implies
that~$\Wn$ has rank~$n$ and so is nonsingular, which contradicts the
hypothesis of this section.  Hence,~\hbox{$\Casi_{\dcom 0 n} = 0$} for the
singular~$\Wn$ case, which together with \eqref{C0seq0} means that a Casimir
that depends on~$\fv^0$ can only be of the form~$\Casi = \af(\fv^0)$.
However, since~$\fv^0$ is not an eigenvector of the~$\W^{(\mu)}$'s, the only
possibility is~$\Casi = \fv^0$, the trivial linear case mentioned in
\secref{cascond}.

The solvability condition \eqref{solvcondz} can thus be rewritten as
\begin{equation}
	\l({\Proj^\nu}_\tau\,{\Wt_\sigma}^{\tau\mu}
		- {\Wt_\sigma}^{\nu\mu}\r) \Casi_{\dcom\mu n} = 0.
	\eqlabel{solvcond}
\end{equation}
An obvious choice would be to
require~\hbox{${\Proj^\nu}_\tau\,{\Wt_\sigma}^{\tau\mu} =
{\Wt_\sigma}^{\nu\mu}$}, but this is too strong.  We will derive a weaker
requirement shortly.

By an argument similar to that of~\secref{nsingWn}, we now assume~$\Casi$ is
of the form
\begin{equation}
	\Casi(\fv^1,\dots,\fv^n) =
	\sum_{i \ge 0} \ag^{(i)}(\fv^1,\dots,\fv^{n-1})\,\af_{i}(\fv^n),
	\eqlabel{singCas}
\end{equation}
where again~$\af_i$ is the $i$th derivative of~$f$ with respect to its
argument.  As in \secref{nsingWn}, we only need to show \eqref{Axeqb}, and
\eqref{cascondsubext} will follow.  The number of independent solutions of
\eqref{Axeqb} is equal of the rank of~$\Wn$.  The choice
\begin{equation}
	\ag^{(0)\nu} = {\Proj^\nu}_{\rho}\,\fv^\rho, \ \ \ \nu=1,\dots,n-1,
	\eqlabel{singsolzero}
\end{equation}
provides the right number of solutions because the rank of~$\Proj$ is equal to
the rank of~$\Wn$.  It also properly specializes to
\eqref{nsingsolzero} when~$\Wn$ is nonsingular, for then~${\Proj^\nu}_\rho =
{\delta^{\,\nu}}_\rho$.

The solvability condition~\eqref{solvcond} with this form for the Casimir
becomes
\begin{equation}
	\l({\Proj^\nu}_\tau\,{\Wt_\sigma}^{\tau\mu}
		- {\Wt_\sigma}^{\nu\mu}\r) \ag^{(i)\nu}_{\dcom\mu} = 0,\ \ \
	i \ge 0.
	\eqlabel{solvcondg}
\end{equation}
For~$i=0$ the condition can be shown to simplify to
\[
	{\Proj^\nu}_\tau\,{\Wt_\sigma}^{\tau\mu}
	= {\Wt_\sigma}^{\nu\tau}\,{\Proj^\mu}_\tau,
\]
or to the equivalent matrix form
\begin{equation}
	\Proj\,\Wt_{(\sigma)} = \Wt_{(\sigma)}\,\Proj,
	\eqlabel{solvcond0}
\end{equation}
since~$\Proj$ is symmetric~\cite{Osta}.

Equation \eqref{Axeqb} becomes
\begin{eqnarray*}
	{\Wn}^{\kappa\mu} \ag^{(0)\nu}_{\dcom\mu\sigma} &=& 0,\\
	{\Wn}^{\kappa\mu} \ag^{(i)\nu}_{\dcom\mu\sigma} &=&
		{\Wt_\sigma}^{\kappa\mu} \ag^{(i-1)\nu}_{\dcom\mu},\ \ \ i > 0.
\end{eqnarray*}
If \eqref{solvcond} is satisfied, we know this has a solution given by
\[
	\ag^{(i)\nu}_{\dcom\lambda\sigma} =
		\Wni_{\lambda\rho}\,{\Wt_\sigma}^{\rho\mu}\,
		\ag^{(i-1)\nu}_{\dcom\mu} + \l({\delta_\lambda}^\mu
		- \Wni_{\lambda\rho}\,\Wn^{\rho\mu}\r)
		\ae^{(i-1)\nu}_{\mu\sigma} ,\ \ \ i > 0,
\]
where~$\ae$ is arbitrary, and~\hbox{$({\delta_\lambda}^\mu -
\Wni_{\lambda\rho}\,\Wn^{\rho\mu})$} projects onto the null space of~$\Wn$.
The left-hand side is symmetric in~$\lambda$ and~$\sigma$, but not the
right-hand side.  We can symmetrize the right-hand side by an appropriate
choice of the null eigenvector,
\[
	\ae^{(i)\nu}_{\lambda\sigma} \ldef
		\Wni_{\sigma\rho}\,{\Wt_\lambda}^{\rho\mu}\,
		\ag^{(i)\nu}_{\dcom\mu},\ \ \ i \ge 0,
\]
in which case
\[
	\ag^{(i)\nu}_{\dcom\lambda\sigma} =
		\coW^\mu_{\lambda\sigma}\,
		\ag^{(i-1)\nu}_{\dcom\mu},\ \ \ i > 0,
\]
where
\begin{equation}
	\coW^\nu_{\lambda\sigma} \ldef
		\Wni_{\sigma\rho}\,{\Wt_\lambda}^{\rho\nu}
		+ \Wni_{\lambda\rho}\,{\Wt_\sigma}^{\rho\nu}
		- \Wni_{\lambda\rho}\,\Wni_{\sigma\kappa}\,
		\Wn^{\rho\mu}\,{\Wt_\mu}^{\kappa\nu}\,,
	\eqlabel{coextension}
\end{equation}
which is symmetric in~$\lambda$ and~$\sigma$.  Equation~\eqref{coextension}
also reduces to \eqref{coextdef} when~$\Wn$ is nonsingular, for then the null
eigenvector vanishes.  The full solution is thus given in the same manner as
\eqref{nonsingsolni} by
\begin{equation}
	\ag^{(i)\nu} = \frac{1}{(i+1)!}\,\,
		\agc^{(i)\nu}_{\tau_1\tau_2\dots\tau_{(i+1)}}\,
		\fv^{\tau_1}\fv^{\tau_2}\cdots\fv^{\tau_{(i+1)}}\,,
		\ \ \ \ i > 0,
	\eqlabel{Cascoeffsing}
\end{equation}
where the constants~$\agc$ are defined by
\begin{equation}
	\agc^{(i)\nu}_{\tau_1\tau_2\dots\tau_{(i+1)}} \ldef
		\coW^{\mu_1}_{\tau_1\tau_2}\,
		\coW^{\mu_2}_{\mu_1\tau_3}\cdots
		\coW^{\mu_{(i-1)}}_{\mu_{(i-2)}\tau_{i}}
		\,\coW^{\mu_{i}}_{\mu_{(i-1)}\tau_{(i+1)}}
		\,{\Proj^{\,\nu}}_{\mu_{i}}\,,
	\eqlabel{agcdefsing}
\end{equation}
and~$\ag^{(0)}$ is given by \eqref{singsolzero}.

The~$\coW$'s must still satisfy the coextension condition \eqref{coextcond}.
Unlike the nonsingular case this condition does not follow directly and is an
extra requirement in addition to the solvability condition \eqref{solvcondg}.
Note that only the~$i=0$ case, Eq. \eqref{solvcond0}, needs to be satisfied,
for then \eqref{solvcondg} follows.  Both these conditions are
coordinate-dependent, and this is a drawback.  Nevertheless, we have found in
obtaining the Casimir invariants for the low-order brackets that if these
conditions are not satisfied, then the extension is a direct sum and the
Casimirs can be found by the method of \secref{Casdirprod}.  However, this has
not been proven rigorously.

\subsection{Examples}
\seclabel{Casex}

We now illustrate the methods developed for finding Casimirs with a few
examples.  First we treat our prototypical case of CRMHD, and give a physical
interpretation of invariants. Then, we derive the Casimir invariants for
Leibniz extensions of arbitrary order. Finally, we give an example involving a
singular~$\Wn$.

\subsubsection{Compressible Reduced MHD}
\seclabel{CRMHDCas}

The~$\W$ tensors representing the bracket for CRMHD (see \secref{CRMHD}) were
given in \secref{matCRMHD}. We have~$n=3$, so from \eqref{Wndef} we get
\[
	\Wn = \l(\begin{array}{cc} 0 & -\crmhdbeta \\ -\crmhdbeta & 0
		 \end{array}\r),\ \ \
	\Wn^{-1} = \l(\begin{array}{cc} 0 & -\crmhdbeta^{-1} \\
		-\crmhdbeta^{-1} & 0 \end{array}\r).
\]
In this case, the coextension is trivial: all three matrices~$\coW^{(\nu)}$
defined by \eqref{coextdef} vanish. Using \eqref{Casform} and
\eqref{nsingsolzero}, with~$\nu=1$ and~$2$, the Casimirs for the solvable part are
\[
	\Casi^1 = \fv^1\,\afii(\fv^3) = \pvel\,\afii(\magf),\ \ \ \Casi^2 =
	\fv^2\,\afiii(\fv^3) = \pres\,\afiii(\magf),
\]
and the Casimir associated with the eigenvector~$\fv^3$ is
\[
	\Casi^3 = \afiv(\fv^3) = \afiv(\magf).
\]
Since~$\Wn$ is nonsingular we also get another Casimir from the semidirect
sum part,
\[
	\Casi^0 = \fv^0\,\afi(\fv^3)
		- \frac{1}{\crmhdbeta}\,\fv^1\,\fv^2\,\afi'(\fv^3)
	= \vort\,\afi(\magf)
		- \frac{1}{\crmhdbeta}\,\pres\,\pvel\,\afi'(\magf).
\]

\inthesis{Have larger section on this.}

The physical interpretation of the invariant~$\Casi^3$ is given in
Morrison~\cite{Morrison1987} and Thiffeault and
Morrison~\cite{Thiffeault1998}.  This invariant implies the preservation of
contours of~$\magf$, so that the value~$\magf_0$ on a contour labels that
contour for all times. This is a consequence of the lack of dissipation and
the divergence-free nature of the velocity. Substituting~\hbox{$\Casi^3(\magf)
= \magf^k$} we also see that all the moments of the magnetic flux are
conserved.  By choosing~\hbox{$\Casi^3(\magf) = \heavyside(\magf(\xv) -
\omega_0)$}, a heavyside function, and inserting into \eqref{formcas}, it
follows that the area inside of any $\magf$-contour is conserved.

To understand the Casimirs~$\Casi^1$ and~$\Casi^2$, we also let
$\afii(\magf)=\heavyside(\magf-\magf_0)$ in~$\Casi^1$. In this case we have
\[
	\Cas^1[\pvel\,;\magf] = \int_{\fdomain}\pvel\,\afii(\magf)\d^2x
	= \int_{\magfcont_0}\,\pvel(\xv)\d^2x,
\]
where~$\magfcont_0$ represents the (not necessarily connected) region of
$\fdomain$ enclosed by the contour $\magf=\magf_0$ and~$\pd\magfcont_0$ is
its boundary. By the interpretation we gave of~$\Casi^3$, the contour
$\pd\magfcont_0$ moves with the fluid. So the total value of~$\pvel$ inside of
a~$\magf$-contour is conserved by the flow. The same is true of the
pressure~$\pres$. (See Thiffeault and Morrison~\cite{Thiffeault1998} for an
interpretation of these invariants in terms of relabeling symmetries, and a
comparison with the rigid body.)

The total pressure and parallel velocity inside of any $\magf$-contour are
preserved.  To understand $\Casi^4$, we use the fact
that~$\vort=\lapl\elecp$ and integrate by parts to obtain
\[
	\Cas^4[\vort,\pvel,\pres,\magf] = -\int_\fdomain
		\l(\grad\elecp\cdot\grad\magf
		+ \frac{\pvel\,p}{\crmhdbeta}\r)\afi'(\magf)\d^2x.
\]
The quantity in parentheses is thus invariant inside of any $\magf$-contour.
It can be shown that this is a remnant of the conservation by the full MHD
model of the cross helicity,
\[
	V = \int_\fdomain {\mathbf{v}}\cdot{\mathbf{B}}\d^2x\,,
\]
at second order in the inverse aspect ratio, while the conservation
of~$\Cas^1[\pvel\,;\magf]$ is a consequence of preservation of this quantity
at first order. Here ${\mathbf{B}}$ is the magnetic field.  The
quantities~$\Cas^3[\magf]$ and~$\Cas^2[\pres\,;\magf]$ they are, respectively,
the first and second order remnants of the preservation of helicity,
\[
	W = \int_\fdomain {\mathbf{A}}\cdot{\mathbf{B}}\d^2x,
\]
where ${\mathbf{A}}$ is the magnetic vector potential.

\subsubsection{Leibniz Extension}
\seclabel{Casleib}

We first treat the nilpotent case.  The Leibniz extension of \secref{Leibniz}
can be characterized by
\begin{equation}
	{\W_\lambda}^{\mu\nu} = {\delta_\lambda}^{\mu+\nu}\,,
	\ \ \ \mu,\nu, \lambda = 1,\dots,n,
	\tag{\ref{eq:sLeib}}
\end{equation}
where the tensor~$\delta$ is the ordinary Kronecker delta.  Upon restricting
the indices to run from~$1$ to~$n-1$ (the tilde notation of \secref{cassoln}),
we have
\[
	\Wn^{\mu\nu} = {\Wt_n}^{\mu\nu} = {\delta_n}^{\mu+\nu}\,,
	\ \ \ \mu,\nu = 1,\dots,n-1.
\]
The matrix~$\Wn$ is nonsingular with inverse equal to
itself:~\hbox{$\Wni_{\mu\nu} =
\delta_{\mu+\nu}^{\,\,n}$}.  The coextension of~$\Wt$ is thus
\[
	\coW^\mu_{\tau\sigma} = \sum_{\nu=1}^{n-1}\Wni_{\tau\nu}\,
		{\Wt_\sigma}^{\nu\mu}
	= \sum_{\nu=1}^{n-1}\delta^n_{\tau+\nu}\,
		{\delta_\sigma}^{\nu+\mu}
	= \delta^{\mu+n}_{\tau+\sigma}\,.
\]
Equation \eqref{agcdef} becomes
\begin{eqnarray*}
	\agc^{(i)\nu}_{\tau_1\tau_2\dots\tau_{(i+1)}} &=&
		\coW^{\mu_1}_{\tau_1\tau_2}\,
		\coW^{\mu_2}_{\mu_1\tau_3}\cdots
		\coW^{\mu_{(i-1)}}_{\mu_{(i-2)}\tau_{i}}\,
		\coW^{\nu}_{\mu_{(i-1)}\tau_{(i+1)}}\nonumber\\
	&=& \delta^{\mu_1+n}_{\tau_1+\tau_2}\,
		\delta^{\mu_2+n}_{\mu_1+\tau_3}\cdots
		\delta^{\mu_{(i-1)+n}}_{\mu_{(i-2)}+\tau_{i}}\,
		\delta^{\nu+n}_{\mu_{(i-1)}+\tau_{(i+1)}}.\nonumber\\
	&=& \delta^{\nu+in}_{\tau_1+\tau_2+\cdots+\tau_{(i+1)}}\,,
		\ \ \ \nu = 1,\dots,n-1,
\end{eqnarray*}
which, as required, is symmetric under interchage of the~$\tau_i$. Using
\eqref{Casform}, \eqref{nsingsolzero}, \eqref{nsingsolone}, and
\eqref{Cascoeff} we obtain the~$n-1$ Casimir invariants
\begin{equation}
	\Casi^\nu(\fv^1,\dots,\fv^n) =
	\sum_{i \ge 0}
	\frac{1}{(i+1)!}\,\,
	{\delta^{\nu+in}_{\tau_1+\tau_2+\cdots+\tau_{(i+1)}}}\,
	\fv^{\tau_1}\cdots\fv^{\tau_{(i+1)}}\,
	\af^\nu_{i}(\fv^n),
	\eqlabel{CasiLeib}
\end{equation}
for~$\nu=1,\dots,n-1$.  The superscript~$\nu$ on~$\af$ indicates that the
arbitrary function is different for each Casimir, and recall the subscript~$i$
denotes the~$i$th derivative with respect to~$\fv^n$.  The~$n$th invariant is
simply~$\Casi^\nu(\fv^n) = \af^n(\fv^n)$, corresponding to the null
eigenvector in the system.  Thus there are~$n$ independent Casimirs, as stated
in \secref{nsingWn}.

For the Leibniz semidirect sum case, since~$\Wn$ is nonsingular, there will be
an extra Casimir given by \eqref{CasiLeib} with~$\nu=0$, and the~$\tau_i$ sums
run from~$0$ to~$n-1$.  This is the same form as the~$\nu=1$ Casimir of the
order~$(n+1)$ nilpotent extension.

For the $i$th term in \eqref{CasiLeib}, the maximal value of any~$\tau_j$ is
achived when all but one (say, $\tau_1$) of the~$\tau_j$ are equal to~$n-1$,
their maximum value.  In this case we have
\[
	\tau_1+\tau_2+\cdots+\tau_{i+1} = \tau_1 + i(n-1) = \nu + i n,
\]
so that~$\tau_1 = i+\nu$.  Hence, the~$i$th term depends only
on~\hbox{$\l(\fv^{\nu+i},\dots,\fv^n\r)$}, and the~$\nu$th Casimir depends
on~\hbox{$\l(\fv^\nu,\dots,\fv^n\r)$}.  Also,
\[
	\max{\l(\tau_1+\cdots+\tau_{i+1}\r)} = (i+1)(n-1) = \nu + i n,
\]
which leads to~\hbox{$\max i = n - \nu - 1$}.  Thus the sum \eqref{CasiLeib}
terminates, as claimed in \secref{nsingWn}.  We rewrite \eqref{CasiLeib} in
the more complete form
\[
	\Casi^\nu(\fv^\nu,\dots,\fv^n) =
	\sum_{k = 1}^{n-\nu}
	\frac{1}{k!}\,\,
	{\delta^{\nu+(k-1)n}_{\tau_1+\tau_2+\cdots+\tau_{k}}}\,
	\fv^{\tau_1}\cdots\fv^{\tau_{k}}\,
	\af^\nu_{k-1}(\fv^n),
\]
for~$\nu=0,\dots,n$.  \tabref{CasimirLeibn5} gives the~$\nu=1$ Casimirs
up to order~$n=5$.

\begin{table}
\caption{Casimir invariants for Leibniz extensions up to order~$n=5$
($\nu=1$).  The primes denote derivatives.}
\tablabel{CasimirLeibn5}
\vskip 1em

\begin{center}
\begin{tabular}{ll} \hline
$n$ & Invariant \\[6pt] \hline

%
%
1 & $\af(\fv^1)$
\\[9pt]

%
%
2 & $\fv^1\af(\fv^2)$
\\[9pt]

%
%
3 & $\fv^1\af(\fv^3) + \frac{1}{2}{(\fv^2)^2}\af'(\fv^3)$
\\[9pt]

%
%
4 & $\fv^1\af(\fv^4) + \fv^2\fv^3\af'(\fv^4)
	+ \frac{1}{3!}(\fv^3)^3\af''(\fv^4)$
\\[9pt]

%
%
5 & $\fv^1\af(\fv^5) + \l(\fv^2\fv^4
		+ \frac{1}{2}(\fv^3)^2\r)\af'(\fv^5)
		+ \frac{1}{2}\fv^3(\fv^4)^2\af''(\fv^5)
		+ \frac{1}{4!}(\fv^4)^4\af'''(\fv^5)$
\\[9pt] \hline

\end{tabular}

\end{center}
\end{table}

\subsubsection{Singular~$\Wn$}
\seclabel{singWnex}

Now consider the~$n=4$ extension from \secref{neqfour}, \caseref{n4-10}c. We
have
\[
	\Wt_{(2)} = \l({\begin{array}{ccc}
		\matone & \matzero & \matzero \\
		\matzero & \matzero & \matzero \\
		\matzero & \matzero & \matzero
	\end{array}}\r),\ \ \ \
	\Wn = \l({\begin{array}{ccc}
		\matzero & \matzero & \matone \\
		\matzero & \matzero & \matzero \\
		\matone & \matzero & \matzero
	\end{array}}\r),
\]
with~$\Wt_{(1)} = \Wt_{(3)} = 0$. The pseudoinverse of~$\Wn$
is~\hbox{$\Wn^{-1}=\Wn$} and the projection operator is
\[
	{\Proj^\nu}_\tau \ldef \Wn^{\nu\kappa}\,\Wni_{\kappa\tau} = 
		\l({\begin{array}{ccc}
			\matone & \matzero & \matzero \\
			\matzero & \matzero & \matzero \\
			\matzero & \matzero & \matone
		\end{array}}\r).
\]
The solvability condition \eqref{solvcond0} is obviously satisfied. We build
the coextension given by \eqref{coextension}, which in matrix form is
\[
	\coW^{(\nu)} = \Wt^{(\nu)}\,\Wn^{-1} + (\Wt^{(\nu)}\,\Wn^{-1})^T
		- \Wn^{-1}\,\Wn\,\Wt^{(\nu)}\,\Wn^{-1},
\]
to obtain
\[
	\coW^{(1)} = \l({\begin{array}{ccc}
		\matzero & \matzero & \matzero \\
		\matzero & \matzero & \matone \\
		\matzero & \matone & \matzero\end{array}}\r),\ \ \
	\coW^{(2)} = \coW^{(3)} = 0.
\]
These are symmetric and obviously satisfy \eqref{coextcond}, so we have a good
coextension. Using \eqref{singCas}, \eqref{singsolzero},
\eqref{Cascoeffsing}, and \eqref{agcdefsing} we can write, for~$\nu=1$ and~$3$,
\begin{eqnarray*}
	\Casi^1 &=& \fv^1\afi(\fv^4) + \fv^2\,\fv^3\afi'(\fv^4),\nonumber\\
	\Casi^3 &=& \fv^3\afii(\fv^4).
\end{eqnarray*}
This extension has two null eigenvectors, so from \secref{Casdirprod} we also
have the Casimir~$\afiii(\fv^2,\fv^4)$. The functions~$\afi$, $\afii$, and
$\afiii$ are arbitrary, and the prime denotes differentiation with respect to
argument.

%
%
%
%
%
%
%
%
%
%
%
%

\section{Casimir Invariants for Low-order Extensions}
\seclabel{caslowdim}

\ifmypprint
{\small
\begin{verbatim}
$Id: sec-casilowext.tex,v 2.1 1998/11/13 08:37:39 jeanluc Exp $
\end{verbatim}
}
\fi

Using the techniques developed so far, we now find the Casimir invariants for
the low-order extensions classified in \secref{lowdimext}.  We first find the
Casimir invariants for the solvable extensions, since these are also
invariants for the semidirect sum case.  Then, we obtain the extra Casimir
invariants for the semidirect case, when they exist.

\subsection{Solvable Extensions}

Now we look for the Casimirs of solvable extensions.
As mentioned in \secref{localcas}, the Casimirs associated with null
eigenvectors (the only kind of eigenvector for solvable extensions) are
actually conserved locally.  We shall still write them in the
form~$\Casi=f(\fv^n)$, where~$\Casi$ is as in
\eqref{formcas}, so they have the correct form as invariants for the
semidirect case of \secref{lowsemicasi} (for which they are no longer locally
conserved).

\subsubsection{n=1}

Since the bracket is Abelian, any function~$\Casi=\Casi(\fv^1)$ is a Casimir.

\subsubsection{n=2}

For the Abelian case we have~$\Casi=\Casi(\fv^1,\fv^2)$.  The only other case
is the Casimir of the Leibniz extension,
\[
	\Casi(\fv^1,\fv^2) = \fv^1\afi(\fv^2) + \afii(\fv^2).
\]

\subsubsection{n=3}
\seclabel{Casn3}

As shown in \secref{neqthree}, there are four cases.  \caseref{n4-00} is the
Abelian case, for which any function~$\Casi=\Casi(\fv^1,\fv^2,\fv^3)$ is a
Casimir.  \caseref{n4-01} is essentially the solvable part of the CRMHD
bracket, which we treated in~\secref{CRMHDCas}.  \caseref{n4-10} is a direct
sum of the Leibniz extension for~$n=2$, which has the bracket
\[
	\lpb(\alpha_1,\alpha_2)\com(\beta_1,\beta_2)\rpb
		= (0,\lpb\alpha_1\com\beta_1\rpb),
\]
with the Abelian algebra~$\lpb\alpha_3\com\beta_3\rpb=0$.  Hence, the Casimir
invariant is the same as for the~$n=2$ Leibniz extension with the
extra~$\fv^3$ dependence of the arbitrary function (see \secref{Casdirprod}).
Finally, \caseref{n4-11} is the Leibniz Casimir.  These results are summarized
in \tabref{Casimirn3}.

Cases~\ref{case:n4-00} and~\ref{case:n4-10} are trivial extensions, that is,
the cocycle appended to the $n=2$ case vanishes.  The procedure of then adding
$\fv^n$ dependence to the arbitrary function works in general.

\begin{table}
\caption{Casimir invariants for solvable extensions of order~$n=3$.}
\tablabel{Casimirn3}
\vskip 1em

\begin{center}
\begin{tabular}{ll} \hline
Case & Invariant \\[6pt] \hline

%
%
1 & $\Casi(\fv^1,\fv^2,\fv^3)$
\\[9pt]

%
%
2 & $\fv^1\afi(\fv^3) + \fv^2\afii(\fv^3) + \afiii(\fv^3)$
\\[9pt]

%
%
3 & $\fv^1\afi(\fv^2) + \afii(\fv^2,\fv^3)$
\\[9pt]

%
%
4 & $\fv^1\afi(\fv^3) + \frac{1}{2}(\fv^2)^2\afi'(\fv^3)
	+ \fv^2\afii(\fv^3) + \afiii(\fv^3)$
\\[9pt] \hline

\end{tabular}
\end{center}

\end{table}

\subsubsection{n=4}
\seclabel{Casn4}

As shown in \secref{neqfour}, there are nine cases to consider.  We
shall proceed out of order, to group together similar Casimir invariants.

Cases~\ref{case:n4-00}a,~\ref{case:n4-01},~\ref{case:n4-10}a,
and~\ref{case:n4-11}a are trivial extensions, and as mentioned in
\secref{Casn3} they involve only addition of $\fv^4$ dependence to
their~$n=3$ equivalents.  \caseref{n4-10}b is a direct sum of two~$n=2$
Leibniz extensions, so the Casimirs add.

\caseref{n4-10}c is the semidirect sum of the $n=2$ Leibniz extension with an
Abelian algebra defined by $\lpb(\alpha_3,\alpha_4)\com(\beta_3,\beta_4)\rpb =
(0,0)$, with action given by
\[
	\rho_{(\alpha_1,\alpha_2)}(\beta_3,\beta_4)
		= (0,\lpb\alpha_1\com\beta_3\rpb).
\]
The Casimir invariants for this extension were derived in \secref{singWnex}.

\caseref{n4-10}d has a nonsingular~$\Wn$, so the techniques of
\secref{nsingWn} can be applied directly.

Finally, \caseref{n4-11}b is the~$n=4$ Leibniz extension, the Casimir
invariants of which were derived in \secref{Casleib}. The invariants are all
summarized in \tabref{Casimirn4}.

\begin{table}
\caption{Casimir invariants for solvable extensions of order~$n=4$.}
\tablabel{Casimirn4}
\vskip 1em

\begin{center}
\begin{tabular}{ll} \hline
Case & Invariant \\[6pt] \hline

%
%
1a & $\Casi(\fv^1,\fv^2,\fv^3,\fv^4)$
\\[9pt]

%
%
1b & $\fv^1\afi(\fv^4) + \fv^2\afii(\fv^4) + \fv^3\afiii(\fv^4)
	+ \afiv(\fv^4)$
\\[9pt]

%
%
2 & $\fv^1\afi(\fv^3) + \fv^2\afii(\fv^3) + \afiii(\fv^3,\fv^4)$
\\[9pt]

%
%
3a & $\fv^1\afi(\fv^2) + \afii(\fv^2,\fv^3,\fv^4)$
\\[9pt]

%
%
3b & $\fv^1\afi(\fv^2) + \fv^3\afii(\fv^4) + \afiii(\fv^2,\fv^4)$
\\[9pt]

%
%
3c & $\fv^1\afi(\fv^4) + \fv^2\fv^3\afi'(\fv^4) + \fv^3\afii(\fv^4)
	+ \afiii(\fv^2,\fv^4)$
\\[9pt]

%
%
3d & $\fv^1\afi(\fv^4) + \frac{1}{2}(\fv^2)^2\afi'(\fv^4) +
	\fv^3\afii(\fv^4) + \fv^2\afiii(\fv^4) + \afiv(\fv^4)$
\\[9pt]

%
%
4a & $\fv^1\afi(\fv^3) + \frac{1}{2}(\fv^2)^2\afi'(\fv^3) +
	\fv^2\afii(\fv^3) + \afiii(\fv^3,\fv^4)$
\\[9pt]

%
%
4b & $\fv^1\afi(\fv^4) + \fv^2\fv^3\afi'(\fv^4)
	+ \frac{1}{3!}(\fv^3)^3\afi''(\fv^4)$\\  &
	$\mbox{} + \fv^2\afii(\fv^4) + \frac{1}{2}(\fv^3)^2\afii'(\fv^4)
	+ \fv^3\afiii(\fv^4) + \afiv(\fv^4)$
\\[9pt] \hline

\end{tabular}

\end{center}
\end{table}

\subsection{Semidirect Extensions}
\seclabel{lowsemicasi}

Now that we have derived the Casimir invariants for solvable extensions, we
look at extensions involving the semidirect sum of an algebra with these
solvable extensions.  We label the new variable (the one which acts on the
solvable part) by~$\fv^0$.  In \secref{nsingWn} we showed that the Casimirs of
the solvable part were also Casimirs of the full extension.  We also concluded
that a necessary condition for obtaining a new Casimir (other than the linear
case~$\Casi(\fv^0) = \fv^0$) from the semidirect sum was that~$\det \W_{(n)}
\ne 0$.  We go through the solvable cases and determine the Casimirs associated
with the semidirect extension, if any exist.

\subsubsection{n=1}

There is only one solvable extension, so upon appending a semidirect part we
have
\[
	\W_{(0)} = \l(\begin{array}{cc}
		\matone & \matzero \\
		\matzero & \matzero
	\end{array}\r),\ \ \ \
	\W_{(1)} = \l(\begin{array}{cc}
		\matzero & \matone \\
		\matone & \matzero
	\end{array}\r).
\]
Since~$\det \W_{(1)}\ne 0$, we expect another Casimir.  In fact this extension
is of the semidirect Leibniz type and has the same Casimir form as the~$n=2$
solvable Leibniz (\secref{Casleib}) extension.  Thus, the new Casimir is
just~$\fv^0\afsd(\fv^1)$.

\subsubsection{n=2}

Of the two possible extensions only the Leibniz one
satisfies~$\det \W_{(2)}\ne 0$.  The Casimir is thus
\[
	\CasiSD = \fv^0\afsd(\fv^2)+\frac{1}{2}(\fv^1)^2\afsd'(\fv^2).
\]

\subsubsection{n=3}

Cases~\ref{case:n4-01} and~\ref{case:n4-11} have a nonsingular~$\W_{(3)}$.
The Casimir for \caseref{n4-01} is
\[
	\CasiSD = \fv^0\afsd(\fv^3) + \fv^1\fv^2\afsd'(\fv^3),
\]
and for \caseref{n4-11} it is of the Leibniz form
\[
	\CasiSD = \fv^0\afsd(\fv^3) + \fv^1\fv^2\afsd'(\fv^3)
		+\frac{1}{3!}(\fv^2)^3\afsd''(\fv^3).
\]

\subsubsection{n=4}

Cases \ref{case:n4-00}b, \ref{case:n4-10}d, and \ref{case:n4-11}b have a
nonsingular~$\W_{(4)}$.  The Casimirs are shown in \tabref{Casimirsemn5}.

\begin{table}
\caption{Casimir invariants for semidirect extensions of order~$n=5$. These
extensions also possess the corresponding Casimir invariants in
\tabref{Casimirn4}.}
\tablabel{Casimirsemn5}
\vskip 1em

\begin{center}
\begin{tabular}{ll} \hline
Case & Invariant \\[6pt] \hline

%
%
1b & $\fv^0\afsd(\fv^4) + \l(\fv^1\fv^3
		+ \frac{1}{2}(\fv^2)^2\r)\afsd'(\fv^4)$
\\[9pt]

%
%
3d & $\fv^0\afsd(\fv^4) + \l(\fv^1\fv^2
		+ \frac{1}{2}(\fv^3)^2\r)\afsd'(\fv^4)
		+ \frac{1}{3!}(\fv^2)^3\afsd''(\fv^4)$
\\[9pt]

%
%
4b & $\fv^0\afsd(\fv^4) + \l(\fv^1\fv^3
		+ \frac{1}{2}(\fv^2)^2\r)\afsd'(\fv^4)
		+ \frac{1}{2}\fv^2(\fv^3)^2\afsd''(\fv^4)
		+ \frac{1}{4!}(\fv^3)^4\afsd'''(\fv^4)$
\\[9pt] \hline

\end{tabular}

\end{center}
\end{table}


%
%
%
%
%
%
%
%
%

\section{Discussion}
\seclabel{conclusion}

\ifmypprint
{\small
\begin{verbatim}
$Id: sec-conclusion.tex,v 2.1 1998/11/13 08:37:51 jeanluc Exp $
\end{verbatim}
}
\fi

Using the tools of Lie algebra cohomology, we have classified low-order
extensions.  We found that there were only a few normal forms for the
extensions, and that they involved no free parameters.  This is not expected
to carry over to higher orders~($n>4$).  The classification includes the
Leibniz extension, which is the maximal extension.  One of the normal forms is
the bracket appropriate to compressible reduced
MHD~\cite{Hazeltine1987,Hazeltine1985b}.

We then developed techniques for finding the Casimir invariants of
Lie--Poisson brackets formed from Lie algebra extensions.  We introduced the
concept of coextension, which allows one to explicitly write down the solution
of the Casimirs.  The coextension for the Leibniz extension can be found for
arbitrary order, so that we were able obtain the corresponding Casimirs in
general.

It would be interesting to generalize the classification scheme presented here
to a completely general form of extension
bracket~\cite{Morrison1980a,Nore1997}. Certainly the type of coordinate
transformations allowed would be more limited, and perhaps one cannot go any
futher than cohomology theory allows.

The interpretation of the Casimir invariants can be pushed further, both in a
mathematical and a physical sense. Mathematically, a precise geometrical
relation between cocycles and the form of the Casimirs could be
formulated. The cocycle and Casimirs should yield information about the
holonomy of the system. For this one must study extensions in the framework of
their principal bundle description~\cite{Azcarraga}. Physically we would like
to attach a more precise physical meaning to these conserved quantities. The
invariants associated with simultaneous eigenvectors can be regarded as
constraining the associated field variable to move with the fluid
elements~\cite{Morrison1987}. The field variable can also be interpreted as
partially labeling a fluid element. Some attempt has been made at formulating
the Casimir invariants of brackets in such a
manner~\cite{Kuznetsov1980,Thiffeault1998}, and an interpretation of cocycles
in the context of dynamical accessiblity has been
offered~\cite{Thiffeault1998diss}.

Sufficient conditions for stability can be obtained via the energy-Casimir
method~\cite{Hazeltine1984,Holm1985,Morrison1987}, or the related technique of
dynamical accessibility~\cite{Morrison1998,Arnold1966b}.  In both these case,
we can make use of the coextension to derive the stability conditions for
Lie--Poisson bracket extensions and a large class of
Hamiltonians~\cite{Thiffeault1998diss}.

\qcomment{Zeitlin truncations}

\begin{ack}

The authors thank Tom Yudichak for his comments and suggestions.  This work
was supported by the U.S. Department of Energy under contract
No.~DE-FG03-96ER-54346.  J-LT also acknowledges support from the Fonds pour
la Formation de Chercheurs et l'Aide \`a la Recherche du Canada.

\end{ack}

\appendix

%
%
%
%
%

\section{Proof of~$\W^{(1)}=I$}
\apxlabel{woneident}

\ifmypprint
{\small
\begin{verbatim}
$Id: apx-woneident.tex,v 2.1 1998/11/13 08:38:25 jeanluc Exp $
\end{verbatim}
}
\fi

Out goal is to demonstrate that through a series of lower-triangular
coordinate transformations we can make~$\W^{(1)}$ (which has an~$n$-fold
degenerate eigenvalue equal to unity) equal to the identity matrix, while
preserving the lower-triangular nilpotent form
of~\hbox{$\W^{(2)},\dots,\W^{(n)}$}.

We first show that we can always make a series of coordinate transformations
to make~\hbox{${\W_\lambda}^{11} = {\delta_\lambda}^1$}.  First note that if
the coordinate transformation~$\M$ is of the form~\hbox{$\M = I + L$},
where~$I$ is the identity and~$L$ is lower-triangular nilpotent,
then~\hbox{$\Wt^{(1)}=\M^{-1}\,\W^{(1)}\,\M$} still has eigenvalue~$1$, and
for~$\mu>1$ the~\hbox{$\Wt^{(\mu)}=\M^{-1}\,\W^{(\mu)}\,\M$} are still
nilpotent.

For~$\lambda>1$ we have
\begin{equation}
	{\overline\W_\lambda}^{11} = {\Wt_\lambda}{}^{11}
		+ {\Wt_\lambda}{}^{1\nu}\,{L_\nu}^1
	= {\Wt_\lambda}{}^{11}
		+ \sum_{\nu=2}^{\lambda-1}{\Wt_\lambda}{}^{1\nu}\,{L_\nu}^1
		+ {L_\lambda}^1,
	\eqlabel{apxcoordtrans}
\end{equation}
where we used~\hbox{${\Wt_\lambda}{}^{1\lambda}=1$}.  Owing to the triangular
structure of the set of equations~\eqref{apxcoordtrans} we can always solve
for the~\hbox{${L_\lambda}^1$} to make~${\overline\W_\lambda}^{11}$ vanish.
This proves the first part.

We now show by induction that if~\hbox{${\W_\lambda}^{11} =
{\delta_\lambda}^1$}, as proved above, then~$\W^{(1)}$ is the identity
matrix. For~\hbox{$\lambda = 1$} the result is trivial. Assume
that~\hbox{${\W_\mu}^{1\nu} = {\delta_\mu}^\nu$}, for~\hbox{$\mu <
\lambda$}. Setting two of the free indices to one, Eq.~\eqref{Wjacob} can be
written
\[
\begin{split}
	{\W_\lambda}^{\mu 1}\,{\W_\mu}^{1\sigma}
	&= {\W_\lambda}^{\mu\sigma}\,{\W_\mu}^{11}\\
	&= {\W_\lambda}^{\mu\sigma}\,{\delta_\mu}^{1}
	= {\W_\lambda}^{1\sigma}\,.
\end{split}
\]
Since~$\W^{(1)}$ is lower-triangular the index~$\mu$ runs from~$2$
to~$\lambda$ (since we are assuming~\hbox{$\lambda > 1$}):
\[
	\sum_{\mu = 2}^{\lambda}
	{\W_\lambda}^{\mu 1}\,{\W_\mu}^{1\sigma}
	= {\W_\lambda}^{1\sigma}\,,
\]
and this can be rewritten, for~$\sigma<\lambda$,
\[
	\sum_{\mu = 2}^{\lambda-1}
	{\W_\lambda}^{\mu 1}\,{\W_\mu}^{1\sigma}
	= 0\,.
\]
Finally, we use the inductive hypothesis
\[
	\sum_{\mu = 2}^{\lambda-1}
	{\W_\lambda}^{\mu 1}\,{\delta_\mu}^{\sigma}
	= {\W_\lambda}^{\sigma 1}= 0\,,
\]
which is valid for~$\sigma < \lambda$.  Hence,~\hbox{${\W_\lambda}^{\sigma 1}
= {\delta_\lambda}^{\sigma}$} and we have proved the result.
(${\W_\lambda}^{\lambda 1}$ must be equal to one since it lies on the diagonal
and we have already assumed degeneracy of eigenvalues.)

\bibliography{journals,books,articles}

\begin{thebibliography}{10}

\bibitem{Arnold}
V.~I. Arnold, {\em Mathematical Methods of Classical Mechanics}, 2nd  ed.
  (Springer-Verlag, New York, 1989).

\bibitem{Meacham1997}
S.~P. Meacham, G.~R. Flierl, and P.~J. Morrison, ``{H}amiltonian moment
  reduction for describing vortices in shear,'' Phys. Fluids {\bf 9},  2310
  (1997).

\bibitem{Morrison1980b}
P.~J. Morrison, ``The {M}axwell--{V}lasov equations as a continuous
  {H}amiltonian system,'' Phys. Lett. A {\bf 80},  383  (1980).

\bibitem{Marsden1982b}
J.~E. Marsden and A. Weinstein, ``The {H}amiltonian structure of the
  {M}axwell--{V}lasov equations,'' Physica D {\bf 4},  394  (1982).

\bibitem{Morrison1980a}
P.~J. Morrison and J.~M. Greene, ``Noncanonical {H}amiltonian density
  formulation of hydrodynamics and ideal magnetohydrodynamics,'' Phys. Rev.
  Lett. {\bf 45},  790  (1980).

\bibitem{Kuznetsov1980}
E.~A. Kuznetsov and A.~V. Mikhailov, ``On the topological meaning of canonical
  {C}lebsch variables,'' Phys. Lett. A {\bf 77},  37  (1980).

\bibitem{Morrison1982}
P.~J. Morrison,  in {\em Mathematical Methods in Hydrodynamics and
  Integrability in Dynamical Systems}, La Jolla Institute, No.~88 in {\em AIP
  Conference Proceedings}, edited by M. Tabor and Y.~M. Treve (American
  Institute of Physics, New York, 1982), pp.\ 13--46.

\bibitem{Olver1982}
P.~J. Olver, ``A nonlinear {H}amiltonian structure for the {E}uler equations,''
  J. Math. Anal. Appl. {\bf 89},  233  (1982).

\bibitem{Marsden1983}
J.~E. Marsden and A. Weinstein, ``Coadjoint orbits, vortices and {C}lebsch
  variables for incompressible fluids,'' Physica D {\bf 7},  305  (1983).

\bibitem{Nore1997}
C. Nore and T.~G. Shepherd, ``A {H}amiltonian weak-wave model for shallow-water
  flow,'' Proc. R. Soc. Lond. A {\bf 453},  563  (1997).

\bibitem{Holmes1983}
P.~J. Holmes and J.~E. Marsden, ``Horseshoes and {A}rnold diffusion for
  {H}amiltonian systems on {L}ie groups,'' Indiana Univ. Math. J. {\bf 32},
  273  (1983).

\bibitem{Morrison1984}
P.~J. Morrison and R.~D. Hazeltine, ``{H}amiltonian formulation of reduced
  magnetohydrodynamics,'' Phys. Fluids {\bf 27},  886  (1984).

\bibitem{Benjamin1984}
T.~B. Benjamin, ``Impulse, flow force, and variational principles,'' IMA J.
  Appl. Math. {\bf 32},  3  (1984).

\bibitem{McLachlan1997}
R.~I. McLachlan, I. Szunyogh, and V. Zeitlin, ``{H}amiltonian
  finite-dimensional models of baroclinic instability,'' Phys. Lett. A {\bf
  229},  299  (1997).

\bibitem{Hazeltine1985}
R.~D. Hazeltine, D.~D. Holm, and P.~J. Morrison, ``Electromagnetic solitary
  waves in magnetized plasmas,'' J. Plasma Physics {\bf 34},  103  (1985).

\bibitem{Kuvshinov1994}
B.~N. Kuvshinov, F. Pegoraro, and T.~J. Schep, ``{H}amiltonian formulation of
  low-frequency, nonlinear plasma dynamics,'' Phys. Lett. A {\bf 191},  296
  (1994).

\bibitem{Hazeltine1987}
R.~D. Hazeltine, C.~T. Hsu, and P.~J. Morrison, ``{H}amiltonian four-field
  model for nonlinear tokamak dynamics,'' Phys. Fluids {\bf 30},  3204  (1987).

\bibitem{Trofimov}
V.~V. Trofimov, {\em Introduction to Geometry of Manifolds with Symmetry}
  (Kluwer, Dordrecht, 1994).

\bibitem{Kuroda1991}
Y. Kuroda, ``On the {C}asimir invariants of {H}amiltonian fluid mechanics,'' J.
  Phys. Soc. Japan {\bf 60},  727  (1991).

\bibitem{Hernandez1998}
B. Hern\'{a}ndez-Bermejo and V. Fair\'{e}n, ``Simple evaluation of {C}asimir
  invariants in finite-dimensional {P}oisson systems,'' Phys. Lett. A {\bf
  241},  135  (1998).

\bibitem{Thiffeault1998}
J.-L. Thiffeault and P.~J. Morrison, ``Invariants and labels in
  {L}ie--{P}oisson systems,'' Ann. N. Y. Acad. Sci. {\bf 867},    (1998), also
  available as IFSR \#815.

\bibitem{Hazeltine1984}
R.~D. Hazeltine, D.~D. Holm, J.~E. Marsden, and P.~J. Morrison, ``Generalized
  {P}oisson brackets and nonlinear {L}iapunov stability --- application to
  reduced {MHD},'' ICPP Proc. (Lausanne) {\bf 2},  204  (1984), also available
  as Institute for Fusion Studies report IFSR \#139.

\bibitem{Holm1985}
D.~D. Holm, J.~E. Marsden, T. Ratiu, and A. Weinstein, ``Nonlinear stability of
  fluid and plasma equilibria,'' Physics Reports {\bf 123},  1  (1985).

\bibitem{Morrison1986}
P.~J. Morrison and S. Eliezer, ``Spontaneous symmetry breaking and neutral
  stability in the noncanonical {H}amiltonian formalism,'' Phys. Rev. A {\bf
  33},  4205  (1986).

\bibitem{Morrison1987}
P.~J. Morrison, ``Variational principle and stability of nonmonotonic
  {V}lasov--{P}oisson equilibria,'' Z. Naturforsch {\bf 42a},  1115  (1987).

\bibitem{Morrison1998}
P.~J. Morrison, ``{H}amiltonian description of the ideal fluid,'' Rev. Modern
  Phys. {\bf 70},  467  (1998).

\bibitem{Chevalley1948}
C. Chevalley and S. Eilenberg, ``Cohomology theory of {L}ie groups and {L}ie
  algebras,'' Trans. Amer. Math. Soc. {\bf 63},  85  (1948).

\bibitem{Knapp}
A.~W. Knapp, {\em {L}ie Groups, {L}ie Algebras, and Cohomology} (Princeton
  University Press, Princeton, N.J., 1988).

\bibitem{Azcarraga}
J.~A. {de Azc\'arraga} and J.~M. Izquierdo, {\em {L}ie Groups, {L}ie Algebras,
  Cohomology and Some Applications in Physics} (Cambridge University Press,
  Cambridge, U.K., 1995).

\bibitem{Marsden1974}
J.~E. Marsden and A. Weinstein, ``Reduction of symplectic manifolds with
  symmetry,'' Rep. Math. Phys. {\bf 5},  121  (1974).

\bibitem{AbrahamMarsden}
R. Abraham and J.~E. Marsden, {\em Foundations of Mechanics}, 2nd  ed.
  (Benjamin/Cummings, Reading, Mass., 1978).

\bibitem{Marsden1982}
J.~E. Marsden, T. Ratiu, and A. Weinstein, ``Semidirect products and reductions
  in mechanics,'' Trans. Amer. Math. Soc. {\bf 281},  147  (1984).

\bibitem{Guillemin}
V. Guillemin and S. Sternberg, {\em Symplectic Techniques in Physics}
  (Cambridge University Press, Cambridge, U.K., 1984).

\bibitem{MarsdenRatiu}
J.~E. Marsden and T.~S. Ratiu, {\em Introduction to Mechanics and Symmetry}
  (Springer-Verlag, Berlin, 1994).

\bibitem{Newcomb1967}
W.~A. Newcomb, ``Exchange invariance in fluid systems,'' Proc. Symp. Appl.
  Math. {\bf 18},  152  (1967).

\bibitem{Bretherton1970}
F.~P. Bretherton, ``A note on {H}amilton's principle for perfect fluids,'' J.
  Fluid Mech. {\bf 44},  19  (1970).

\bibitem{Padhye1996a}
N. Padhye and P.~J. Morrison, ``Fluid element relabeling symmetry,'' Phys.
  Lett. A {\bf 219},  287  (1996).

\bibitem{Lie}
S. Lie, {\em Theorie der Transformationsgruppen}, 2nd  ed. (B. G. Teubner,
  Leipzig, 1890), reprinted by Chelsea, New York (1970).

\bibitem{Berezin1967}
F.~A. Berezin, ``Some remarks about the associated envelope of a {L}ie
  algebra,'' Func. Anal. Appl. {\bf 1},  91  (1967).

\bibitem{Arnold1966a}
V.~I. Arnold, ``Sur la g\'eom\'etrie diff\'erentielle des groupes de {L}ie de
  dimension infinie et ses applications \`a l'hydrodynamique des fluides
  parfaits,'' Ann. Inst. Fourier, Grenoble {\bf 16},  319  (1966).

\bibitem{Kirillov1962}
A.~A. Kirillov, ``Unitary representations of nilpotent {L}ie groups,'' Russian
  Math. Surveys {\bf 17},  53  (1962).

\bibitem{Kostant1966}
B. Kostant, ``Orbits, symplectic structures and representation theory,'' Proc.
  US--Japan Seminar on Diff. Geom., Kyoto. Nippon Hyronsha, Tokyo, 77  (1966).

\bibitem{Souriau}
J.-M. Souriau, {\em Structure des Syst\`emes Dynamiques} (Dunod, Paris, 1970).

\bibitem{AMR}
R. Abraham, J.~E. Marsden, and T. Ratiu, {\em Manifolds, Tensor Analysis, and
  Applications}, 2nd  ed. (Springer-Verlag, New York, 1988).

\bibitem{Audin}
M. Audin, {\em Spinning Tops} (Cambridge University Press, Cambridge, U.K.,
  1996).

\bibitem{Strauss1977}
H.~R. Strauss, ``Dynamics of high $\beta$ tokamaks,'' Phys. Fluids {\bf 20},
  1354  (1977).

\bibitem{Zeitlin1992}
V. Zeitlin, ``On the structure of phase-space, {H}amiltonian variables and
  statistical approach to the description of two-dimensional hydrodynamics and
  magnetohydrodynamics,'' J. Phys. A {\bf 25},  L171  (1992).

\bibitem{Hazeltine1985b}
R.~D. Hazeltine, M. Kotschenreuther, and P.~J. Morrison, ``A four-field model
  for tokamak plasma dynamics,'' Phys. Fluids {\bf 28},  2466  (1985).

\bibitem{CecileII}
Y. Choquet-Bruhat and C. {DeWitt-Morette}, {\em Analysis, Manifolds, and
  Physics. Part {II}: 92 Applications} (Elsevier, Amsterdam, 1982).

\bibitem{Weiss}
E. Weiss, {\em Cohomology of Groups} (Academic Press, New York, 1969).

\bibitem{Marsden1984}
J.~E. Marsden and P.~J. Morrison, ``Noncanonical {H}amiltonian field theory and
  reduced {MHD},'' Contemp. Math. {\bf 28},  133  (1984).

\bibitem{Suprunenko}
D.~A. Suprunenko and R.~I. Tyshkevich, {\em Commutative Matrices} (Academic
  Press, New York, 1968).

\bibitem{Jacobson}
N. Jacobson, {\em {L}ie Algebras} (Dover, New York, 1962).

\bibitem{Parthasarathy1976}
K.~R. Parthasarathy and K. Schmidt, ``A new method for constructing
  factorisable representations for current groups and current algebras,''
  Commun. Math. Phys. {\bf 50},  167  (1976).

\bibitem{Thiffeault1998diss}
J.-L. Thiffeault, Ph.D. thesis, The University of Texas at Austin, 1998,
  available as Institute for Fusion Studies Report \#847.

\bibitem{Osta}
A. Ostaszewski, {\em Advanced Mathematical Methods} (Cambridge University
  Press, Cambridge, U.K., 1990).

\bibitem{Arnold1966b}
V.~I. Arnold, ``Sur un principe variationnel pour les \'ecoulements
  stationaires des liquides parfaits et ses applications aux probl\`emes de
  stabilit\'e non lin\'eaires,'' Journal de M\'ecanique {\bf 5},  29  (1966).

\end{thebibliography}

\end{document}